\documentclass[preprint,aps,11pt,eqsecnum,nofootinbib]{revtex4}
\usepackage{mathtools}
\usepackage{graphics}
\usepackage{graphicx}
\usepackage{dcolumn}
\usepackage{bm}
\usepackage{dsfont} 
\usepackage{amsmath,amssymb}
\usepackage{hyperref}
\usepackage{tabularx}
\usepackage{epstopdf}
\usepackage[usenames]{color}
\hypersetup{
   colorlinks=true,
    linkcolor=blue,
   filecolor=magenta,      
    urlcolor=blue,
    citecolor=blue
}

\newcommand\be{\begin{equation}}
\newcommand\ba{\begin{eqnarray}}
\newcommand\ee{\end{equation}}
\newcommand\ea{\end{eqnarray}}
\newcommand\bw{\begin{widetext}}
\newcommand\ew{\end{widetext}}

\newcommand{\nn}{\nonumber}

\newcommand{\tmop}[1]{\ensuremath{\operatorname{#1}}}

 \hypersetup{breaklinks=true}

\begin{document}
\title{
Gravitational wave energy-momentum tensor and radiated power in a strongly curved background}

\author{Yuchen Du}
\affiliation{
 Department of Physics, University of Virginia, Charlottesville, Virginia 22904, USA
}

\author{Diana Vaman}
\affiliation{
 Department of Physics, University of Virginia, Charlottesville, Virginia 22904, USA
}

\author{Kent Yagi}
\affiliation{
 Department of Physics, University of Virginia, Charlottesville, Virginia 22904, USA
}

\date{\today}

\begin{abstract}Allowing for the possibility of extra dimensions, there are two paradigms: either the extra dimensions are hidden from observations by being compact and small as in Kaluza-Klein scenarios,  or the extra dimensions are large/non-compact and undetectable due to a large warping as in the Randall-Sundrum scenario.  In the latter case, the five-dimensional background has a large curvature, and Isaacson's  construction of the gravitational energy-momentum tensor, which relies on the assumption that the wavelength of the metric fluctuations is much smaller than the curvature length of the background spacetime,  cannot be used.  In this paper, we construct  the gravitational energy-momentum tensor in a strongly curved background such as Randall-Sundrum. We perform a scalar-vector-tensor decomposition of the metric fluctuations with respect to the $SO(1,3)$ background isometry  and construct the covariantly-conserved gravitational energy-momentum tensor out of the gauge-invariant metric fluctuations.  We give a formula for the power radiated by gravitational waves and verify it in known cases.  In using the gauge-invariant metric fluctuations to construct   the gravitational energy-momentum tensor we follow  previous work  done in cosmology.  Our framework has applicability beyond the Randall-Sundrum model.  
\end{abstract}

\maketitle
{
   \hypersetup{linkcolor=blue}
   \tableofcontents

}

\section{Introduction}

In this paper, we address the energy-momentum tensor of the gravitational waves in the context of a strongly curved background.
Historically, gravitational waves were understood as ripples across spacetime, with the wavelength of the ripples much smaller than the curvature length of the background.  The most commonly used definition of the energy-momentum tensor of the gravitational waves is due to Isaacson~\cite{Isaacson:1968hbi,Isaacson:1968zza}. In a couple of seminal papers, Isaacson performed an expansion of the Einstein equations  to the lowest order in non-linearities and interpreted the terms quadratic in fluctuations as an energy-momentum source due to the gravitational field, backreacting on the spacetime geometry.  With the image of ripples propagating across spacetime implying a separation of scales between the high-frequency gravitational waves and the large scale on which the background is changing,  Isaacson added an averaging to his definition of the energy-momentum tensor
\be
\langle T_{\mu\nu}(x) \rangle_{\rm I}=\int d^d x' \sqrt{g} f(x,x') v_\mu^{\mu'}(x,x') v_\nu^{\nu'}(x,x') T_{\mu'\nu'}(x')\,,\label{isaacsonav}
\ee
where the integration region is defined by the choice of the compact support function $f(x,x')$, centered at $x$,  such that it has a characteristic size smaller than the curvature scale of the background,  but larger than the wavelength of the radiation.  Furthermore,  in order for the outcome of the integration to be a tensor,  the integrand $T_{\mu'\nu'}$ needs to be contracted with the bitensors $v_{\mu}^{\mu'}$ and $v_{\nu}^{\nu'}$ which transform as vectors under coordinate transformations performed at either $x$ or $x'$.  On the one hand, the small wavelength assumption means that covariant derivatives commute.  On the other hand, the averaging \eqref{isaacsonav}  brings with it the freedom to perform integration by parts\footnote{ E.g.  $\langle h_{\mu\sigma}{}^{;\rho} h_{\nu \rho}{}^{;\sigma} \rangle_{\rm I} = -\langle h_{\mu\sigma} h_{\nu\rho}{}^{;\sigma\rho}\rangle_{\rm I}$. After such integrations by parts are performed,  further simplifications arise as a result of either applying  a gauge fixing condition or through the use of the equations of motion. }. Together they imply that the simplified expression of the (quasi-local) energy-momentum tensor  
\be 
\langle T_{\mu\nu}\rangle_{\rm I}=\frac{1}{4} \langle h_{\rho\sigma;\mu} h^{\rho\sigma}{}_{;\nu}\rangle_{\rm I}\label{i1}
\ee
 is  background-covariantly conserved in vacuum,  and gauge-invariant, which, of course, is a desired feature of any definition of the energy-momentum tensor.  Another definition of the gravitational field energy-momentum tensor $T^{\mu\nu}_{\rm{LL}}$ which is due to Landau and Lifshitz \cite{LL} has the advantage of being conserved $\partial_{\mu} T^{\mu\nu}_{\rm{LL}}=0$. However, it suffers from two major drawbacks: it is not a tensor (hence it is often referred to as the Landau-Lifshitz pseudo-tensor) and it is not gauge-invariant.   

There is at least one situation of interest when the approximations used by Isaacson are not applicable, that is, gravitational waves in extra dimensions. Gravitational waves from binary black holes and neutron stars detected by LIGO and Virgo have been used to test strong-field gravity in various ways~\cite{LIGOScientific:2016lio,Yunes:2016jcc,LIGOScientific:2019fpa,LIGOScientific:2021sio,Berti:2018cxi,Berti:2018vdi}. One  such test is to probe the existence of extra dimensions. For example, the presence of a large extra dimension modifies how the gravitational wave amplitude falls off as a function of the distance it traveled, so the luminosity distance measured by gravitational waves would be inconsistent with those from electromagnetic counterparts if one assumes 4d General Relativity~\cite{LIGOScientific:2018dkp,Deffayet:2007kf,Pardo:2018ipy}. In~\cite{Du:2020rlx} we studied a simple Kaluza-Klein model and showed that the luminosity of gravitational waves emitted from a binary black hole is smaller than that of the 4d case, leading to a relatively large phase shift that is inconsistent with observations.

One paradigm of large extra dimensions is the Randall-Sundrum model,  with a 3-brane curving the 5d spacetime around it until it looks like a slab of Anti-de Sitter space~\cite{Randall:1999vf}\footnote{{Other applications of gravitational waves in Randall-Sundrum model have been studied e.g. in~\cite{Yagi:2011yu,Yu:2016tar,Visinelli:2017bny}.}}.  Matter sources are localized on the brane. The background geometry has $SO(1,3)$ isometry and the fifth dimension is warped
\be 
ds^2 = \exp(-2\kappa |y|) dx^\mu dx^\nu \eta_{\mu\nu}+d y^2\,,\label{rs}
\ee
with $\kappa$ proportional to the inverse curvature length.
Gravity is localized near the brane and deviations from four-dimensional Newtonian potential are parametrized in terms of $\kappa$ as~\cite{Randall:1999vf,Garriga:1999yh} 
\be
V_{{\rm N}}(r)= - \frac{GM}{r}\bigg(1+\frac{2}{3\kappa^2 r^2}\bigg).
\ee
Given that the Newtonian potential has been probed by Cavendish-type torsion scale experiments to micrometer scale~\cite{Adelberger:2006dh,Kapner:2006si} this means that the curvature length of the background  must be smaller than this scale.  We are thus looking at a strongly  curved background,  where its curvature length is smaller than the typical wavelength of the gravitational waves generated by a black-hole binary source,  which is in the $10^2-10^4$ km range.  
In this case,  the underlying assumptions behind the well-known formula \eqref{i1} are no longer valid and we need  a new approach.

Our paper proposes a definition of the energy-momentum tensor of the gravitational field which does not rely on the WKB approximation nor the spatial averaging introduced by Isaacson,  which cannot be used in the case of strongly curved  backgrounds\footnote{See e.g.~\cite{Stein:2010pn,Saffer:2017ywl} for other works on computing the energy-momentum tensor for gravitational waves in theories beyond General Relativity using the Isaacson averaging.}.  Instead, we are constructing the energy-momentum tensor from gauge-invariant metric fluctuations. Our procedure is similar to~\cite{Abramo:1997hu}, which dealt with the energy-momentum tensor for cosmological perturbations, though how one foliates the spacetime and decomposes the metric fluctuations is different.  Namely, within the cosmological context, it is natural to foliate the 4d spacetime with fixed-time 3d spatial slices.  Since we are interested in the case of the extra dimensions,  for one large extra dimension, $y$,  we foliate the 5d spacetime with fixed-$y$ 4d spacetime slices, and we similarly decompose the metric fluctuations in scalar-vector-tensors (SVT) with respect to the $SO(1,3)$ isometry group.  The gravitational energy-momentum tensor is constructed out of the gauge-invariant fluctuations. The resulting expression is manifestly gauge-invariant, and it is background-covariantly conserved. Without the benefit of the averaging procedure, the expression is quite involved.  Nonetheless, it can be simplified significantly when computing the radiated power (radiated energy per unit time) asymptotically, far away from the sources.

In studying physical problems in the Randall-Sundrum model, a common approach is to use the reduced 4d Einstein's equation supplemented by the Israel's junction condition.  E.g. in \cite{Shiromizu:1999wj, Sasaki:1999mi, Kinoshita:2005nx, Maartens:2010ar, Garcia-Aspeitia:2013jea, Garcia-Aspeitia:2020snv} the 5d effects are encoded in various additional terms in the 4d reduced Einstein's equation, relative to the usual one,   and the additional junction condition. This method treats the brane and the bulk differently and it can be effective when studying problems on the brane.  However, separating the brane from the bulk seems less appropriate when studying the propagation of the gravitational waves, which propagate equally on the brane and into the bulk. The physical picture can be murky since the meaning of those additional terms in the reduced equation is not very intuitive. 
In the literature,  in order to solve the reduced 4d equations, various  terms are  dropped for practical purposes, though the reason behind this is not often clear. Our work comes directly from a 5d setup which treats the bulk and the brane on an equal footing, and has a clear physical picture. The gravitational energy-momentum tensor we calculated here can be used in  applications other than computing the radiated power.  Lastly, due to the gauge-invariant nature of the method we used here, our work has a larger applicability outside the Randall-Sundrum set-up. 

The paper is organized as follows.  In Section \ref{sec:GW_EM_tensor} we give our main formulae for the gravitational energy-momentum tensor \eqref{2nd}, and radiated power \eqref{radpow0} and \eqref{radpow}. In Section \ref{sec:GW} we discuss gravitational waves in a curved maximally symmetric spacetime such as Anti-de  Sitter and in Randall-Sundrum geometry.  In Section \ref{sec:GW_EM_gauge_inv} we perform various SVT decompositions for the metric fluctuations in 4d and 5d flat spacetimes, and in the 5d Randall-Sundrum background. In each case, we construct the gravitational energy-momentum tensor and  give a formula for the radiated power. As we will see, one of the main differences with respect to previous results in the literature \cite{Cai:2021jbi} is that the radiated power is expressed not only in terms of the tensor metric fluctuations. This is to be expected given how the graviton degrees of freedom are accounted for in the various SVT decompositions.  
In Appendix \ref{only tensor for non-compact geometry} we are explicitly verifying that our approach yields the expected result for the power radiated by gravitational waves away from a binary source in flat 4d spacetime.  Appendix \ref{allmodescompact} deals with a compact extra dimension flat 5d scenario and recovers a previous result for the radiated power, using the approach presented in Section  \ref{sec:GW_EM_tensor}. Technical details are relegated to the other appendices.

\section{Gravitational wave energy-momentum tensor in a curved background}
\label{sec:GW_EM_tensor}

We begin by considering  a curved background $\bar g_{\mu\nu}$, with a non-vanishing cosmological constant $\Lambda$, solution to the source-free Einstein equations
\ba
\bar G_{\mu\nu}+\bar g_{\mu\nu}\Lambda \equiv \bar R_{\mu\nu}-\frac 12 \bar g_{\mu\nu} (\bar R-2\Lambda)=0.
\ea
Next consider another metric,
\ba
g_{\mu\nu}=\bar g_{\mu\nu}+h_{\mu\nu},\label{one}
\ea
which is a  solution to the source-free Einstein equations as well:
\ba
G_{\mu\nu}+ g_{\mu\nu}\Lambda\equiv R_{\mu\nu}-\frac 12  g_{\mu\nu} (R-2\Lambda)=0.
\ea
Note that \eqref{one} is exact, in other words $h_{\mu\nu}$ represents the difference between two spacetime metrics. Expanding in $h_{\mu\nu}$ leads to the following definition of the energy-momentum tensor ${\cal T}_{\mu\nu}$ of the gravitational field   \cite{Abbott:1981ff}
\ba
\delta^{(1)} G_{\mu\nu}+\Lambda h_{\mu\nu}= -\bigg(\delta^{(2)} G_{\mu\nu}+\delta^{(3)} G_{\mu\nu}+{\cal O}(h^4)\bigg)\equiv{\cal T}_{\mu\nu}\label{deltag}
\ea
where $\delta^{(1)} G_{\mu\nu}$ is linear in the difference between the two metrics $h_{\mu\nu}$, $\delta^{(2)} G_{\mu\nu}$ is quadratic, 
 etc.

The linearized Einstein tensor evaluates to:
\ba
&&\delta^{(1)}G_{\mu\nu}=\delta^{(1)} R_{\mu\nu}-\frac 12 \bar g_{\mu\nu} \delta^{(1)} R_{\rho\sigma} \bar g^{\rho\sigma}+\frac 12 \bar g_{\mu\nu} \bar R_{\rho\sigma} h^{\rho\sigma}-\frac 12 h_{\mu\nu} \bar R 
\nn\\
&=&\frac 12\bigg(-\bar\Box h_{\mu\nu}-h_{;\mu;\nu}+h_{\rho\nu;\mu}{}^{;\rho}+h_{\rho\mu;\nu}{}^{;\rho}-\bar g_{\mu\nu}(
-\bar\Box h+h_{\rho\sigma}{}^{;\rho;\sigma}) +\frac{2\Lambda}{d-2}(\bar g_{\mu\nu} h - d h_{\mu\nu})\bigg)\,,\nn\\\label{eq2.5}
\ea
where {$d$ is the number of spacetime dimension and} we used that
\ba
\bar R_{\mu\nu}=\frac{2}{d-2}\Lambda\bar g_{\mu\nu}\,.\label{barricci}
\ea

Reshuffling the background-covariant derivatives and using \eqref{barricci},
one can show that
\ba
\bar\nabla^\mu \bigg(\delta^{(1)} G_{\mu\nu}+\Lambda h_{\mu\nu}\bigg)=0\,,\label{g1id}
\ea
for any two-index symmetric tensor $h_{\mu\nu}$.

This in turn implies that ${\cal T}_{\mu\nu}$ is a background-conserved tensor \cite{Abbott:1981ff}
\ba
\bar\nabla^\mu{\cal T}_{\mu\nu}=0\,.
\ea
Not only that, but ${\cal T}_{\mu\nu}$ is invariant under background linearized gauge transformations $\delta h_{\mu\nu}=\bar\nabla_{\mu} \xi_\nu+\bar\nabla_{\nu}\xi_\mu$ since the left hand side of \eqref{deltag} is invariant under these transformations.

Furthermore consider a background that admits a time-like Killing vector (e.g. for the RS model, such a Killing vector would be $\partial_0=k^\mu\partial_\mu$)
\ba
\bar\nabla_\mu k_\nu + \bar\nabla_\nu k_\mu=0\,.
\ea
Then 
\ba
V^\mu\equiv {\cal T}^{\mu\nu}k_\nu
\ea
is a background-conserved vector
\ba
\bar \nabla_\mu  V^\mu=0\,\label{consv}.
\ea
This implies a conservation law:
\ba
0=\int d^d x \sqrt{-\bar g} \,\bar\nabla_\mu V^\mu =\int d^d x \partial_\mu(\sqrt{-\bar g} \, V^\mu).
\ea

The presence of sources alters slightly the previous scenario. 
From
\ba
\delta^{(1)} G_{\mu\nu}+\Lambda h_{\mu\nu}= T_{\mu\nu}-\bigg(\delta^{(2)} G_{\mu\nu}+\delta^{(3)} G_{\mu\nu}+{\cal O}(h^4)\bigg)\equiv T_{\mu\nu}+{\cal T}_{\mu\nu}\label{deltag1},
\ea
{where $T_{\mu\nu}$ is the matter energy-momentum tensor,} using \eqref{g1id} we find 
the conservation law obeyed by the total (sources plus gravitational field) energy-momentum tensor 
\be
\bar \nabla^\mu({\cal T}_{\mu\nu}+T_{\mu\nu})=0.
\ee
Given a time-like Killing vector $k_\mu$, one can construct a conserved current
\be
V^\mu\equiv k_\nu ( T^{\mu\nu}+{\cal T}^{\mu\nu}), \qquad \bar\nabla_\mu V^\mu=0.
\ee
The total energy in some region of space $M$ is 
\be
E= \int_M d^{d-1} x \sqrt{-\bar g} V^0=\int_M d^{d-1} x \sqrt{-\bar g} k_\mu (T^{0\mu}+{\cal T}^{0\mu}).\label{energy1}
\ee
The rate of change of the energy in this region of space can be expressed in terms of the flux of $V^i$ through the boundary:
\be
\frac {dE}{dt}=\int_M  d^{d-1} x\partial_{0}(\sqrt{-\bar g} V^0)=-\int_M  d^{d-1} x \partial_i (\sqrt{-\bar g} V^i)=-\int_{\partial M} d^{d-2}x \sqrt{-\bar g} n_i V^i
\ee
where $n^i$ is an outward pointing, unit vector on the boundary. 

If there are no sources on the boundary of the spatial region $M$, then the radiated power through the boundary $\partial M$ is given by:
\ba
P=-\frac{dE}{dt}=\int_{\partial M}  d^{d-2}x \sqrt{-\bar g} n_i k_\mu {\cal T}^{\mu i}.\label{radpow0}
\ea

Furthermore, assuming that the sources are generating gravitational waves and that the period of the gravitational waves is $T$,  we will compute the averaged radiated power through the boundary $\partial M$ which we take to be asymptotically far away from all sources. Thus
\ba
\langle P\rangle = \frac 1T \int_{0}^T dt\int_{\partial M} d^{d-2}x \sqrt{-\bar g} n_i k_\mu {\cal T}^{\mu i}.\label{radpow}
\ea

This expression is background gauge-independent since as we have already discussed ${\cal T}_{\mu\nu}$ is invariant under background gauge transformations.

In general, though, we are interested in problems where the metric $g_{\mu\nu}$ is a small perturbation of some exact background metric, due to sources and gravitational waves. Then the metric $g_{\mu\nu}=\bar g_{\mu\nu}+h_{\mu\nu}$ is typically solved in perturbation theory,  with $\bar g_{\mu\nu}$ an exact background, and with $h_{\mu\nu}$ expanded in a perturbative series
\ba
h_{\mu\nu}=\epsilon h^{(1)}_{\mu\nu}+\epsilon^2 h^{(2)}_{\mu\nu}+\dots\,,
\ea
where $\epsilon$ is some small expansion parameter (e.g. in thinking about the gravitational waves sourced by a binary the small parameter could be the post-Newtonian expansion parameter,  $\epsilon = |\vec v|/c$ where $\vec v$ is the velocity of binary sources).
Then,  the Einstein equation can be solved order by order in $\epsilon$. To the lowest orders in perturbation theory,  setting the sources to zero for clarity, we have
\ba
&&\delta^{(1)} G_{\mu\nu}[h^{(1)}]-\Lambda h^{(1)}_{\mu\nu}=\frac 12\bigg(-\bar\Box h^{(1)}_{\mu\nu}-h^{(1)}_{;\mu;\nu}+h^{(1)}_{\rho\nu;\mu}{}^{;\rho}+h^{(1)}_{\rho\mu;\nu}{}^{;\rho}-\bar g_{\mu\nu}(
-\bar\Box h^{(1)}+h^{(1)}_{\rho\sigma}{}^{;\rho;\sigma})\bigg)-\Lambda h^{(1)}_{\mu\nu}\nn\\
&&\qquad\qquad\qquad\qquad\;\;\;\;=0\,,
\\
&&\delta^{(1)} G_{\mu\nu}[h^{(2)}]-\Lambda h^{(2)}_{\mu\nu}=-\delta^{(2)} G_{\mu\nu}[h^{(1)}]\,.\label{delta2G}
\ea
One way to interpret the equation \eqref{delta2G} is that the metric fluctuation $h^{(1)}_{\mu\nu}$, solution to the linearized equation of motion, backreacts on the background geometry, with the right-hand side of \eqref{delta2G} playing the role of an energy-momentum tensor source:
\ba
\epsilon^2 {\cal T}_{\mu\nu}=-\epsilon^2 \delta^{(2)} G_{\mu\nu}[h^{(1)}]+{\cal O}(\epsilon^3)\,.
\ea
Using the results derived in  Appendix \ref{appendixa}, to leading order in $\epsilon$, the energy-momentum tensor of the gravitational field  takes the form\footnote{It it important that in solving for $h^{(1)}_{\mu\nu}$ we consistently keep all the terms of the same order in $\epsilon$. For example, in solving for fluctutations sourced by a binary to leading order in the velocity expansion in \cite{Du:2020rlx},   the spatial fluctuations $h^{(1)}_{IJ}$ received two contributions,  both of the same order in $\epsilon$: one  contribution from the linearized Einstein equation sourced by the matter energy-momentum tensor and a second contribution,   from the backreaction of the Coulombic part of $h^{(1)}_{00}$.}
\ba
{\cal T}_{\mu\nu}&=&-\frac 12 h^{(1)}{}^{\alpha\beta}(h^{(1)}_{\mu\nu;\alpha;\beta}-h^{(1)}_{\nu\alpha;\mu;\beta}
-h^{(1)}_{\mu\alpha;\nu;\beta}+h^{(1)}_{\alpha\beta;\mu;\nu})
+\frac 12 h^{(1)}_{\nu\beta;\alpha}(h^{(1)}_\mu{}^{\alpha;\beta}-h^{(1)}_\mu{}^{\beta;\alpha})\nn\\
&&-\frac 14 h^{(1)}_{\alpha\beta;\mu}h^{{(1)}\alpha\beta}{}_{;\nu}-\frac 14 (h^{(1)}_{\nu\alpha;\mu}+h^{(1)}_{\mu\alpha;\nu})(h^{{(1)};\alpha}-2 h^{(1)}_{\beta}{}^{\alpha}{}^{;\beta})+\frac 14 h^{(1)}_{\mu\nu}{}^{;\alpha}(h^{(1)}{}_{;\alpha} -2
h^{(1)}_{\alpha\beta}{}^{;\beta})\nn\\
&&+\frac 14 \bar g_{\mu\nu} \bigg(h^{{(1)}\alpha\beta}(h^{(1)}{}_{;\alpha;\beta}+h^{(1)}_{\alpha\beta}{}^{;\gamma}{}_{;\gamma}-2h^{(1)}_{\alpha \gamma}{}^{;\gamma}{}_{;\beta}
)
-\frac 12 h^{(1)}{}_{;\alpha} h^{{(1)};\alpha}-2h^{(1)}_{\alpha\beta}{}^{;\alpha} h^{{(1)}\beta\gamma}{}_{;\gamma} \nn\\
&&+2 h^{(1)}{}_{;\alpha}    h^{{(1)}\alpha\beta}{}_{;\beta}    -h^{(1)}_{\alpha\gamma;\beta} h^{{(1)}\alpha\beta;\gamma}
+
\frac 32 h^{(1)}_{\alpha\beta;\gamma}h^{(1)\alpha\beta;\gamma}   \bigg)\,.\label{2nd}
\ea
 Given that ${\cal T}_{\mu\nu}$ is the right-hand side of \eqref{delta2G}, the same argument of Abbott and Deser \cite{Abbott:1981ff}, which we reviewed earlier,  applies:  the gravitational energy-momentum tensor \eqref{2nd} is background covariantly-conserved. 
However, due to the perturbative expansion we have just performed,  this expression is no longer invariant under background gauge transformations.  As noticed by \cite{Abramo:1997hu},   we can remedy this: by using only the gauge-invariant pieces of the metric fluctuation $h_{\mu\nu}^{(1)}$,  the gravitational energy-momentum tensor defined in \eqref{2nd} becomes manifestly gauge-invariant. We will elaborate on this in the next sections.

We can compare \eqref{2nd}  with known expressions of the energy-momentum tensor in flat space: replace all the background covariant derivatives with partial derivatives, choose the Lorenz gauge $\partial_\mu h^{(1)\mu\nu}=0$, and fix the remaining gauge freedom by setting $h^{(1)}=0$.  If an averaging is performed as in \cite{Isaacson:1968zza} then one can do integration by parts to take advantage of the gauge choice. Lastly, using the equation of motion of the linearized, gauge-fixed fluctuations $\Box h_{\mu\nu}^{(1)}=0$, the energy-momentum tensor simplifies to 
\ba
{\cal T}_{\mu\nu}=\frac 14\langle  {h^{(1)\rho\sigma}}_{;\mu} h^{(1)}_{\rho\sigma;\nu}\rangle\,_{\rm I},\label{isaac}
\ea
where the brackets denote the averaging done by Isaacson \cite{Isaacson:1968zza}.

If the background is  curved,  choose instead the de Donder gauge $\bar \nabla^\mu h^{(1)}_{\mu\nu}=0$. Under the assumption that the metric fluctuation varies on a scale $\lambda$ (e.g. $\bar \nabla_. h^{(1)}_{..}\sim 1/\lambda$), while the background metric varies on a scale $L$ ($\bar R\sim 1/L^2$, where $L$ is a curvature scale) such that $\lambda\ll L$, then we can commute the background covariant derivatives,  just like we would commute partial derivatives (since the error made is of the order $\lambda^2/L^2$).  Note that the same assumptions would render the cosmological constant term $\Lambda h^{(1)}_{..} h^{(1)}_{..}$ irrelevant to the order we are working because $\bar \nabla_{.} h^{(1)}_{..}\bar \nabla_. h^{(1)}_{..}\sim 1/\lambda^2$ while $\Lambda h^{(1)}_{..} h^{(1)}_{..}\sim 1/L^2$, and therefore it is suppressed by $\lambda^2/L^2$ relative to the former terms. If an averaging is performed, as in \cite{Isaacson:1968zza}, then we can integrate by parts under the averaging sign and arrive at \eqref{isaac}, where the derivatives are background-covariant.\footnote{ The boundary terms vanish because the averaging function is chosen to vanish at the boundary of the integration region.  Also, equally important for the averaging procedure performed in curved backgrounds are the bitensors which, when contracted with ${\cal T}_{\mu\nu}$, render the integrand a background scalar, and make possible the integration by parts. } (See for example the equations (5.37-5.39) in \cite{Flanagan:2005yc}.)

If the curvature scale of the spacetime is small,  the wavelength of the gravitational waves must be even smaller in order for the approximations and averaging performed by Isaacson \cite{Isaacson:1968zza} (see also Chapter 35 in Misner, Thorne and Wheeler \cite{Misner:1973prb}) to be applicable. This is certainly not the case for the Randall-Sundrum background, 
\ba
ds^2= dy^2 + \exp(-2\kappa |y|)\,dx^\mu dx^\nu \eta_{\mu\nu}, \qquad \mu,\nu=0,1,2,3,
\ea
where $\bar R \sim \kappa^2$, and where $\kappa$ is constrained by corrections to Newton's law to be such that $\kappa r\gg1 $ for $r\sim 1 \mu{\rm m}$ in a Cavendish-type experiment. In this scenario, the curvature scale is $1/\kappa\ll 1\mu{\rm m}$, while for the gravitational waves detected by LIGO the wavelength  $\lambda\sim 10^2-10^4$  km is much larger than the curvature scale.

Nonetheless, the formula derived earlier for the radiated power \eqref{radpow}, with the gravitational field energy-momentum tensor given by \eqref{2nd}, can still be used in a Randall-Sundrum setup.

One of the goals for the next sections is to bring \eqref{2nd} and \eqref{radpow} to a more manageable form.

\section{Gravitational waves in a curved spacetime: $AdS_5$ and Randall-Sundrum}
\label{sec:GW}
\noindent
Consider a background $\bar g_{\mu\nu}$, perturbed by gravitational waves $h_{\mu\nu}^{(1)}$,  and set matter sources to zero ($T_{\mu\nu}=0$). Allowing for a non-vanishing cosmological constant, the background satisfies the Einstein equations
\ba
\bar R_{\mu\nu}-\frac 12 \bar g_{\mu\nu} (\bar R-2\Lambda)=0,\qquad \bar R=\frac {2d }{d-2}\Lambda\,,
\ea
where $d$ is the number of spacetime dimensions. 
The linearized Einstein equations can be written in a simpler form in terms of 
\ba
\psi_{\mu\nu}\equiv h_{\mu\nu}^{(1)}-\frac 12 \bar g_{\mu\nu} h^{(1)}
\ea
as
\ba
\bar\Box \psi_{\mu\nu}+\bar g_{\mu\nu}\bar\nabla^\rho\bar\nabla^\sigma \psi_{\rho\sigma}-\bar\nabla^\rho\bar\nabla_\mu \psi_{\rho\nu}-\bar\nabla^\rho\bar\nabla_\nu \psi_{\rho\mu}+\frac{4\Lambda}{d-2}\psi_{\mu\nu}=0\,.
\ea
This can be further manipulated into
\ba
\bar\Box \psi_{\mu\nu}+\bar g_{\mu\nu}\bar\nabla^\rho\bar\nabla^\sigma \psi_{\rho\sigma}-\bar\nabla_\mu\bar\nabla^\rho \psi_{\rho\nu}-\bar\nabla_\nu\bar\nabla^\rho \psi_{\rho\mu}-2\bar R_{\sigma(\mu\nu)\rho}\psi^{\rho\sigma}=0\,.
\ea
Choosing the de Donder gauge
($
\bar\nabla^\mu\psi_{\mu\nu}=0
$)
leads to
\ba
\bar\Box \psi_{\mu\nu}-2\bar R_{\sigma(\mu\nu)\rho}\psi^{\rho\sigma}=0\,.\label{wave}
\ea
At this point one could follow Isaacson and use the WKB approximation for the gravitational waves (assume that the wavelength is much shorter than the background curvature length) and drop the curvature term from \eqref{wave} and approximate \eqref{wave} by $\bar \Box \psi_{\mu\nu} \approx \bar g^{\rho\sigma}\partial_\rho\partial_\sigma \psi_{\mu\nu}\approx 0$. 

However, we are interested in cases when this approximation is invalid,  and therefore we refrain from ignoring the curvature and Christoffel contributions.  For concreteness let us consider a maximally symmetric background:
\ba
\bar R_{\sigma\mu\nu\rho}=\frac{2\Lambda}{(d-1)(d-2)}(\bar g_{\sigma\nu}\bar g_{\rho\mu}-\bar g_{\mu\nu}\bar g_{\sigma\rho}).
\ea
Substituting into \eqref{wave} leads to
\ba
\bar\Box \psi_{\mu\nu}-\frac{4\Lambda}{(d-1)(d-2)}\psi_{\mu\nu}+\frac{4\Lambda}{(d-1)(d-2)}\bar g_{\mu\nu} \psi_\rho^\rho=0.
\ea
If, as we do in flat space, we fix the residual gauge freedom by imposing tracelessness 
$
\psi^\rho_\rho=0$,
then the linearized Einstein equation,  in the now transverse (de Donder) and traceless gauge,  reads
\ba
\bar\Box \psi_{\mu\nu}-\frac{4\Lambda}{(d-1)(d-2)}\psi_{\mu\nu}=0.
\ea
Despite the apparent simplicity of this equation,  the various components of the metric fluctuation remain coupled.
An alternative approach which leads to decoupled equations of motion starts by decomposing the metric in scalar, vector, tensor (SVT) fluctuations with respect to background isometries.  As a bonus, we will be able to extract the gauge-invariant metric fluctuations, and use them to construct the gravitational energy-momentum tensor according to \eqref{2nd}. We will discuss this at length in section IV. 

In the remaining parts of this section we discuss plane waves (vacuum gravitational wave solutions) in 5d Anti de-Sitter (AdS) and Randall-Sundrum geometries, and construct spherical wave solutions relevant for gravitational waves far away from sources.

\subsection{Vacuum solutions (plane waves)}

Consider the 5d metric fluctuations
\ba
h^{(1)}_{MN} dx^M dx^N= h^{(1)}_{yy} dy^2+ 2 h^{(1)}_{y\mu} dy dx^\mu+h^{(1)}_{\mu\nu} dx^\mu dx^\nu\,,
\ea
where $M,N = 0,1,2,3,5$ while $\mu,\nu=0,1,2,3$ and $y \equiv x^5$. The background AdS metric in the Poincare patch can be written as
\ba
\bar g_{MN}dx^M dx^N= dy^2+ e^{-2 \kappa y}\eta_{\mu\nu}dx^\mu dx^\nu, \qquad \kappa>0 \,,
\ea
and the background Randall-Sundrum metric was given earlier in \eqref{rs}: $\bar g_{MN}dx^M dx^N= dy^2+ e^{-2 \kappa |y|}dx^\mu dx^\nu$. Next we 
decompose the metric fluctuations into scalar, vector, and tensor fluctuations with respect to the 4d Lorentz isometries:
\ba
h^{(1)}_{MN} dx^M dx^N= 2\phi dy^2+2(\partial_\mu B-S_\mu) dx^\mu dy+
(\partial_\mu\partial_\nu E+ 2\eta_{\mu\nu}\psi+ \partial_\mu F_\nu+\partial_\nu F_\mu+f_{\mu\nu})dx^\mu dx^\nu\,,\nn\\\label{rsdecomp0}
\ea
where $
\eta^{\mu\nu}\partial_\mu S_\nu=0, \; \eta^{\mu\nu}\partial_\mu F_\nu=0, \;
\eta^{\mu\nu}\partial_\mu f_{\nu\rho}=0, \; f_{\mu\nu}\eta^{\mu\nu}=0.
$ 
When performing a gauge transformation $
\delta_\xi h^{(1)}_{MN}=\bar\nabla_M \xi_N+\bar\nabla_N\xi_M
$ 
we can decompose the gauge parameter in a similar way
$
\xi^M =(\xi^{(T)}_\mu +\partial_\mu \xi^{(L)}, \xi)$, with $\partial_\mu \xi^{(T)\,\mu}=0.$ 
The tensor metric fluctuations are gauge invariant \cite{Mukohyama:2000ui}:
$
\delta_\xi f_{\mu\nu}=0\,.
$
Given a monochromatic plane wave $\exp(i k_\mu x^\mu)$, with $k^\mu$ a time-like 4-vector ($k^2=k_\mu k_\nu \eta^{\mu\nu} <0$), we can define three space-like vectors $\epsilon^{(p)}_\mu$, transverse to $k_\mu$ and to each other
\ba
\epsilon_\mu^{(p)} k_\nu\eta^{\mu\nu}=0, \qquad \epsilon^{(p)}_\mu \epsilon^{(q)}_\nu \eta^{\mu\nu}=\delta ^{pq}, \qquad p,q=1,2,3.
\ea
The metric tensor fluctuations can be written as 
\ba
f_{\mu\nu}= \epsilon^{(p)}_\mu \epsilon^{(q)}_\nu e^{i k_\lambda x^\lambda} f^{pq}(y), \qquad f^{pq}\delta^{pq}=0\,,
\ea
where $f^{pq}$ obey the following decoupled equations:  for (i) AdS$_5$
\ba
\bigg[\frac{d^2}{dy^2}-4 \kappa^2 -k^2 e^{2 \kappa y}\bigg] f^{pq}(y)=0, \label{ttf}
\ea
and for (ii) Randall-Sundrum \cite{Randall:1999vf,Garriga:1999yh}\footnote{These authors did not perform an SVT decomposition, rather they chose the so-called Randall-Sundrum gauge $h_{yy}=h_{\mu y}=0, h^\mu_\mu=0, \partial_\mu h^{\mu\nu}=0$ which essentially projects onto the tensor fluctuations. See also \cite{Kakushadze:2000rz} regarding comments about the implementation of the Randall-Sundrum gauge. Of note is that in order to reach this gauge in general one needs to perform gauge transformations that will change the position of the brane at $y=0$.}
:
\ba
\bigg[\frac{d^2}{dy^2}-4 \kappa^2 +4\kappa \delta(y) -k^2 e^{2 \kappa| y|} \bigg]f^{pq}(y)=0 \label{ttfrs}\,.
\ea

The equation \eqref{ttf} admits two linearly independent solutions, expressed in terms of Bessel functions:
\ba
f^{pq}=c^{pq} J_2(e^{\kappa y} \sqrt{-k^2/\kappa^2})+d^{pq} Y_2(e^{\kappa y} \sqrt{-k^2/\kappa^2}), \qquad c^{pq}\delta_{pq}=0, \qquad d^{pq}\delta_{pq}=0\,.\label{solttf}
\ea
This solution exhibits oscillatory (wave-like) behavior in $y$ as well, with an amplitude which decreases with $y$. Of the two Bessel functions, only $Y_2(\sqrt{-k^2/\kappa^2} \exp(\kappa y))$ blows up in the interior of AdS$_5$, for $y\to \infty$. 
If instead we were solving in the WKB limit to leading order we would begin with the ansatz
\ba
f\sim \exp(i S), \qquad \frac{d}{dy} S\gg1  \qquad  \frac{d^2}{dy^2}S\ll1 \, .
\ea
Then the equation \eqref{ttf} simplifies to
\ba
\frac{d}{dy} S=\pm \sqrt{-k^2} e^{\kappa y}\,.
\ea
The WKB phase is
\ba
S=\pm\exp(\kappa y)\frac{\sqrt{-k^2}}{\kappa}.
\ea
This captures the asymptotic (large argument) behavior of the Bessel functions. The WKB solution  is a good approximation only deep in the interior of AdS$_5$ space, as long as $\sqrt{-k^2/\kappa^2} \exp(\kappa y)\gg 1$. 
If $k^2=0$, the solutions to \eqref{ttf}, $\exp(\pm 2\kappa y)$, blow up either at the boundary $y\to-\infty$, or deep in the interior of AdS$_5$. Consequently, there are no normalizable zero-modes in AdS, but there is one discrete normalizable zero-mode in Randall-Sundrum. Similarly, for $k^2>0$ there are no normalizable solutions.

The solution to \eqref{ttfrs} takes a similar form to \eqref{solttf}, 
\ba
&&\!\!\!\!\!\!\!\!\!\!f^{pq}=c_+^{pq} J_2(e^{\kappa y} \sqrt{-k^2/\kappa^2})+d_+^{pq} Y_2(e^{\kappa y} \sqrt{-k^2/\kappa^2}), \qquad c_+^{pq}\delta_{pq}=0, \qquad d_+^{pq}\delta_{pq}=0, \qquad y>0\,,\nn\\
&&\!\!\!\!\!\!\!\!\!\!f^{pq}=c_-^{pq} J_2(e^{-\kappa y} \sqrt{-k^2/\kappa^2})+d_-^{pq} Y_2(e^{-\kappa y} \sqrt{-k^2/\kappa^2}), \qquad c_-^{pq}\delta_{pq}=0, \qquad d_-^{pq}\delta_{pq}=0, \qquad y<0\,,\nn\\
\ea
and satisfy the additional matching condition
\ba
\frac{d}{dy} f^{pq}\bigg|_{y\to 0^+}-\frac{d}{dy} f^{pq}\bigg|_{y\to 0^-}=-4\kappa f^{pq}(0).
\ea

\subsection{Retarded Green's functions}

The equation of motion of the tensor mode fluctuation is related to the equation of motion of a massless, minimally coupled scalar field  $\varphi$ in AdS$_5$
\ba
\Box_{5d, AdS}\,\varphi= e^{4\kappa y} \bigg[\frac{d}{dy} e^{-4\kappa y}\frac{d}{dy} -k^2 e^{-2\kappa y}\bigg]\varphi=\bigg[\frac{d^2}{dy^2} +4 \kappa \frac{d}{dy} -k^2 e^{2\kappa y}\bigg]\varphi=0\,,
\ea
or Randall-Sundrum
\ba
\Box_{5d, RS}\,\varphi= e^{4\kappa |y|} \bigg[\frac{d}{dy} e^{-4\kappa |y|}\frac{d}{dy} -k^2 e^{-2\kappa |y|}\bigg]\varphi=\bigg[\frac{d^2}{dy^2} -4 \kappa\, \text{sign}(y)\,\frac{d}{dy} -k^2 e^{2\kappa |y|}\bigg]\varphi=0\,,
\ea
through the following scaling: $\varphi= \exp(2\kappa y) f$ or $\varphi= \exp(2\kappa|y|) f$, for AdS$_5$ or Randall-Sundrum respectively, where $f$ is a placeholder for $f^{pq}$ in \eqref{ttf} or \eqref{ttfrs}.  In writing the above equations, we have Fourier-transformed along the 4d $x^\mu$ coordinates.
The scalar Green's function satisfies
\ba
\Box_{5d, \text{AdS/RS}}\,G_\text{scalar}(y,y';k^\mu)= (- \bar g)^{-1/2} \delta(y-y')\,,\label{scalar}
\ea
where $\Box_{5d, \text{AdS/RS}} $ refers to the scalar d'Alembertian in the curved geometries, and $-\bar g=-\det(\bar g)$ is $ \exp(-8\kappa y)$ for AdS$_5$ and 
$ \exp(-8\kappa |y|)$ for Randall-Sundrum.
The corresponding Green's function for the tensor mode metric fluctuations \ba
G(y,y';k^\mu)=\exp(-2\kappa|y|) \exp(-2\kappa |y'|)G_{\text{scalar}}(y,y';k^\mu) \label{scalar-tensor} 
\ea
obeys
\ba
\bigg[\partial_y^2   - 4 \kappa^2  -k^2 e^{2 \kappa | y |}\bigg]
    G(y ,y';k^\mu) & = & \delta (y - y')\,,
    \ea
    for AdS$_5$, and 
 \ba
    \bigg[\partial_y^2   - 4 \kappa^2 + 4 \kappa \delta (y) -k^2 e^{2 \kappa | y |}\bigg]
    G(y ,y';k^\mu) & = & \delta (y - y')\,,
  \ea
  for Randall-Sundrum.
There are several methods we can use to construct the retarded propagator (or retarded Green's function).
Starting from the Euclidean propagator, we can arrive at the retarded propagator by analytical continuation \cite{Barnes:2010jp}. We can use (if known) the position-space Euclidean propagator as follows. For example, in 4d flat space, the Euclidean propagator is $1/(4\pi^2 (t_E^2+r^2))$, where $t_E$ is the Euclidean time. Switching to Minkowski signature, the retarded propagator is obtained from $i\theta(t)/(4\pi^2)\bigg(1/(-(t-i\epsilon)^2+r^2)-1/(-(t+i\epsilon)^2+r^2)\bigg)$, with $\epsilon\to 0$ and where $\theta(t)$ is the Heaviside step function. The $i\epsilon$ prescription identifies the two terms as Wightman two-point functions, with the retarded propagator written as the difference of the two Wightman two-point functions times the step-function $\theta(t)$. The 4d flat spacetime retarded propagator evaluates to $\theta(t)/(2\pi^2) \, \epsilon/(\epsilon^2+(r^2-t^2)^2)$ which in the limit $\epsilon\to 0$ yields $(1/2 \pi) \theta(t) \delta(t^2-r^2)=1/(4 \pi r)\theta(t) \delta(t-r)$. Alternatively, we can start from the momentum-space Euclidean propagator, which in 4d flat space is $1/(k^2)$ and obtain the momentum-space retarded propagator by doing the analytical continuation $1/(-(k^0+i\epsilon)^2+\vec k\cdot\vec k)$. Then Fourier-transform to position space and arrive at the result quoted earlier,  $1/(4 \pi r)\theta(t) \delta(t-r)$. While the defining feature of the 4d flat space retarded propagator is its support on the forward light-cone, this feature is lost in flat odd-dimension spacetimes, when the retarded propagator has support inside the forward light-cone (as expected, based on causality arguments).

The Euclidean boundary-to-bulk scalar propagator for AdS$_5$, from the boundary point  $(t_E'=0, \vec r\,{}'=0, y')$ with $e^{\kappa y'}=\varepsilon\ll1$ to some point in the bulk $(t_E,\vec r, y)$, is given by \cite{Witten:1998qj,{DHoker:1999mqo}}:
\ba
G_{\text{scalar,Eucl AdS}}=\frac{6 \kappa^3 e^{4\kappa y}}{\pi^2 (e^{2\kappa y}+\kappa^2 (r^2+t_E^2))^4}\varepsilon^4\,.
\ea
Then the corresponding retarded propagator, derived as described above, is (see also the Appendix C in \cite{Barnes:2010jp}):
\ba 
G_{\text{ret, scalar, AdS}}=\lim_{\epsilon\to 0}\frac{48 \kappa^3 e^{4\kappa y} \epsilon \left(e^{2\kappa y} +\kappa^2(r^2-t^2)\right)^3}{\pi^2\left(\left(e^{2\kappa y} +\kappa^2(r^2-t^2)\right)^2+\epsilon^2\right)^4}\varepsilon^4\theta(t)\,.
\ea
This leads to the tensor mode boundary-to-bulk retarded propagator 
\ba
G_{\text{ret, AdS}}=\lim_{\epsilon\to 0}\frac{48 \kappa^3 e^{2\kappa y} \epsilon \left(e^{2\kappa y} +\kappa^2(r^2-t^2)\right)^3}{\pi^2\left(\left(e^{2\kappa y} +\kappa^2(r^2-t^2)\right)^2+\epsilon^2\right)^4}\varepsilon^2\theta(t)\,,
\ea
which is proportional to $\partial_\Delta^3 \delta(\Delta)$, with $\Delta=e^{2\kappa y} +\kappa^2(r^2-t^2)$.

For the Randall-Sundrum background, the tensor mode retarded propagator was derived by Garriga and Tanaka \cite{Garriga:1999yh}.
The idea behind their formula is that the Green's function can be written in terms of eigenfunctions of the corresponding  differential operator. For the Randall-Sundrum geometry we have the following eigenvalue problem: 
\ba [- k^2 + e^{2 \kappa | y |} \partial_y (e^{- 4 \kappa | y |} \partial_y)]
     f_{(\lambda)} = - \lambda^2 f_{(\lambda)}\,, 
     \ea
or, equivalently,
\ba 
e^{2 \kappa | y |} \partial_y (e^{- 4 \kappa | y |} \partial_y) f_{(q)} =
     - q^2  f_{(q)}\,, \label{rseigen}
     \ea
     where we  defined $q = \sqrt{\lambda^2 - k^2}$ and with $-\lambda^2$ being the eigenvalues. 
 The Euclidean signature Green's function in momentum ($k$-) space has the generic form
 \ba
 G(y,y')=\int\!\!\!\!\!\!\!\!\!\sum_q \frac{f_{(q)}(y) f_{(q)}^*(y')}{k^2+q^2}e^{-2\kappa |y|}e^{-2\kappa|y'|}\,,
 \ea
 where one sums over the discrete eigenvalues and integrates over the continuum ones.
 The retarded Green's function is obtained by doing the analytic continuation
 \ba
 G_{\text{ret,RS}}(y,y')=\int\!\!\!\!\!\!\!\!\!\sum_q \frac{f_{(q)}(y) f_{(q)}^*(y')}{(-k^0+i\epsilon)^2 +\vec k\,{}^2+q^2}e^{-2\kappa |y|}e^{-2\kappa|y'|}\,.
 \ea
 The eigenvalue problem \eqref{rseigen} has one discrete $q=0$ mode, the 
  bound state being $f_{(0)} = \sqrt{\kappa}$,  and a continuum set of modes for $q>0$ 
  \ba
 f_{(q)} = \sqrt{\frac{q}{2 \kappa (1 + \alpha^2_{(q)})}} e^{2 \kappa | y |}
     \left[ J_2 \left( \frac{q}{\kappa} e^{\kappa | y |} \right) + \alpha_{(q)} Y_2 \left(
     \frac{q}{\kappa} e^{\kappa | y |} \right) \right] \,,
     \ea
  where $\alpha_{(q)} = - {J_1 (q / \kappa)}/{Y_1 (q / \kappa)}$ was determined from the matching condition at $y=0$. These modes obey the normalization conditions:
  \ba
 \int_{-\infty}^\infty {dy} \,e^{- 2 \kappa | y |}  f_{(0)} (y)
    f_{(0)} (y) = 1\,,\nn\\
 \int_{-\infty}^\infty {dy}\, e^{- 2 \kappa | y |}   f_{(q)} (y)
     f_{(q')} (y) = \delta (q - q')\,.
     \ea
 Putting everything together one arrives at the result of \cite{Garriga:1999yh}
 \ba
G_{\text{ret, RS}}(x^\mu, y ; x'{}^\mu, y')& =& \int \frac{d^4 k}{(2 \pi)^4} e^{i k \cdot
     (x - x')} \left[ \frac{\kappa }{-(k^0+i\epsilon)^2+\vec k^2} \right.\nn\\
     &&+\left. \int_0^\infty dq 
     \frac{1}{-(k^0+i\epsilon)^2+\vec k^2
     + q^2}  f_{(q)} (y) f_{(q)} (y') \right]e^{-2\kappa |y|}e^{-2\kappa|y'|}\label{gretrs}\,.
     \ea
     From a 4d perspective, the 5d bound state is a massless mode, while the 5d continuum states are massive modes.

\subsection{Static, spherically-symmetric solutions}
If we consider a static source, point-like and localized at the boundary, $j(x')= M \delta^3 (\vec x\,{}') \delta(y'-(\ln\varepsilon)/\kappa)$ with $\varepsilon\ll1$, the tensor fluctuations in AdS$_5$ are of the form
\ba
\int d^5 x' G_{\text{ret, AdS}}(x',x)j(x')&=&
 \int dt'  M \frac{48 \kappa^3 e^{2\kappa y} \epsilon \left(e^{2\kappa y} +\kappa^2 r^2-\kappa^2(t-t')^2\right)^3}{\pi^2\left(\left(e^{2\kappa y} +\kappa^2 r^2-\kappa^2 (t-t')^2\right)^2+\epsilon^2\right)^4}\varepsilon^2\theta(t-t')\nn\\
 &=& M \frac{15\kappa^2 }{2 \pi^2}\frac{e^{2 \kappa y}}{(e^{2\kappa y}+\kappa^2 r^2)^{7/2}} \,\varepsilon^2\,.\label{statads}
\ea
 Next, assume that a point-like static source, localized at $y=0$ in the Randall-Sundrum geometry sources the tensor modes equation \eqref{ttfrs}. We are doing a similar calculation to the one done earlier in AdS, but now we are using the retarded propagator \eqref{gretrs}. First, the integral over $t'$ sets $k^0=0$. The integral over $\vec k$ results in an exponential suppression factor $\exp(-q r)$.\footnote{Use that $(1/2 \pi^2) \int_0^\infty dk \,\sin(kr) \,k/(r (k^2+q^2))=\exp(-q r)/(4\pi  r)$. This expression is the familiar  Yukawa-type static Green's function of massive modes in flat 4d space. This result is an intermediate step in our 5d Randall-Sundrum calculation, where the $q$-modes appear massive from a 4d perspective.} We were unable to perform the last integral, over $q$, analytically. However, we come close for large enough $r$. Then the exponential suppression $\exp(-qr)$ factor localizes 
the integral over $q$  in the small $q$-range. Using the small argument expansion of the Bessel functions $J_1(q/\kappa)$, $Y_1(q/\kappa)$, and performing the $q$-integral results in the following solution: 
\ba
&&\int d^5 x' G_{\text{ret, RS}}(x',x)j(x')\simeq M\kappa^2\bigg[\bigg(\frac{2 \kappa^2r^2+3 e^{2\kappa |y|} }{8 \pi  \left(e^{2\kappa |y|}+\kappa^2 r^2\right)^{3/2}}\nn\\
&&+\frac{15 e^{4\kappa |y|} \log \left(\sqrt{\frac{\kappa^2 r^2}{e^{2\kappa |y|}}+1}+\frac{\kappa r}{e^{\kappa |y|}}\right)}{16 \pi  \left(e^{2\kappa |y|}+\kappa^2 r^2\right)^{7/2}}+\frac{2 \kappa^4 r^4+9 \kappa^2 r^2 e^{2\kappa |y|}-8e^{4\kappa |y|}}{16 \pi  \kappa r \left(e^{2\kappa |y|}+\kappa^2 r^2\right)^3}\bigg)e^{-2\kappa |y|}\bigg]\,,\label{statrs}
\ea
where the zero mode contribution was canceled by part of the massive mode contribution.

We would like to point out that in using the Randall-Sundrum geometry as a model for large extra dimensions, we are already requiring that $\kappa r\gg1$. This is exactly the regime when our small-argument approximation for $J_1(q/\kappa)$ and $Y_1(q/\kappa)$ is applicable, since on the one hand $r\kappa\gg1$ and on the other hand $q r <{\cal O}(1)$ due to the exponential suppression factor. Put together this implies that  $q/\kappa\ll1$, thus justifying our  small argument expansion of the Bessel functions $J_1(q/\kappa)$, $Y_1(q/\kappa)$. In evaluating the integrals in \eqref{statrs} we did not make any further approximations to the other two Bessel functions $J_2(q e^{\kappa|y|}/\kappa )$ and $Y_2(q e^{\kappa|y|}/\kappa )$. 

For $y=0$, which would correspond to both source and fluctuation on the brane, and to leading order in $r$ this approximates to $M\,\kappa/(4\pi r)(1+1/(2\kappa^2 r^2))$ \cite{Garriga:1999yh}.

Stripping off the factor of $M$ from the expressions in (\ref{statads}, \ref{statrs}) we get the Green's functions for the time-independent Laplacian operators. This can be explicitly verified. For example, we can show that the action of $\partial_y^2+ e^{2\kappa y}\vec\nabla^2-4\kappa^2$ on the right hand side of \eqref{statads} is zero when $y$ is not on the boundary, and the action of $\partial_y^2+ e^{2\kappa |y|}\vec\nabla^2-4\kappa^2$ on the right hand side of \eqref{statrs} is zero for $y\neq 0$.  We can also show by using Gauss' theorem that the delta-function source term in the Green's function equation is accounted for appropriately. Using \eqref{scalar-tensor} together with (\ref{statads}, \ref{statrs})   we find the static scalar Green's function $G_{\text{scalar}}$ of either AdS$_5$ or Randall-Sundrum spacetimes, solution to 
\ba
\frac{1}{\sqrt{-\bar g}} \partial_I (\sqrt{-\bar g} g^{IJ} \partial_J) G_{\text{ scalar}}(y,\vec r; y'=0,\vec r\,'=0)=\frac{\delta^3(\vec r)\delta (y)}{\sqrt{-\bar g}},\qquad  I,J=1,2,3,y\,.
\ea
Then we can integrate over the spatial coordinates $\vec r, y$. Using the analogy of  cylindrical coordinates in flat space, we compute the flux through the surface at infinity; there are two regions: one at fixed, large $r$  with $y$ integrated over (this is like integrating over the length of the 3d cylinder in our analogy) and the other surface with $r$ integrated over and fixed, large $|y|$  (this is like integrating over the two caps of the cylinder). With an infinitely long cylinder we only need to compute the flux through the sides of the cylinder. For the Randall-Sundrum case, truncating to the leading order term in \eqref{statrs}, the flux through the side of the cylinder yields 
\ba
2\times 4\pi \kappa^2 \int_0^\infty dy \bigg[e^{-4\kappa y}r^2 e^{2\kappa y} \frac{\partial}{\partial r}\frac{(2 \kappa^2 r^2+ 3   e^{2\kappa y}  )}{
     8 \pi ( e^{2\kappa y}+r^2 )^{3/2}}\bigg]\bigg|_{r=R_\infty} =\frac{\kappa^3 R_\infty^3}{(1+\kappa^2 R_\infty^2)^{3/2}}\,,
     \ea
which, in the limit $\kappa r\gg 1$ when \eqref{statrs} is applicable, gives the expected result.

\subsection{Spherical waves}\label{Sphrs}

To illustrate the propagation of gravitational waves in AdS$_5$, assuming that a periodic source is at the boundary, we consider solving the tensor modes equation, with a point-like periodic source $j(x')=M \delta^3 (\vec x\,{}') \delta(y'-(\ln\varepsilon)/\kappa) e^{i\omega t'}$: 
\ba
\int d^5 x'  G_{\text{ret, AdS}}(x',x)j(x')&=&
 \int dt'  M \kappa^3\frac{48 \kappa^3 e^{2\kappa y} \epsilon \left(e^{2\kappa y} +\kappa^2 r^2-\kappa^2 (t-t')^2\right)^3}{\pi^2\left(\left(e^{2\kappa y} +\kappa^2 r^2-\kappa^2(t-t')^2\right)^2+\epsilon^2\right)^4}\varepsilon^2\theta(t-t')\nn\\
 &&\!\!\!\!\!\!\!\!\!\!\!\!\!\!\!\!\!\!\!\!\!\!\!\!\!\!\!\!\!\!\!\!= M \frac{1}{2\pi^2}\frac{15 \kappa^2  + 15 i {\text R} \kappa \omega - 6 {\text R}^2 \omega^2 - 
   i {\text R}^{3} \omega^3\kappa^{-1}}{8 {\text R}^{7}}\, e^{i  \omega (t-R/\kappa)}\,\varepsilon^2\,,\label{adssphw}
   \ea
where ${\text R}=\sqrt{\exp(2\kappa y)+\kappa^2 r^2}.$
If we were to perform a WKB approximation, the leading-order WKB approximation would have captured only $e^{i \omega (t-{\text R}/\kappa)}$ part of the exact result \eqref{adssphw}.

Similarly, a periodic, point-like source localized at $y=0$ which sources the tensor mode equation \eqref{ttfrs} yields the following fluctuation 
\ba
\int d^5 x' G_{\text{ret, RS}}(x',x)j(x')\simeq \frac{\kappa^2M}{2\pi^2}\left(\frac{2 \kappa^2 r^2+3 e^{2 \kappa |y|}}{\left(\kappa^2 r^2+e^{2 \kappa |y|}\right)^{3/2}}+\frac{i \omega e^{2\kappa |y|}}{\kappa(\kappa^2 r^2+e^{2 \kappa |y|})}\right) e^{-2 \kappa|y|}e^{i \omega \left(t-\sqrt{r^2+e^{2 \kappa |y|}/\kappa^2}\right)}\nn\\\label{sphrs}\!\!\!\!\!\!
\ea
under the same assumption that the distance $r$ (measured along the brane) from the source is sufficiently large such that the $q$-integral is localized at small values of $q$, and where we kept terms up to first order in $\omega$.


\section{The gravitational energy-momentum tensor and the radiated power in terms of 
gauge-invariant fluctuations}
\label{sec:GW_EM_gauge_inv}

In this section we give explicit expressions for the energy-momentum tensor of gravitational waves and for the power radiated away from a source by gravitational waves.  Similar to the approach of \cite{Abramo:1997hu}, our expressions are made manifestly gauge-invariant by using the gauge-invariant part of the metric fluctuations, which is found through a scalar-vector-tensor (SVT) decomposition. We study three cases: (A) 4d flat spacetime, (B) flat 5d spacetime with one compact dimension, and (C) Randall-Sundrum.  Further checks on our results can be found in Appendices \ref{only tensor for non-compact geometry} and \ref{allmodescompact}.

\subsection{4d flat spacetime}\label{et4d}

\noindent As we have seen, in general,  the metric fluctuations obey coupled equations of motion.  One way to decouple them is to use the symmetries of the background.  In \cite{Abramo:1997hu}, Abramo et al.  considered the following scenario, which is relevant for cosmological backgrounds:
\ba
\bar g_{\mu\nu} dx^\mu dx^\nu= -dt^2+a^2(t)\delta_{ij} dx^i dx^j. \nn
\ea
Given the rotational isometries of the background, they decomposed the metric fluctuation $h_{\mu\nu}$ in components which transform as scalars, vectors and tensors under the rotation group $SO(3)$.
For simplicity, we review and adapt their analysis in the context of flat 4d spacetime and set $a(t)=1$. 

A small difference between our paper and \cite{Abramo:1997hu} is that we set up the perturbative expansion
\be
g=\bar g+h^{(1)}+h^{(2)}+\dots\label{eq41}
\ee
with the background metric $\bar g$ an exact solution to the Einstein equation, 
whereas for \cite{Abramo:1997hu} the perturbative set-up has $q=q_0+\delta q$ with the background defined to be the homogeneous part of the metric $q_0=\langle q\rangle$ (the brackets denote spatial averaging of the metric $q$ on a fixed time slice).  In their case,  the background is only an approximate solution of the Einstein equation and it receives contributions from the backreaction due to the gravitational wave fluctuations $\delta q$. 
On the other hand,  we simply work perturbatively around the exact background $\bar g$,  and account for the backreaction of $h^{(1)}$ as a source term for the $h^{(2)}$ equation of motion.

We begin by decomposing the metric fluctuations in  representations of the $SO(3)$ rotation group:
\ba h_{\mu \nu} = \left(\begin{array}{cc}
     2 \phi & \partial_j B + S_j\\
     \partial_i B + S_i & 2 \psi \delta_{i j} + \partial_i \partial_j E +
     \partial_i F_j + \partial_j F_i + f_{i j}
   \end{array}\right), \label{eq4.2}
\ea
where $S_i, F_i $ and $f_{ij} $ are transverse: $\partial_i S^i=\partial_i F^i=\partial_i f^{ij}=0$, and $f_{ij}$ is traceless: $f_{ij}\delta^{ij}=0$.

\noindent The following expressions  $\Phi, \Psi, {\bf S}_i, f_{ij}$ are gauge invariant:
\begin{eqnarray}
  \Phi & = & \phi - \partial_0 (B - \tfrac 12\partial_0 E), \nn\\
  \Psi & = & \psi, \nn\\
  {\bf S}_{i} & = & S_{i} - \partial_0 F_{i}.
\label{gi4dso3}
\end{eqnarray}
This can be derived and explicitly verified by considering a linearized gauge transformation $\delta_\xi h_{\mu\nu}=\partial_\mu \xi_\nu+\partial_\nu \xi_\mu$ and substituting  a similar $SO(3)$ scalar-vector decomposition of the gauge parameter $\xi^\mu=(\xi, \xi^{(T)\,i}+\partial^i \xi^{(L)})$, with $\partial_i \xi^{(T)\,i}=0.$

Starting from $g=\bar g+ h^{(1)}+h^{(2)}+\dots$ we perform gauge transformations 
 \be
g \rightarrow \tilde g=\exp(\mathcal L_\xi) g=\bar g +h^{(1)}_{g.i.}+ h^{(2)}_{g.i.}+\dots 
\ee
 such that $h^{(1)}_{g.i.}=\mathcal L_\xi  \bar g+ h^{(1)}$ is expressed through the gauge-invariant terms given in \eqref {gi4dso3}.
E.g. with
\be
\xi^{(L)}=-\frac12 E, \qquad \xi_i^{(T)}=-F_i, \qquad \xi=B-\frac 12 \dot E
\ee
we have:
\be h_{\mu \nu}|_{g.i.} = \left(\begin{array}{cc}
     2 \Phi & {\bf S}_j\\
     {\bf S}_i & 2 \Psi \delta_{i j} +  {f}_{i j}
   \end{array}\right) .
\label{hgi4}
\ee
This was referred to as a "longitudinal gauge" in  \cite{Abramo:1997hu}. There are other  choices for the gauge-parameter $\xi^\mu$ which lead to different ways of expressing the metric fluctuations through  the gauge-invariant fluctuations \eqref{gi4dso3}.  We will nontheless use the longitudinal gauge in what follows. 
Similarly,  to second order in perturbations we arrange for $ h^{(2)}_{g.i.}$ to be written in terms of gauge invariant pieces.   
Since Einstein's equations are gauge-invariant one can easily verify that the linearized Einstein equation
in the absence of sources
\be
\delta^{(1)} G_{\mu\nu}(h^{(1)})=0 \nn
\ee
can also be packaged, as expected,  only in terms of the gauge-invariant functions given in equation \eqref{gi4dso3}:
\be
\delta^{(1)} G_{\mu\nu}(h^{(1)}|_{g.i.})=0\,.
\ee

At second order, the Einstein equation can be written as:
\ba
\delta^{(1)} G_{\mu\nu}(h^{(2)})=-\delta^{(2)} G_{\mu\nu}(h^{(1)},h^{(1)})\,.\label{eqh2o}
\ea
While the  Einstein equation is gauge-invariant,  when casting it in the form given in \eqref{eqh2o}, with a non-vanishing right-hand-side  quadratic in the linearized fluctuations, both sides of equation are gauge-dependent. However, by using the metric fluctuations $h|_{g.i.}$ both sides of the second-order Einstein equation are now written in a gauge-invariant form,
\be
\delta^{(1)} G_{\mu\nu}(h^{(2)}|_{g.i.})=-\delta^{(2)} G_{\mu\nu}(h^{(1)}|_{g.i.},h^{(1)}|_{g.i.})\,.\label{eqh2n}
\ee

  This backreaction of the gauge-invariant linearized fluctuations (i.e. gravitational waves) is interpreted as the energy-momentum tensor of the gravitational waves.  And since it is expressed in terms of the gauge-invariant fluctuations \eqref{gi4dso3} it is gauge-invariant by construction.

Consider next  the Einstein equations in the presence of matter sources:
\ba
G_{\mu \nu} = T_{\mu \nu}.
\ea
We perform the same SVT decomposition for both sides,
\begin{eqnarray*}
  G_{\mu \nu} & = & \left(\begin{array}{cc}
    G_{00} & \partial_j G^{(L)}_0 + G^{(T)}_{j 0}\\
    \partial_i G^{(L)}_0 + G^{(T)}_{i 0} & 2 G^{(Y)} \delta_{i j} + \partial_i
    \partial_j G^{(LL)} + \partial_i G^{(LT)}_j + \partial_j G^{(LT)}_i +
    G^{(TT)}_{i j}
  \end{array}\right)\,,\\
  T_{\mu \nu} & = & \left(\begin{array}{cc}
    T_{00} & \partial_j T^{(L)}_0 + T^{(T)}_{j 0}\\
    \partial_i T^{(L)}_0 + T^{(T)}_{i 0} & 2 T^{(Y)} \delta_{i j} + \partial_i
    \partial_j T^{(L L)} + \partial_i T^{(LT)}_j + \partial_j T^{(LT)}_i +
    T^{(TT)}_{i j}
  \end{array}\right)\,.
\end{eqnarray*}
The linearized equations of motion for the scalar gauge-invariant fluctuations $\Phi$ and $\Psi$ come from the components $\delta^{(1)} G_{00}\,, \delta^{(1)}  G^{(L)}_0\,, \delta^{(1)}G ^{(Y)}$ and $\delta^{(1)}G^{(L L)}$,
\begin{eqnarray}
\delta^{(1)}  G_{00} & = & - 2 \delta^{i j} \partial_i \partial_j \Psi =   T_{00}\,,\\
 \delta^{(1)}  G^{(L)}_0 & = & - 2 \partial_0 \Psi =  T^{(L)}_0\,,\\
2 \delta^{(1)}   G^{(Y)} & = & - 2 \partial_0^2 \Psi - \delta^{i j} \partial_i \partial_j
  \Phi + \delta^{i j} \partial_i \partial_j \Psi =  2 T^{(Y)},\\
 \delta^{(1)}  G^{(L L)} & = & \Phi - \Psi =  T^{(L L)}\,.
\end{eqnarray}
The equations of motion for the transverse-vector gauge-invariant fluctuations ${\bf S}_i$ come from the components $\delta^{(1)}G^{(T)}_{i 0}$ and
$\delta^{(1)}G^{(LT)}_i$:
\begin{eqnarray}
\delta^{(1)}   G^{(T)}_{i 0} & = & - \frac{1}{2} \delta^{j k} \partial_j \partial_k
  {\bf S}_i =   T^{(T)}_{i 0}\,,\\
\delta^{(1)}   G_i^{(L T)} & = & - \frac{1}{2} \partial_0 {\bf S}_i =  T^{(LT)}_i\,.
\end{eqnarray}
The equations of motion for the transverse-traceless-tensor gauge-invariant fluctuations ${f}_{i j}$ come from $G^{(TT)}_{i j}$:
\ba
\delta^{(1)} G^{(TT)}_{i j} = \frac{1}{2} \partial_0^2{f}_{i j} - \frac{1}{2}
   \delta^{p q} \partial_p \partial_q {f}_{i j} =   T^{(TT)}_{i j}.
\ea
We can quickly count the  degrees of freedom by considering the equations of motion in 
vacuum:
\begin{eqnarray}
  \delta^{i j} \partial_i \partial_j \Psi & = & 0\, , \qquad
  \partial_0 \Psi  =  0\, ,\qquad
  \Phi  =  \Psi \,, \\
  \delta^{j k} \partial_j \partial_k {\bf S}_i & = & 0\, ,\qquad
  \partial_0 {\bf S}_i =  0\, , \\
  &&
  \Box {f}_{i j}  =  0\,.
\end{eqnarray}
 The scalar and vector fluctuations are not dynamical,  unlike  the tensor modes
$f_{i j}$. Since the tensors are transverse $\partial_i f^{ij}=0$ and traceless $\delta_{ij} f^{ij}=0$, this matches the counting of the degrees of freedom for a 4d graviton. 

After fixing the gauge in \eqref{hgi4},
the gravitational energy-momentum tensor can be computed from \eqref{2nd}. Since \eqref{2nd} is quadratic in the metric fluctuations which are now in the gauge \eqref{hgi4}, we indicate which fluctuations are contributing to the various terms in the
energy-momentum tensor as follows:
\ba \mathcal{T}_{\mu \nu} =\mathcal{T}^{(S)}_{\mu \nu} +\mathcal{T}^{(V)}_{\mu
   \nu} +\mathcal{T}^{(T)}_{\mu \nu} +\mathcal{T}^{(S V)}_{\mu \nu}
   +\mathcal{T}^{(S T)}_{\mu \nu} +\mathcal{T}^{(V T)}_{\mu \nu} \,,\ea
where $\mathcal{T}^{(S)}_{\mu \nu}$, $\mathcal{T}^{(V)}_{\mu \nu}$,
$\mathcal{T}^{(T)}_{\mu \nu} $ are terms involving scalar, vector, tensor
modes only and $\mathcal{T}^{(S V)}_{\mu \nu}$, $\mathcal{T}^{(S T)}_{\mu
\nu}$, $\mathcal{T}^{(V T)}_{\mu \nu}$ are terms that mix different modes.
We will focus on $\mathcal{T}_{0i}$ since it is needed to compute the radiated power: 
\begin{eqnarray}
  \mathcal{T}_{0 i}^{(S)} & = & 2 (\partial_0 \Psi) \partial_i \Phi - 4
  \partial_0 (\Psi \partial_i \Psi)\nonumber\\
  \mathcal{T}_{0 i}^{(V)} & = & \frac{1}{2}  {\bf S}^j \partial_0
  (\partial_i {\bf S}_j + \partial_j {\bf S}_i)\nonumber\\
  \mathcal{T}_{0 i}^{(T)} & = & \frac{1}{2}  {f}^{j k} \partial_0
  \partial_k {f}_{i j} - \frac{1}{2}  {f}^{j k} \partial_0
  \partial_i {f}_{j k} - \frac{1}{4} \partial_0 {f}_{j k}
  \partial_i {f}^{j k}\nonumber\\
  \mathcal{T}_{0 i}^{(S V)} & = & - {\bf S}_i \partial_0 \partial_0 \Psi -
  \Psi \partial_j \partial^j {\bf S}_i - {\bf S}^j \partial_j \partial_i
  \Phi\nonumber\\
  &  & + \frac{1}{2} \partial_j \Phi (\partial_i {\bf S}^j + \partial^j
  {\bf S}_i) + \frac{1}{2} \partial_j \Psi (\partial_i {\bf S}^j -
  \partial^j {\bf S}_i)\nonumber\\
  \mathcal{T}_{0 i}^{(S T)} & = & {f}_{i j} \partial_0 \partial^j \Psi -
  \frac{1}{2} \partial_0 {f}_{i j} \partial^j \Phi - \frac{1}{2}
  \partial_0 {f}_{i j} \partial^j \Psi\nonumber\\
  \mathcal{T}_{0 i}^{(V T)} & = & - \frac{1}{2} \partial_0 ({\bf S}^j
  \partial_0 {f}_{i j}) + \frac{1}{2} (\partial_0 {\bf S}^j)
  \partial_0 {f}_{i j} + \frac{1}{2}  {f}_{j k} \partial^j
  (\partial_i {\bf S}^k - \partial^k {\bf S}_i)\nonumber\\
  &  & + \frac{1}{2}  (\partial^j {\bf S}^k) \partial_k {f}_{i j} -
  \frac{1}{2}  (\partial_j {f}_{i k}) \partial^j {\bf S}^k.\label{T0i4d}
\end{eqnarray}
As shown in appendix \ref{appendix: 4dSVanalysis}, the energy-momentum tensor for gravitational waves will only receive contributions from the tensor mode. Thus we have 
\begin{eqnarray}
  \mathcal{T}_{0 i} & = & \frac{1}{2}  {f}^{j k} \partial_0 \partial_k
  {f}_{i j} - \frac{1}{2}  {f}^{j k} \partial_0 \partial_i
 {f}_{j k} - \frac{1}{4} \partial_0 {f}^{j k} \partial_i
  {f}_{j k}.
\end{eqnarray}

We can now compute the averaged radiated power
\ba
\langle P\rangle =\frac 1T  \int_0^T dt  \int d\Omega_2 R_\infty^2 \frac{x^i}{R_\infty} {\cal T}_{0i}\,,
\ea
where $T$ is the period of the gravitational waves and we substituted the normal unit vector as $n^i=x^i/R_\infty$.
Far away from the sources, the waves are spherical waves\footnote{One may wonder if, indeed, the transverse traceless tensor modes $f_{\mu\nu}$  which are the result of applying a projector which is local in momentum space and non-local in position space, are indeed spherical waves asymptotically far from sources. In  Appendix \ref{app:non-local} we solve the $SO(1,3)$ gauge-invariant fluctuations due to a static source.  In Appendix \ref{only tensor for non-compact geometry} we solve for the $SO(1,2)$ gauge-invariant fluctuations asymptotically far away from a binary source.  In either case the gauge-invariant fluctuations retain the generic feature of falling off with $1/r$, where $r$ is the distance to the source,  and are spherical waves in the second case.}
\ba
f_{ij}\sim \frac{\sin[\omega(t-R_\infty)]}{R_\infty}\,.
\ea
To leading order in $1/R_\infty$, the spatial derivatives can be replaced by 
\ba
\partial_i f_{jk}\sim \frac{x_i}{R_\infty} \partial_{R_\infty}  f_{jk}\sim - \frac{x_i}{R_\infty} \partial_{0}  f_{jk}\,.
\ea
Next we note that ${\cal T}_{0i}$ can be written as
\ba
 {\cal T}_{0i}=\frac 14\partial_0  f^{jk} \partial_i  f_{jk} - \frac 12\partial_i( f^{jk}\partial_0  f_{jk} )+\frac 12 \partial_k( f^{jk}\partial_0 f_{ij})\,.\label{two}
\ea
Asymptotically far away from the sources, the last two terms in \eqref{two} will average to zero as we will now show.  Consider one of those terms and start by  trading off the spatial derivative for a time derivative 
\ba 
 \int_0^T dt \int d\Omega_2 R_\infty^2 \frac{ x^i}{R_\infty} \partial_i ( f^{jk}\partial_0 f_{jk})=-
\int_0^T dt \int d\Omega_2 R_\infty^2 \partial_0( f^{jk}\partial_0 f_{jk})+{\cal O}\left(\frac 1{R_{\infty}}\right)\,.\label{av4d}
\ea
This vanishes since the integral in \eqref{av4d} is the integral of a  total derivative,  and the integrand is a periodic function with period $T$.
Therefore the averaged radiated power simplifies to
\ba
\label{eq:power-4d}
\langle P\rangle =\frac 1T  \int_0^T dt  \int d\Omega_2 R_\infty^2 \frac{x^i}{R_\infty}\frac 14 \partial_0 f^{jk} \partial_i  f_{jk} 
= \frac 1T  \int_0^T dt  \int d\Omega_2 R_\infty^2 \frac 14 \partial_0 f^{jk} \partial_0 f^{jk}\,.
\ea
This is a familiar result, which in the literature is obtained after going to the transverse-traceless gauge (see \cite{Ashtekar:2017wgq} for disambiguation regarding the various meanings of the "transverse traceless gauge"), as it is usually done for $4d$ flat spacetime gravitational waves, and performing the Isaacson average discussed in Section I. 

However we have arrived at it in a different way: we used only the gauge-invariant parts of the metric fluctuation to turn \eqref{2nd} into a manifestly gauge-invariant expression, and we only performed a time average over the period of the gravitational waves.

For yet another take on the same problem, in Appendix \ref{so12} we perform an $SO(1,2)$ SVT decomposition of the metric fluctuations and in  and Appendix \ref{only tensor for non-compact geometry} we solve explicitly for the gauge-invariant SVT components asymptotically far away from a binary source. Then  using the gauge-invariant metric fluctuations in the gravitational energy-momentum tensor \eqref{2nd} and the formula for the radiated power \eqref{radpow} we recover the known expression for the radiated power.

\subsection{5d flat spacetime}\label{sec:5d_flat}

\noindent Anticipating further applications to  models of extra dimensions  such as Kaluza-Klein theories (small extra dimensions)  or the Randall-Sundrum model (large extra dimension), next we will decompose the metric fluctuations about a 5d background into SVT components, with respect to the $SO(1,3)$ Lorentz group. In this section we have in mind a 5d flat spacetime, with one compact dimension $x^5\sim x^5+l$; this breaks the isometry group from $SO(1,4)$ to $SO(1,3)$.  
We proceed then with the following SVT decomposition:
\ba h_{MN} = \left(\begin{array}{cc}
     2 \psi \eta_{\mu \nu} + \partial_{\mu} \partial_{\nu} E + \partial_{\mu}
     F_{\nu} + \partial_{\nu} F_{\mu} + f_{\mu \nu} & \;\;\; \partial_{\mu} B +
     S_{\mu}\\
     \partial_{\nu} B + S_{\nu} & \;\;\;2 \phi
   \end{array}\right) \,,
\ea
where we have introduced the 5d indices $M, N= 0,1,2,3,5 $,  and $\partial_\mu S^\mu=\partial_\mu f^{\mu\nu}=\partial_\mu F^\mu=0$ and $f_{\mu\nu}\eta^{\mu\nu}=0$.\footnote{We performed a similar decomposition in 4d non-compact flat spacetime in Appendix \ref{so12} and Appendix \ref{only tensor for non-compact geometry}.}

The gauge-invariant metric fluctuations are $\Phi$, $\Psi$, ${\bf S}_{\mu}$ and ${f}_{\mu \nu}$, where
\begin{eqnarray}
  \Phi & = & \phi - \partial_5 (B - \tfrac12\partial_5 E)\,,\nn\\
  \Psi & = & \psi\,, \nn\\
  {\bf S}_{\mu} & = & S_{\mu} - \partial_5 F_{\mu}\,.
\end{eqnarray}
We gauge-fix such that the metric fluctuations contain only these gauge-invariant components:
\ba h_{M N} |_{g.i.}= \left(\begin{array}{cc}
     2 \Psi \eta_{\mu \nu} +  {f}_{\mu \nu} & {\bf S}_\nu\\
    {\bf S}_\mu & 2 \Phi
   \end{array}\right) \,.\label{hgi5}
\ea

Consider next the Einstein equations $ G_{M N} =8\pi G_{\text{5d}} T_{M N} $, where $G_{\text{5d}} $ is the 5d Newton's constant. To streamline our equations we adopt the same convention and set $8\pi G_{\text{5d}}=0 $. We perform the SVT decomposition
\begin{eqnarray}
  G_{M N} & = & \left(\begin{array}{cc}
    2 G^{(Y)} \eta_{\mu \nu} + \partial_{\mu} \partial_{\nu} G^{(LL)} +
    \partial_{\mu} G^{(L T)}_{\nu} + \partial_{\nu} G^{(L T)}_{\mu} +
    G^{(TT)}_{\mu \nu} & \;\;\;\; \partial_{\nu} G^{(L)}_5 + G^{(T)}_{\nu 5}\\
    \partial_{\mu} G^{(L)}_5 + G^{(T)}_{\mu 5} &  G_{55}
  \end{array}\right)\,,\nn\\
  T_{M N} & = & \left(\begin{array}{cc}
    2 T^{(Y)} \eta_{\mu \nu} + \partial_{\mu} \partial_{\nu} T^{(LL)} +
    \partial_{\mu} T^{(L T)}_{\nu} + \partial_{\nu} T^{(L T)}_{\mu} +
    T^{(TT)}_{\mu \nu} & \;\;\;\;\partial_{\nu} T^{(L)}_5 + T^{(T)}_{\nu 5}\\
    \partial_{\mu} T^{(L)}_5 + T^{(T)}_{\mu 5} &  T_{55}\,.
  \end{array}\right)\,.
\end{eqnarray}
The equations of motion for the scalar fluctuations $\Phi$ and $\Psi$ arise from
\begin{eqnarray}
 \delta^{(1)} G_{55} & = & 3 \eta^{\alpha \beta} \partial_{\alpha} \partial_{\beta} \Psi =
 T_{55}\,, \nn\\
 \delta^{(1)} G^{(L)}_5 & = & - 3 \partial_5 \Psi = T^{(L)}_5\,,\nn\\
2\delta^{(1)}  G^{(Y)} & = & 3 \partial_5^2 \Psi + \eta^{\alpha \beta} \partial_{\alpha}
  \partial_{\beta} \Phi + 2 \eta^{\alpha \beta} \partial_{\alpha}
  \partial_{\beta} \Psi = 2 T^{(Y)}\,, \nn\\
 \delta^{(1)} G^{(L L)} & = & - \Phi - 2 \Psi =  T^{(LL)}\,.
\end{eqnarray}
The equations of motion for the transverse vector fluctuations ${\bf S}_{\mu}$ arise from
\begin{eqnarray}
 \delta^{(1)} G^{(T)}_{\mu 5} & = & - \frac{1}{2} \eta^{\alpha \beta} \partial_{\alpha}
  \partial_{\beta}{\bf S}_{\mu} =  T^{(T)}_{\mu 5}\,,\\
\delta^{(1)}  G_{\mu}^{(L T)} & = & \frac{1}{2} \partial_5 {\bf S}_{\mu} = 
  T^{(L T)}_{\mu}\,.
\end{eqnarray}
Lastly, the equations of motion for the transverse traceless tensor fluctuations ${f}_{\mu \nu}$ come from $\delta^{(1)}G^{(TT)}_{\mu \nu}$:
\ba 
\delta^{(1)}G^{(TT)}_{\mu \nu} = - \frac{1}{2} \partial_5^2 {f}_{\mu \nu} -
   \frac{1}{2} \eta^{\alpha \beta} \partial_{\alpha} \partial_{\beta}
 {f}_{\mu \nu} =T^{(TT)}_{\mu \nu} \,.\ea
We can quickly count the degrees of freedom by considering the vacuum equations of motion
\begin{eqnarray}
  \eta^{\alpha \beta} \partial_{\alpha} \partial_{\beta} \Psi & = & 0\,,\qquad
  \partial_5 \Psi  =  0\,, \qquad
  \Phi  =  - 2 \Psi\,,\nn\\
  \eta^{\alpha \beta} \partial_{\alpha} \partial_{\beta}{\bf S}_{\mu} & = &
  0\,,\qquad
  \partial_5 {\bf S}_{\mu}  =  0\,,\nn\\
   \partial_5^2 {f}_{\mu \nu} &+& \eta^{\alpha \beta} \partial_{\alpha}
\partial_{\beta}{f}_{\mu \nu}  =  0\,.
\end{eqnarray}
When $\Psi$ (and therefore $\Phi$) is $x^5$-independent, ${\Psi}$ describes a 4d massless scalar, which has 1 degree of freedom.  For $x^5$-independent vector fluctuations, ${\bf S}_{\mu}$ describes a 4d massless vector which has 2 degrees of freedom. The $x^5$-independent tensor ${f}_{\mu \nu}$ describes a 4d massless graviton which has 2
  degrees of freedom.  This is  the scenario for Kaluza-Klein reduction,  when 5d gravity reduces to a 4d Einstein-Maxwell-dilaton theory.\footnote{For a more careful analysis of the zero-mode case leading to the same conclusion, see Appendix \ref{ZM}.}
Otherwise,  for $x^5$-dependent fluctuations, only the tensor ${f}_{\mu \nu}$ is non-zero and describes a 4d massive graviton, which has
 five degrees of freedom.

Next, we construct the energy-momentum tensor of gravitational waves.
We follow the same procedure as in the previously discussed 4d case. We write the
energy-momentum tensor for gravitational waves as
\ba \mathcal{T}_{MN} =\mathcal{T}^{(S)}_{MN} +\mathcal{T}^{(V)}_{MN} +\mathcal{T}^{(T)}_{MN} +\mathcal{T}^{(S V)}_{MN}
   +\mathcal{T}^{(S T)}_{MN} +\mathcal{T}^{(V T)}_{MN} \,,\ea
where $\mathcal{T}^{(S)}_{MN}$, $\mathcal{T}^{(V)}_{MN}$,
$\mathcal{T}^{(T)}_{MN} $ are terms involving scalar, vector, tensor
modes only and $\mathcal{T}^{(S V)}_{MN}$, $\mathcal{T}^{(S T)}_{MN}$, $\mathcal{T}^{(V T)}_{MN}$ are terms that mix different modes. Based on our earlier counting of degrees of freedom, the massive modes contribute only to ${\cal T}_{MN}^{(T)}$. 
We will focus on $\mathcal{T}_{0i}$ since it needed to compute the radiated power at infinity.  We do not need $\mathcal{T}_{05}$ due to the periodicity of the fifth dimension. 
\allowdisplaybreaks
\begin{eqnarray}
  \mathcal{T}^{(S)}_{0 i} & = & - \partial_0 \Psi \partial_i \Phi - \partial_i
  \Psi \partial_0 \Phi + \partial_0 \Phi \partial_i \Phi - 2 \partial_0 \Psi
  \partial_i \Psi - 2 \partial_0 (\Phi \partial_i \Phi) - 4 \partial_0 (\Psi
  \partial_i \Psi)\nn\\
  \mathcal{T}^{(V)}_{0 i} & = & - \frac{1}{2} \partial_{\alpha} ({\bf S}_0
  \partial^{\alpha} {\bf S}_i) + \frac{1}{2}  {\bf S}_0 \partial_{\alpha}
  \partial^{\alpha} {\bf S}_i + \frac{1}{2} \partial^{\alpha}
  ({\bf S}_{\alpha} (\partial_0 {\bf S}_i + \partial_i {\bf S}_0))\nn\\
  &  & + \frac{1}{2} \partial_0 {\bf S}_{\alpha} \partial_i
 {\bf S}^{\alpha} - \partial_0 ({\bf S}_{\alpha} \partial_i
 {\bf S}^{\alpha})\nn\\
  \mathcal{T}^{(T)}_{0 i} & = & - \frac{1}{2}  (\partial_5 ({f}_{0
  \alpha} \partial_5 {f}_i^{\alpha}) + \partial_{\beta} ({f}_{0
  \alpha} \partial^{\beta}{f}^{\alpha}_i) - {f}_{0 \alpha} 
  (\partial_5 \partial^5 + \partial_{\beta} \partial^{\beta})
 {f}_i^{\alpha})\nn\\
  &  & + \frac{1}{2} \partial^{\alpha}  \partial^{\beta} ({f}_{0 \beta}
 {f}_{i \alpha}  - {f}_{\alpha \beta} {f}_{0 i}) +
  \frac{1}{2} \partial^{\alpha} ({f}_{\alpha \beta} (\partial_0
  {f}_i^{\beta} + \partial_i {f}_0^{\beta}))\nn\\
  &  & + \frac{1}{4} \partial_0 {f}_{\alpha \beta} \partial_i
 {f}^{\alpha \beta} - \frac{1}{2} \partial_0 ({f}_{\alpha \beta}
  \partial_i {f}^{\alpha \beta})\nn\\
  \mathcal{T}^{(S V)}_{0 i} & = & \partial_5 ({\bf S}_i \partial_0 \Psi +
{S}_0 \partial_i \Psi + \Phi \partial_i {\bf S }_0 + \Phi \partial_0
  {\bf S}_i) - \frac{1}{2}  (\partial_0 {\bf S}_i + \partial_i
  {\bf S}_0) \partial_5 (\Phi + 2 \Psi)\nn\\
  \mathcal{T}^{(S T)}_{0 i} & = & - \frac{1}{2} \partial_5 (\Phi \partial_5
 {f}_{0 i}) - \frac{1}{2} \Phi (\partial_5 \partial^5 +
  \partial_{\alpha} \partial^{\alpha}) {f}_{0 i}\nn\\
  &  & + \frac{1}{2} \partial_{\alpha} (\Phi \partial^{\alpha} {f}_{0
  i} - \Phi \partial_0 {f}_i^{\alpha} - \Phi \partial_i
  {f}_0^{\alpha}) + \partial^{\alpha} ({f}_{i \alpha} \partial_0
  \Psi +{f}_{0 \alpha} \partial_i \Psi - {f}_{0 i}
  \partial_{\alpha} \Psi)\nn\\
  \mathcal{T}^{(V T)}_{0 i} & = & - \partial^{\alpha} ({\bf S}_{\alpha}
  \partial_5 {f}_{0 i}) + \frac{1}{2} \partial_5 ( {\bf S}_{\alpha}
  \partial_0 {f}_i^{\alpha} +{\bf S}_{\alpha} \partial_i
  {f}_0^{\alpha})\nn\\
  &  & + \frac{1}{2} \partial^{\alpha} ({\bf S}_0 \partial_5 {f}_{i
  \alpha} + {\bf S}_i \partial_5 {f}_{0 \alpha} - {f}_{0 i}
  \partial_5 {\bf S}_{\alpha})
\end{eqnarray}
These expressions can be greatly simplified under certain conditions. For example let us assume that all the source terms  have compact support and are localized at $x^5 = 0$. We will extract all the parts of $\mathcal{T}_{0i}$ that  give a non-vanishing contribution when computing the radiated power. Since the sources have compact support and are localized at $x^5 = 0$,  at spatial infinity the fluctuations will take  the following form, to leading order in $1/r$:  (i) spherical waves for the zero-modes (ii) exponentially suppressed with $r$ for the massive modes. For a binary source this behavior is: (i) $\exp(2 i \Omega(t-r))/r $, where $\Omega$ is the frequency of the binary sources,  for the zero-modes, (ii) exponentially suppressed $\exp( 2i \Omega t) \exp(i\, 2\pi n x^5/l) \exp(- r\sqrt{(2\pi n /l)^2-4\Omega^2})/r$, where $l$ is the periodicity of the fifth dimension, and $n$ is an integer, for the massive modes \cite{Du:2020rlx}.    Because the fifth dimension is periodic,  and we integrate over $x^5$ in computing the radiated power, we can drop any term that has only one derivative with respect to $x^5$. Furthermore,  because we compute the power at spatial infinity  we only need  the leading order
in ${1}/{r}$ for any fluctuation. As a consequence,  we can trade $\partial_i$ for $n_i \partial_0$ for the zero modes just like in previous section \ref{et4d}. Even though the massive modes do not depend on time through the combination $t-r$, given that the $\partial_r$ derivative must act on the exponential or else it will give a contribution which vanishes at spatial infinity, we can still trade $\partial_r$ for $\partial_0$ (appropriately multiplied by a frequency and $n$ dependent numerical factor).   Under time averaging,  any term that is a total 
derivative with respect to time  drops out.  
We are left with:
\begin{eqnarray}
  \mathcal{T}^{(S)}_{0 i} & \rightarrow & - \partial_0 \Psi \partial_i \Phi -
  \partial_i \Psi \partial_0 \Phi + \partial_0 \Phi \partial_i \Phi - 2
  \partial_0 \Psi \partial_i \Psi\nn\\
  \mathcal{T}^{(V)}_{0 i} & \rightarrow & \frac{1}{2} \partial_0
{\bf S}_{\alpha} \partial_i {\bf S}^{\alpha}\nn\\
  \mathcal{T}^{(T)}_{0 i} & \rightarrow & \frac{1}{4} \partial_0
{f}_{\alpha \beta} \partial_i {f}^{\alpha \beta}\nn\\
  \mathcal{T}^{(S V)}_{0 i} & \rightarrow & 0\nn\\
  \mathcal{T}^{(S T)}_{0 i} & \rightarrow & 0\nn\\
  \mathcal{T}^{(V T)}_{0 i} & \rightarrow & 0.
\end{eqnarray}
Combining all the parts,  the formula for the radiated power simplifies to
\begin{eqnarray}
 \langle P  \rangle &=&\frac{1}{T}  \int_0^T dt  \int dx^5 \int d\Omega_2 R_\infty^2  n^i \mathcal{T}_{0 i} \bigg|_{r=R_\infty} \nn\\
&= &\frac{1}{T}  \int_0^T dt  \int dx^5  \int d\Omega_2 R_{\infty}^2 \frac{x^i}{R_{\infty}} \bigg( - \partial_0 \Psi \partial_i \Phi -
  \partial_i \Psi \partial_0 \Phi + \partial_0 \Phi \partial_i \Phi - 2
  \partial_0 \Psi \partial_i \Psi\nn\\
  &  & \qquad\qquad\qquad \qquad\qquad + \frac{1}{2} \partial_0 {\bf S}_{\alpha} \partial_i
 {\bf S}^{\alpha} + \frac{1}{4} \partial_0 {f}_{\alpha \beta}
  \partial_i {f}^{\alpha \beta}\bigg) \bigg|_{r=R_\infty}\nn\\
  &=&\frac{l}{T}  \int_0^T dt   \int d\Omega_2 R_{\infty}^2  \bigg(  6 \partial_0 \Psi \partial_0 \Psi 
  + \frac{1}{2} \partial_0 {\bf S}_{\alpha} \partial_0
 {\bf S}^{\alpha}
+ \frac{1}{4} \partial_0 {f}_{\alpha \beta}
  \partial_0 {f}^{\alpha \beta}\bigg|_{\text{massless}}\bigg) \bigg|_{r=R_\infty}\nn\\
  &&+\frac{1}{T}  \int_0^T dt \int dx^5   \int d\Omega_2 R_{\infty}^2 
  \bigg(\frac{1}{4} \partial_0 {f}_{\alpha \beta}
 \partial_r{f}^{\alpha\beta}\bigg|_{\text{massive}}\bigg)\bigg|_{r=R_\infty}\,,\label{5dradpow}
\end{eqnarray}
where we have highlighted the contributions of the massless and massive 
sectors and we have used that asymptotically far from the sources the scalars are related by the vacuum equation $\Phi+2\Psi=0$.  Due to the exponential suppression with $r$ in the massive mode sector for source frequencies $\Omega>2\pi/l$, it is reasonable to approximate \eqref{5dradpow} by keeping  only the massless sector contribution. We complete the check of the formula for the radiated power \label{5drad} in Appendix \ref{allmodescompact} by concretely solving for the gauge-invariant fluctuations and computing the luminosity of a binary source. We show that we reproduce previous results in the literature.


\subsection{Randall-Sundrum model}\label{sec:RS}

\noindent 
For the Randall-Sundrum model we start with the Einstein's equation,
\ba
R_{MN}-\frac 12 g_{MN}(R-2\Lambda)+\frac 12 \lambda \frac{\sqrt{-\det({}^*g_{\mu\nu})}}{\sqrt{-\det(g_{MN})}}\delta_M^\mu\delta_N^\nu {}^*g_{\mu\nu}\delta(y)=T_{MN}\,,\label{rseinstein}
\ea
where $T_{MN}$ are the matter sources,  ${}^*g_{\mu\nu}$ is the pull-back of the bulk metric to the 3-brane located at  $y=0$, and where the brane tension  $\lambda$ and the cosmological constant $\Lambda$ are tuned such that  
\ba
\Lambda=-6\kappa^2\;\;\; \text{and}\;\;\; \lambda=12\kappa\, .
\ea
Given the $SO(1,3)$ isometry of the background metric, $ds^2 = dy^2 + e^{- 2\kappa | y |} \eta_{\mu \nu}
dx^{\mu} dx^{\nu},$  we start by decomposing  the metric perturbation into scalar-vector-tensor fluctuations as follows:
\ba h_{MN} = \left(\begin{array}{cc}
      2 \psi \eta_{\mu \nu} + \partial_{\mu} \partial_{\nu} E +
     \partial_{\mu} F_{\nu} + \partial_{\nu} F_{\mu} + f_{\mu \nu} & \;\;\;
     \partial_{\mu} B + S_{\mu}\\
     \partial_{\nu} B + S_{\nu} & \;\;\;2 \phi
   \end{array}\right)\,. \label{Rsdecomp}
   \ea

The gauge invariant fluctuations are $\Phi$, $\Psi$, ${\bf S}_{\mu}$
and ${f}_{\mu \nu}$, where
\begin{eqnarray}
  \Phi & = & \phi - \partial_y \bigg( B - \tfrac 12 e^{-2 \kappa |y|} \partial_y (e^{2\kappa |y|} E)\bigg)\,,\nn\\
  \Psi & = & \psi - \tfrac 12 (\partial_y e^{-2 \kappa |y|}) (B - \tfrac 12 e^{-2 \kappa |y|} \partial_y  (e^{2\kappa|y|} E))\,,\nn\\
  {\bf S}_{\mu} & = & S_{\mu} - e^{-2\kappa|y|} \partial_y (e^{2\kappa |y|} F_{\mu}) \,.
\end{eqnarray}
 Next we perform the same SVT decomposition on the left-hand-side of Einstein's equation \eqref{rseinstein} which we denote here by $\mathcal E_{MN}$
\ba
  \!\!\!\!\!\!\!\!\!\mathcal E_{M N} & = & \left(\begin{array}{cc}
2 \mathcal E^{(Y)} \eta_{\mu \nu} + \partial_{\mu} \partial_{\nu}
    \mathcal E^{(LL)} + \partial_{\mu} \mathcal E^{(L T)}_{\nu} + \partial_{\nu} \mathcal E^{(L
    T)}_{\mu} + \mathcal E^{(T)}_{\mu \nu}& \;\;\;\partial_{\nu} \mathcal E^{(L)}_y +
    \mathcal E^{(T)}_{\nu y}\\
    \partial_{\mu} \mathcal E^{(L)}_y + \mathcal E^{(T)}_{\mu y} & \;\;\; \mathcal E_{yy}
  \end{array}\right)\,,
\ea
and to the matter sources on the right-hand-side of \eqref{rseinstein} (note that we included the brane contribution in $\mathcal E_{MN}$):
\ba
  \!\!\!\!\!\!\!\!\!T_{M N} & = & \left(\begin{array}{cc}
2 T^{(Y)} \eta_{\mu \nu} + \partial_{\mu} \partial_{\nu}
    T^{(LL)} + \partial_{\mu} T^{(L T)}_{\nu} + \partial_{\nu} T^{(L
    T)}_{\mu} + T^{(T)}_{\mu \nu} & \;\;\; \partial_{\nu} T^{(L)}_y +
    T^{(T)}_{\nu y}\\
   \partial_{\mu} T^{(L)}_y + T^{(T)}_{\mu y} &  T_{yy}
  \end{array}\right)\,.
\ea
Then we expand in fluctuations and write the linearized equations of motion for the gauge-invariant fluctuations. 
The linearized equations of motion for the scalars $\Phi$ and $\Psi$ arise from the components $\delta^{(1)}E_{yy}$,
$\delta^{(1)}\mathcal E^{(L)}_y$, $\delta^{(1)}\mathcal E^{(Y)}$and $\delta^{(1)}\mathcal E^{(LL)}$:
\begin{eqnarray}
  \delta^{(1)} \mathcal E_{yy} & =&e^{2\kappa|y|} \bigg[3e^{2\kappa|y|}\eta^{\alpha\beta}\partial_\alpha\partial_\beta +12 e^{\kappa|y|} (e^{-\kappa|y|})'\partial_y-12 e^{2\kappa|y|} ((e^{-\kappa|y|})')^2 +4e^{\kappa|y|} (e^{-\kappa|y|})''\nn\\
&
+&\frac 23 \lambda \delta(y)+\frac83 \Lambda \bigg]\Psi
+
\bigg[-\frac 23 \lambda\delta(y)-\frac 23 \Lambda -12 e^{2\kappa|y|}((e^{-\kappa|y|})')^2-4e^{\kappa|y|}(e^{-\kappa|y|})''\bigg]\Phi\nn\\
&=&e^{2\kappa|y|} \bigg[3e^{2\kappa|y|}\eta^{\alpha\beta}\partial_\alpha\partial_\beta -12\kappa\text{sign}(y)\partial_y-
24\kappa^2 \bigg]\Psi -12\kappa^2\Phi
   =  T_{yy}\,,\nn\\
\delta^{(1)}  \mathcal E^{(L)}_y & = &
\bigg[-3e^{2\kappa|y|}\partial_y +6e^{3\kappa|y|}(e^{-\kappa|y|})'\bigg]\Psi+3e^{\kappa|y|}(e^{-\kappa|y|})'\Phi\nn\\
&=&
e^{2\kappa|y|}\bigg[-3\partial_y -6\kappa\,\text{sign}(y)\bigg]\Psi-3\kappa\,\text{sign}(y)\Phi=
   T^{(L)}_y\,,\nn\\
2 \delta^{(1)} \mathcal E^{(Y)} & = &\bigg[
2e^{2\kappa|y|} \eta^{\alpha\beta}\partial_\alpha\partial_\beta+3\partial_y^2-4e^{\kappa|y|}(e^{-\kappa|y|})''+
\frac 13 \lambda\delta(y)+\frac 43\Lambda\bigg]\Psi
\nn\\&+&e^{-2\kappa|y|}\bigg[\eta^{\alpha\beta}\partial_\alpha\partial_\beta+\frac 16\lambda\delta(y)+\frac 23 \Lambda-2e^{\kappa|y|}(e^{-\kappa|y|})''-6e^{2\kappa|y|}((e^{-\kappa|y|})')^2-3e^{\kappa|y|}(e^{-\kappa|y|})'\partial_y\bigg]\Phi\nn\\
 &=&\bigg[
2e^{2\kappa|y|} \eta^{\alpha\beta}\partial_\alpha\partial_\beta+3\partial_y^2-12\kappa^2+12\kappa\,\delta(y)\bigg]\Psi\nn\\
&+&e^{-2\kappa|y|}\bigg[\eta^{\alpha\beta}\partial_\alpha\partial_\beta+3\kappa\,\text{sign}(y)\partial_y-12\kappa^2+6\kappa\delta(y)\bigg]\Phi
= 2 T^{(Y)}\,,\nn\\
\delta^{(1)}  \mathcal E^{(L L)} & = & - \Phi - 2 e^{2\kappa|y|}\Psi =  T^{(LL)}\,,
\end{eqnarray}
where we used primes to denote differentiation with respect to $y$.
The linearized equations of the vector fluctuation ${\bf S}_{\mu}$ come from the components $\delta^{(1)}{\cal E}^{(T)}_{\mu y}$
and $\delta^{(1)} \mathcal E^{(L T)}_{\mu}$:
\begin{eqnarray}
\delta^{(1)}  \mathcal E^{(T)}_{\mu y} & = & 
\bigg[-\frac 12 e^{2\kappa|y|}\eta^{\alpha\beta}\partial_\alpha\partial_\beta-e^{\kappa|y}(e^{-\kappa|y|})''-3e^{2\kappa|y|}
((e^{-\kappa|y|})')^2-\frac 23 \lambda\delta(y)-\frac 23 \Lambda\bigg]{\bf S}_\mu\nn\\
& =&\bigg[-\frac 12 e^{2\kappa|y|} \eta^{\alpha\beta}\partial_\alpha\partial_\beta-6\kappa\delta(y)\bigg]{\bf S}_\mu=
  T^{(T)}_{\mu y}\,,\nn\\
\delta^{(1)} \mathcal E_{\mu}^{(L T)} & = &  \bigg[\frac 12\partial_y -e^{-\kappa|y|}(e^{-\kappa |y|})'\bigg] {\bf S}_\mu=
 \bigg[\frac 12 \partial_y -\kappa\text{sign}(y)\bigg] {\bf S}_\mu= T^{(L T)}_{\mu}\,.
\end{eqnarray}
The  tensor ${f}_{\mu \nu}$ equation of motion is:
\ba
\delta^{(1)}\mathcal E^{(T)}_{\mu \nu}& = &
\bigg[-\frac 12  e^{2\kappa|y|}\eta^{\alpha\beta}\partial_\alpha\partial_\beta --\frac 12 \partial_y^2  -\frac 16\lambda\delta(y)-\frac 23 \Lambda -2e^{2\kappa|y|} ((e^{-\kappa|y|})')^2\bigg]f_{\mu\nu}\nn\\
&=&
\bigg[-\frac 12e^{2\kappa|y|} \eta^{\alpha\beta}\partial_\alpha\partial_\beta
-\frac 12 \partial_y^2+2\kappa^2-2\kappa \delta(y)\bigg]f_{\mu\nu} = 
   T^{(T)}_{\mu \nu}\,.
   \ea
and originates in $\delta^{(1)}{\cal E}_{\mu\nu}$. In the absence of matter sources we recognize here our earlier equation \eqref{ttfrs}.

In vacuum, the 
set of equations obeyed by the gauge-invariant fluctuations reduces to
\begin{eqnarray}
 && \eta^{\alpha \beta} \partial_{\alpha} \partial_{\beta} \Psi 
  = 0\,,\qquad \partial_y \Psi = 0\,,\qquad
  \Phi + 2 e^{2\kappa |y|}  \Psi  =  0\,,\nn\\
&&
 e^{2\kappa|y|} \eta^{\alpha\beta}\partial_\alpha\partial_\beta {\bf S}_\mu=-12\kappa\delta(y){\bf S}_\mu\,,\qquad\bigg[\frac 12 \partial_y -\kappa\text{sign}(y)\bigg] {\bf S}_\mu=0 \,,\nn\\
&&\bigg[-\frac 12 e^{2\kappa|y|}\eta^{\alpha\beta}\partial_\alpha\partial_\beta
-\frac 12 \partial_y^2+2\kappa^2-2\kappa \delta(y)\bigg]f_{\mu\nu} = 0\,.\label{rsvac}
\end{eqnarray}
The linearized vector vacuum equations admit no solution due to the delta-function present on the right-hand-side of the equations \eqref{rsvac} and the absence of any $\partial_y$ derivatives, which imply that ${\bf S}_\mu$ vanishes on the brane. The tensor equation however does not suffer from this problem and admits solutions. The scalar equations admit solutions for null 4-momenta, but the $\Psi$ scalar metric fluctuations are $|y|$-independent, and the $\Phi$ fluctuations are growing with $|y|$.  Both fluctuations are non-normalizable, exponentially growing with $|y|$ relative to the background metric. 

We are now turning our attention to constructing the gravitational energy-momentum tensor ${\cal T}_{\mu\nu}$.  It is quadratic in fluctuations: we denote the scalar, vector and tensor contributions by $ \mathcal{T}^{(S)}_{\rho\sigma},\mathcal{T}_{\rho\sigma}^{(V)},\mathcal{T}_{\rho\sigma}^{(T)}$ and the mixed contributions by  $\mathcal{T}^{(SV)}_{\rho\sigma}$, for the mixed scalar-vector contribution,  etc.  We give each one of these expressions below:
	\begin{align}
	\mathcal{T}^{(S)}_{\rho\sigma}=	& \eta_{\rho \sigma } \bigg[-4 e^{-\kappa|y|} (e^{-\kappa|y|})'\Phi \partial_{y}\Phi - 4 e^{\kappa|y|}   (e^{-\kappa|y|})'  \Psi \partial_{y}\Phi  + 6\partial_{y}\Psi \partial_y\Phi \nonumber \\ 
	&	 - 4   e^{-\kappa|y|} (e^{-\kappa|y|})' \Phi \partial_{y}\Phi
  - 12 \Phi^2 ( (e^{-\kappa|y|})')^2
-  8e^{2\kappa|y|}   ((e^{-\kappa|y|})')^2 \Phi \Psi  \nonumber \\ 
		& + 8 e^{4\kappa|y|}((e^{-\kappa|y|})')^2 { \Psi^2}{} -  {8  e^{\kappa|y|}(e^{-\kappa|y|})' \Phi \partial_{y}\Psi}{} - 3  \partial_{y}\Phi \partial_{y}\Psi\nn\\
	&	+ {16 e^{\kappa|y|}  (e^{-\kappa|y|})' \Phi \partial_{y}\Psi}{} -  {8 e^{3\kappa |y|} (e^{-\kappa|y|})'\Psi \partial_{y}\Psi}- 4  e^{-\kappa|y|} (e^{-\kappa|y|})''  \Phi^2 \nonumber \\ 
		& -  {8 e^{3\kappa|y|} (e^{-\kappa|y|})'' } \Psi^2 + 2 \Phi \partial_{y}\partial_{y}\Psi + {4 e^{2\kappa|y|}  \Psi \partial_{y}\partial_{y}\Psi}{} 
		 + \eta^{\alpha \beta }  \partial_{\alpha }\Phi \partial_{\beta }\Phi -  {\eta^{\alpha \beta }e^{2\kappa|y|} \partial_{\alpha }\Phi \partial_{\beta }\Psi}{} \nonumber \\ 
		& + {3 e^{4\kappa|y|}  \eta^{\alpha \beta } \partial_{\alpha }\Psi \partial_{\beta }\Psi}{} + \eta^{\alpha \beta } \Phi \partial_{\beta }\partial_{\alpha }\Phi
		 + {e^{2\kappa|y|}\eta^{\alpha \beta }\Psi \partial_{\beta }\partial_{\alpha }\Phi}{} + {4 e^{4\kappa|y|} \eta^{\alpha \beta } \Psi \partial_{\beta }\partial_{\alpha }\Psi}{}\bigg] \nonumber \\ 
		& -  \partial_{\rho }\Phi \partial_{\sigma }\Phi - e^{2\kappa |y|}(  {\partial_{\rho }\Psi \partial_{\sigma }\Phi}{} +  {\partial_{\rho }\Phi \partial_{\sigma }\Psi}{} )-  {6 e^{4\kappa|y|} \partial_{\rho }\Psi \partial_{\sigma }\Psi}{} 
		- 2 \Phi \partial_{\sigma }\partial_{\rho }\Phi -  {4e^{4\kappa|y|} \Psi \partial_{\sigma }\partial_{\rho }\Psi}{}\,,
\nn\\
&- \eta_{\rho\sigma} \lambda \delta (y)\bigg[\tfrac{1}{4} \Phi^2  + 3\Phi \Psi\bigg]\,, \\
	\mathcal{T}_{\rho\sigma}^{(V)}=	&\eta_{\rho \sigma } \bigg[- \tfrac 32{e^{2\kappa|y|}  {\bf S}_{\alpha } {\bf S}^{\alpha } ((e^{-\kappa|y|})')^2}{} -  {3 e^{\kappa|y|} (e^{-\kappa|y|})'   {\bf S}^{\alpha } \partial_{y}{\bf S}_{\alpha }}{}
 -  \tfrac 32{ e^{\kappa|y|}  {\bf S}_{\alpha } {\bf S}^{\alpha } (e^{-\kappa|y|})''}{ }\nn\\
&+ \tfrac 34{e^{2\kappa|y|} \eta^{\beta \gamma }   \partial_{\beta }{\bf S}^{\alpha } \partial_{\gamma }{\bf S}_{\alpha }}{ }
+ \tfrac 12 {e^{2\kappa|y|} \eta^{\beta \gamma } {\bf S}_{\alpha } \partial_{\gamma }\partial_{\beta }{\bf S}^{\alpha }}{ }-\tfrac 14 e^{2\kappa|y|} {\partial_{\alpha }{\bf S}^{\beta } \partial_{\beta }{\bf S}^{\alpha }}{} 
-\tfrac12\lambda \delta (y) S_{\alpha } S^{\alpha } \bigg]\nonumber \\ 
		& 
		 +\tfrac 12 e^{2\kappa|y|} {{\bf S}^{\alpha } \partial_{\alpha }(\partial_{\rho }{\bf S}_{\sigma }+ \partial_{\sigma }{\bf S}_{\rho }}{} ) 
		- \tfrac 12 e^{2\kappa|y|} { \eta^{\alpha \beta } \partial_{\alpha }{\bf S}_{\rho } \partial_{\beta }{\bf S}_{\sigma }}{}
		 -  \tfrac 12 e^{2\kappa|y|}{\partial_{\rho }{\bf S}^{\alpha } \partial_{\sigma }{\bf S}_{\alpha }}{} 
		-  e^{2\kappa|y|}{{\bf S}_{\alpha } \partial_{\sigma }\partial_{\rho }{\bf S}^{\alpha }}{}\nn\\
	\mathcal{T}_{\rho\sigma}^{(T)}=	& \eta_{\rho \sigma } \bigg[
- \tfrac32 e^{3\kappa|y|}  (e^{-\kappa|y|})' f^{\alpha \beta }\partial_{y}f_{\alpha \beta } + \tfrac38 e^{2\kappa|y|}   \eta^{\alpha \beta }  \eta^{\gamma \kappa }  \partial_{y}f_{\alpha \gamma } \partial_{y}f_{\beta \kappa }{}  
+ \tfrac38 e^{4\kappa|y|} { \eta^{\gamma \kappa }   \partial_{\gamma }f^{\alpha \beta } \partial_{\kappa }f_{\alpha \beta }}{}\nonumber \\ 
& + e^{4\kappa|y|} ((e^{-\kappa|y|})')^2  f_{\alpha \beta } f^{\alpha \beta } {} 
+ \tfrac 14 e^{2\kappa|y|} {f^{\alpha \beta }  \partial_{y}^2 f_{\alpha \beta }}{} 
 -  \tfrac 12 e^{3\kappa|y|} {f_{\alpha \beta } f^{\alpha \beta } (e^{-\kappa|y|})''} \nn\\
& + \tfrac 14 e^{4\kappa|y|} {f_{\alpha \beta }  \eta^{\gamma \kappa }  \partial_{\kappa }\partial_{\gamma }f^{\alpha \beta }}{}
- \tfrac 14 e^{4\kappa|y|} \partial_{\beta }f^{\alpha \gamma } \partial_{\gamma }f_{\alpha }{}^{\beta }{} \bigg]\nonumber \\ 
		& + e^{3\kappa|y|}(e^{-\kappa|y|})'({f_{\sigma }{}^{\alpha } \partial_{y}f_{\rho \alpha }}{} +{f_{\rho }{}^{\alpha } \partial_{y}f_{\sigma \alpha }}{} )
		 -  \tfrac 12 e^{2\kappa|y|}{ \eta^{\alpha \beta } \partial_{y}f_{\rho \alpha } \partial_{y}f_{\sigma \beta }}{} -  {2 e^{4\kappa|y|} ((e^{-\kappa|y|})')^2  f_{\rho \alpha } f_{\sigma }{}^{\alpha }}{} \nonumber \\ 
		&+ e^{4 \kappa|y|}\bigg[
\tfrac 12  {\partial_{\alpha }f_{\sigma }{}^{\beta } \partial_{\beta }f_{\rho }{}^{\alpha }}{}
		- \tfrac 12 {f^{\alpha \beta } \partial_{\beta }\partial_{\alpha }f_{\rho \sigma }}{} + \tfrac12  {f_{\alpha }{}^{\beta } \partial_{\beta }\partial_{\rho }f_{\sigma }{}^{\alpha }}{} 
		+\tfrac 12  {f_{\alpha }{}^{\beta } \partial_{\beta }\partial_{\sigma }f_{\rho }{}^{\alpha }}{}  \nn\\&
		 - \tfrac 12 { \eta^{\beta \gamma } \partial_{\beta }f_{\rho }{}^{\alpha } \partial_{\gamma }f_{\sigma \alpha }}{ } 
		 -  \tfrac 14 {\partial_{\rho }f^{\alpha \beta } \partial_{\sigma }f_{\alpha \beta }}{}
		 -  \tfrac 12 {f_{\alpha \beta } \partial_{\sigma }\partial_{\rho }f^{\alpha \beta }}{}\bigg]\nn\\
&- \tfrac14 \eta_{\rho \sigma} {\lambda \delta (y) f_{\alpha \beta } f^{\alpha \beta } }{}\,,\\
	\mathcal{T}_{\rho\sigma}^{(SV)}=	&e^{2\kappa|y|} {\eta_{\rho \sigma } {\bf S}^{\alpha } \partial_{y}\partial_{\alpha }\Psi}{} + 
e^{2\kappa|y|} {\eta^{\alpha \beta } \eta_{\rho \sigma } \partial_{y}{\bf S}_{\beta } \partial_{\alpha }\Psi}{}  \nn\\
&+ \Phi \partial_{y}\partial_{\rho }{\bf S}_{\sigma } + e^{2\kappa|y|} {{\bf S}_{\sigma } \partial_{y}\partial_{\rho }\Psi}{} 
		 + \Phi \partial_{y}\partial_{\sigma }{\bf S}_{\rho } + e^{2\kappa|y|}{{\bf S}_{\rho } \partial_{y}\partial_{\sigma }\Psi}{}
		 + \tfrac{1}{2} \partial_{y}\Phi \partial_{\rho }{\bf S}_{\sigma } + {2e^{\kappa|y|} (e^{-\kappa|y|})' \Phi \partial_{\rho }{\bf S}_{\sigma }}{} \nn\\
&+ {2e^{3\kappa|y|}  (e^{-\kappa|y|})' \Psi \partial_{\rho }{\bf S}_{\sigma }}{}
		 - e^{2\kappa|y|} {\partial_{y}\Psi \partial_{\rho }{\bf S}_{\sigma }}{} + e^{2\kappa|y|}{\partial_{y}{\bf S}_{\sigma } \partial_{\rho }\Psi}{} + \tfrac{1}{2} \partial_{y}\Phi \partial_{\sigma }{\bf S}_{\rho } 
		 + {2e^{\kappa|y|}  (e^{-\kappa|y|})'\Phi \partial_{\sigma }{\bf S}_{\rho }}{}\nn\\
& + {2 e^{3\kappa|y|} (e^{-\kappa|y|})' \Psi \partial_{\sigma }{\bf S}_{\rho }}{} -  e^{2\kappa|y|} {\partial_{y}\Psi \partial_{\sigma }{\bf S}_{\rho }}{}
		 + e^{2\kappa|y|}{\partial_{y}{\bf S}_{\rho } \partial_{\sigma }\Psi}{}\,,\\
	\mathcal{T}_{\rho\sigma}^{(ST)}=&	\tfrac 12e^{2\kappa|y|} {f^{\alpha \beta } \eta_{\rho \sigma } \partial_{\beta }\partial_{\alpha }\Phi}{} +
\tfrac{1}{2} \partial_{y}\Phi \partial_{y}f_{\rho \sigma } -  \partial_{y}f_{\rho \sigma } \partial_{y}\Phi - 4e^{2\kappa|y|}  ((e^{-\kappa|y|})')^2 f_{\rho \sigma } \Phi{} \nonumber \\ 
		& -  {8 e^{4\kappa|y|} ((e^{-\kappa|y|})')^2 f_{\rho \sigma } \Psi}{} 
+ {4 e^{3\kappa|y|} (e^{-\kappa|y|})'  f_{\rho \sigma }  \partial_{y}\Psi}{} -  \Phi \partial_{y}\partial_{y}f_{\rho \sigma } 
		 + \tfrac 12 e^{2\kappa|y|} {\eta^{\alpha \beta } \partial_{\alpha }\Phi \partial_{\beta }f_{\rho \sigma }}{} \nn\\
&-  e^{4\kappa|y|}{\eta^{\alpha \beta } \partial_{\alpha }\Psi \partial_{\beta }f_{\rho \sigma }}{}
		-  e^{4\kappa|y|} {\eta^{\alpha \beta } \Psi \partial_{\beta }\partial_{\alpha }f_{\rho \sigma }}{}  -  \tfrac 12 e^{2\kappa|y|} {\partial_{\alpha }\Phi \partial_{\rho }f_{\sigma }{}^{\alpha }}{} + e^{4\kappa|y|} {f_{\sigma }{}^{\alpha } \partial_{\rho }\partial_{\alpha }\Psi}{} \nn\\&-  \tfrac 12 e^{2\kappa|y|}{\partial_{\alpha }\Phi \partial_{\sigma }f_{\rho }{}^{\alpha }}{} 
		 + e^{4\kappa|y|}{f_{\rho }{}^{\alpha } \partial_{\sigma }\partial_{\alpha }\Psi}{}\nn\\
&+ \lambda \delta (y) f_{\rho \sigma } \bigg[\tfrac 12 \Phi + {2\Psi}{}\bigg]\,,\\
	\mathcal{T}_{\rho\sigma}^{(VT)}=	& e^{2\kappa|y|}\bigg[- \tfrac 12\eta_{\rho \sigma } {\eta^{\beta \gamma }  \partial_{y}f_{\alpha \beta } \partial_{\gamma }{\bf S}^{\alpha }}{}-
 { {\bf S}^{\alpha } \partial_{y}\partial_{\alpha }f_{\rho \sigma }}{} -  \tfrac12{f^{\alpha \beta } \eta_{\rho \sigma } \partial_{y}\partial_{\beta }{\bf S}_{\alpha }}{} 
		 + \tfrac12{{\bf S}^{\alpha } \partial_{y}\partial_{\rho }f_{\sigma \alpha }}{} \nn\\&+
 \tfrac12 {{\bf S}^{\alpha } \partial_{y}\partial_{\sigma }f_{\rho \alpha }}{} + e^{\kappa|y|}(e^{-\kappa|y|})' {{\bf S}^{\alpha } \partial_{\alpha }f_{\rho \sigma }}{}
		 -  e^{\kappa|y|}(e^{-\kappa|y|})'{f_{\sigma }{}^{\alpha }\partial_{\alpha }{\bf S}_{\rho }}{} - 
 e^{\kappa|y|} (e^{-\kappa|y|})' {f_{\rho }{}^{\alpha } \partial_{\alpha }{\bf S}_{\sigma }}{} \nonumber \\ 
		& -  \tfrac12 {\eta^{\alpha \beta } \partial_{y}{\bf S}_{\alpha } \partial_{\beta }f_{\rho \sigma }}{} + \tfrac12 {\eta^{\alpha \beta } \partial_{y}f_{\sigma \alpha } \partial_{\beta }{\bf S}_{\rho }}{}
		 + \tfrac 12{\eta^{\alpha \beta } \partial_{y}f_{\rho \alpha } \partial_{\beta }{\bf S}_{\sigma }}{}  
		 + \tfrac12 {\partial_{y}{\bf S}_{\alpha } \partial_{\rho }f_{\sigma }{}^{\alpha }}{} + 
\tfrac 12 {\partial_{y}{\bf S}_{\alpha } \partial_{\sigma }f_{\rho }{}^{\alpha }}{}	\bigg]\,,
	\end{align}
where the 4d indices $\alpha, \beta$ on the vector and tensor fluctuations have been raised with the Minkowski metric e.g. ${\bf S}^\alpha=\eta^{\alpha\beta} {\bf S}_\beta$. The delta-function terms arise due to the presence of the brane. 

Under the same assumption that all sources have compact support, let us extract those parts in $\mathcal{T}_{\mu\nu}$ which are relevant for computing the radiated power. For $\kappa r\gg1$ we derived in Section \ref{Sphrs} the profile of spherical waves in Randall-Sundrum background.  We see that the same arguments we have been using earlier still apply.  First,  for the radiated power we only need ${\cal T}_{0i}$ if we chose to compute the rate of the energy flux through the surface of an infinitely long cylinder (we are thus keeping $r=R_\infty$ large and fixed and integrating over $y$).  This is merely a convenient choice of the surface enclosing the sources,  and keeping with our assumption that we are measuring the radiated power far away from the sources. Second,  the relevant terms in ${\cal T}_{0i}$ which contribute to the averaged radiated power can be be found by (i) dropping all total $\partial_i$ derivatives (since these can be turned in $(x^i/r) \partial_r$ in $\mathcal T_{0i}^{(S)}$ and  $\mathcal T_{0i}^{(V)}$; the $\partial_r$ derivative must  act on the phase of the spherical wave otherwise  it will lead to a flux which vanishes asymptotically far away ($r=R_\infty \to\infty$); at this point $\partial_r$ can be converted into $\partial_t$; lastly the time average will set this term to zero); similarly, total derivatives can be dropped from $\mathcal T_{0i}^{(T)}$; (ii) dropping all the terms which are odd in $y$ such as single $\partial_y$ derivatives.  What is left is
\allowdisplaybreaks
\begin{eqnarray}
  \mathcal{T}^{(S)}_{0 i} & \rightarrow & \partial_0 \Phi \partial_i \Phi - 2 e^{4 \kappa | y |}
  \partial_0 \Psi \partial_i \Psi - e^{2 \kappa | y |} (\partial_0 \Psi
  \partial_i \Phi + \partial_0 \Phi \partial_i \Psi)\nn\\
  \mathcal{T}^{(V)}_{0 i} & \rightarrow &  \tfrac{1}{2} {\bf S}_0  e^{2 \kappa | y |}\eta^{\alpha\beta}\partial_\alpha\partial_\beta {\bf S}_i +\tfrac{1}{2} e^{2 \kappa | y |}\partial_0
  {\bf{S}}_{\alpha} \partial_i {\bf{S}}^{\alpha}\,,\nn\\
  \mathcal{T}^{(T)}_{0 i} & \rightarrow &  - \tfrac{1}{2} e^{- 2 \kappa | y |} \partial_y (e^{2 \kappa
  | y |} f_{0 \alpha}) \partial_y (e^{2 \kappa | y |} f_i^{\alpha}) -
  \tfrac{1}{2} e^{4 \kappa | y |} f_0^{\alpha} \Box^{(4 d)} f_{i \alpha}
  + \tfrac{1}{4} e^{4 \kappa | y |} \partial_0 f_{\alpha \beta}
  \partial_i f^{\alpha \beta}\,,\nn\\
  \mathcal{T}^{(S V)}_{0 i} & \rightarrow & 0\,,\nn\\
  \mathcal{T}^{(S T)}_{0 i} & \rightarrow & \frac{1}{2} \partial_y \Phi \partial_y f_{0
  i} - \partial_y f_{0 i} \partial_y \Phi - 4 \kappa^2 f_{0 i} \Phi\nn\\
  &  & - 8 \kappa^2 e^{2 \kappa | y |} f_{0 i} \Psi - 4 \kappa\, \text{sign} (y)\,
  e^{2 \kappa | y |} f_{0 i} \partial_y \Psi - \Phi \partial_y^2 f_{0
  i} + \frac{1}{2} e^{2 \kappa | y |} \eta^{\alpha \beta} \partial_{\alpha}
  \Phi \partial_{\beta} f_{0 i}\nn\\
  &  & - e^{4 \kappa | y |} \eta^{\alpha \beta} \partial_{\alpha} \Psi
  \partial_{\beta} f_{0 i} - e^{4 \kappa | y |} \Psi \eta^{\alpha\beta}\partial_\alpha\partial_\beta f_{0 i} -
  \frac{1}{2} e^{2 \kappa | y |} \partial_{\alpha} \Phi \partial_0
  f_i^{\alpha} + e^{4 \kappa | y |} f_i^{\alpha} \partial_0 \partial_{\alpha}
  \Psi\nn\\
  &  & - \frac{1}{2} e^{2 \kappa | y |} \partial_{\alpha} \Phi \partial_i
  f_0^{\alpha} + e^{4 \kappa | y |} f_0^{\alpha} \partial_i \partial_{\alpha}
  \Psi\nn\\
  &  & + \left( 2 \Psi f_{0 i} + \frac{1}{2} \Phi f_{0 i} \right) \lambda
  \delta (y)\,,\nn\\
  \mathcal{T}^{(VT)}_{0 i} & \rightarrow & 0\,.\label{caltrs}
\end{eqnarray}

The luminosity (radiated power) of the gravitational waves is obtained by substituting the expressions in \eqref{caltrs} into \eqref{radpow}
\ba
\langle P\rangle&=&\frac{1}{T}\int_0^T dt \int_{-\infty}^\infty dy \,e^{-4\kappa|y|}\int d\Omega_2\,R_{\infty}^2e^{2\kappa|y|}\bigg[{\mathcal T}_{0i}^{(S)}+{\mathcal T}_{0i}^{(V)}+{\mathcal T}_{0i}^{(T)}+{\mathcal T}_{0i}^{(SV)}+{\mathcal T}_{0i}^{(VT)}+{\mathcal T}_{0i}^{(ST)}\bigg]\,.\nn\\
\ea

Lastly, this expression can be further simplified: since we are asymptotically far away from sources we will use the vacuum relations satisfied by the scalars: $\Phi+2\exp(2\kappa|y|) \,\Psi=0$ and ignored the vector contribution since it does not couple to matter on the brane:
\ba
\langle P\rangle&=&\frac{1}{T}\int_0^T dt \int_{-\infty}^\infty dy \,e^{-4\kappa|y|}\int d\Omega_2\,R_{\infty}^2e^{2\kappa|y|} \bigg[\,6 e^{4\kappa|y| }\dot \Psi\dot \Psi
\nn\\
&&+e^{4\kappa|y|}\frac{x^i}{R_\infty} \bigg(- \tfrac{1}{2} e^{- 6 \kappa | y |} \partial_y (e^{2 \kappa
  | y |} f_{0 \alpha}) \partial_y (e^{2 \kappa | y |} f_i^{\alpha}) 
-
  \tfrac{1}{2}  f_0^{\alpha} \eta^{\mu\nu}\partial_\mu\partial_\nu  f_{i \alpha}
  + \tfrac{1}{4}  \partial_0 f_{\alpha \beta}
  \partial_i f^{\alpha \beta}\bigg)\nn\\
&&+e^{4\kappa|y|}\frac{x^i}{R_\infty} \bigg(e^{-4\kappa|y|}\partial_y(e^{2\kappa|y|}\Psi)\partial_y f_{0i}
+\eta^{\alpha\beta}( 
- 2\partial_\alpha \Psi\partial_\beta f_{0i}-\Psi\partial_\alpha\partial_\beta f_{0i})\nn\\
&&+\eta^{\alpha\beta}\partial_\alpha\partial_i(\Psi f_{0\beta}) +2e^{-2\kappa|y|} \Psi \partial_y^2 f_{0i}-4\kappa\text{sign}(y) e^{-2\kappa|y|}f_{0i}\partial_y\Psi
 +12\kappa
  \delta (y)\Psi f_{0 i} 
\bigg)
\bigg]\,.\nn\\\label{powerrs}
\ea

\section{Conclusions}

In this paper we constructed  the gravitational wave energy-momentum tensor in a strongly curved background.
The main take-away is that the quasi-local expression ubiquitous in the literature
$\langle T_{\mu\nu}\rangle_{\text I}=(1/4)\langle h_{\rho\sigma;\mu} h^{\rho\sigma}{}_{;\nu}\rangle_{\text I}$, where the subscript ``I" denotes an averaging procedure devised by Isaacson,  is valid when
 the background metric satisfies certain conditions
and those conditions fail to hold in a strongly curved background. 
The expression $\langle T_{\mu\nu}\rangle_{\text I}=(1/4)\langle h_{\rho\sigma;\mu} h^{\rho\sigma}{}_{;\nu}\rangle_{\text I}$ assumes the integration (i.e. averaging) over a region of space-time,   smaller in size than the curvature length of the background $R\sim 1/L^2$  and larger than the wavelength $\lambda$ of the gravitational fluctuation $h_{\mu\nu}$. These assumptions imply that $L>\lambda$,  inside the averaging region the metric is almost flat, curvature effects are negligible and background-covariant dervatives commute.  Then the gravitational radiation backreaction on the background geometry which is given by
\ba
{\cal T}_{\mu\nu}&=&-\frac 12 h^{(1)}{}^{\alpha\beta}(h^{(1)}_{\mu\nu;\alpha;\beta}-h^{(1)}_{\nu\alpha;\mu;\beta}
-h^{(1)}_{\mu\alpha;\nu;\beta}+h^{(1)}_{\alpha\beta;\mu;\nu})
+\frac 12 h^{(1)}_{\nu\beta;\alpha}(h^{(1)}_\mu{}^{\alpha;\beta}-h^{(1)}_\mu{}^{\beta;\alpha})\nn\\
&&-\frac 14 h^{(1)}_{\alpha\beta;\mu}h^{{(1)}\alpha\beta}{}_{;\nu}-\frac 14 (h^{(1)}_{\nu\alpha;\mu}+h^{(1)}_{\mu\alpha;\nu})(h^{{(1)};\alpha}-2 h^{(1)}_{\beta}{}^{\alpha}{}^{;\beta})+\frac 14 h^{(1)}_{\mu\nu}{}^{;\alpha}(h^{(1)}{}_{;\alpha} -2
h^{(1)}_{\alpha\beta}{}^{;\beta})\nn\\
&&+\frac 14 \bar g_{\mu\nu} \bigg(h^{{(1)}\alpha\beta}(h^{(1)}{}_{;\alpha;\beta}+h^{(1)}_{\alpha\beta}{}^{;\gamma}{}_{;\gamma}-2h^{(1)}_{\alpha \gamma}{}^{;\gamma}{}_{;\beta}
)
-\frac 12 h^{(1)}{}_{;\alpha} h^{{(1)};\alpha}-2h^{(1)}_{\alpha\beta}{}^{;\alpha} h^{{(1)}\beta\gamma}{}_{;\gamma} \nn\\
&&+2 h^{(1)}{}_{;\alpha}    h^{{(1)}\alpha\beta}{}_{;\beta}    -h^{(1)}_{\alpha\gamma;\beta} h^{{(1)}\alpha\beta;\gamma}
+
\frac 32 h^{(1)}_{\alpha\beta;\gamma}h^{(1)\alpha\beta;\gamma}   \bigg)\,,\label{getc}
\ea
in the de Donder gauge $h^{(1)}_{\mu\nu}{}^{;\mu}=0$,  simplifies to the point where it is given by 
\be
\langle {\cal T}_{\mu\nu}\rangle_{\text I} =\frac{1}{4}\langle h^{(1)}_{\rho\sigma;\mu} h^{(1)\,\rho\sigma}{}_{;\nu}\rangle_{\text I}\label{isaacson1}
\ee 
after integration by parts (e.g.  $\langle h^{(1)}_{\mu\nu;\alpha;\beta} h^{(1)\,\alpha\beta}\rangle_{\text I}=  - \langle h^{(1)}_{\mu\nu;\alpha} h^{(1)\,\alpha\beta}{}_{;\beta} \rangle_{\text I} $) and use of the linearized equations of motion obeyed by $h^{(1)}_{\mu\nu}$.   The other advantage of Isaacson's averaging is that the resulting expression is manifestly gauge-invariant. This can be seen by substituting the gauge-transformed fluctuations $\delta h^{(1)}_{\mu\nu}=\xi_{(\mu;\nu)}$
in \eqref{isaacson1}, integrating by parts and using the linearized equations of motion. Of course, one of the main conditions for defining a good gravitational wave energy-momentum tensor is gauge independence.

However, if the background geometry varies on scales shorter than the wavelength of the radiation, we cannot rely on the averaging procedure of Isaacson.  This is the case, in particular, for the Randall-Sundrum  model of large extra-dimensions.
The background curvature length must be smaller than $\mu m$ in order to confine gravity near the brane and hide any deviations from Newton's law at sub-$\mu m$ scales.  A similar situation arises in a cosmological scenario \cite{Abramo:1997hu} when the curvature scale is smaller than the wavelength of the gravitational fluctuations.  

Without the benefit of the averaging, not only can we not simplify the expression in \eqref{getc} further, but we are looking at a gauge-dependent quantity.
Our resolution to the problem of defining a gauge-invariant gravitational wave energy-momentum  tensor in a strongly curved background mirrors that of Abramo et al.  \cite{Abramo:1997hu}. Namely, we use only the gauge-invariant parts of the linearized  fluctuations $h^{(1)}_{\mu\nu}$ in the expression \eqref{getc}. This definition can be used in {any} background geometry that invalidates the assumptions made by Isaacson, and it does reduce to the usual expression \eqref{isaacson1} in flat backgrounds. 
To drive the message home, take for example flat space as a background, and consider the linearized fluctuations
$h^{(1)}_{\mu\nu}$ as we did in Section \ref{et4d}.  We further decompose the metric fluctuations  according to how they transform under the rotation group, $SO(3)$, i.e. scalars, vectors, tensors, as in equation \eqref{eq4.2}. This is known as performing a scalar-vector-tensor (SVT) decomposition of the linearized metric fluctuations. The linear combinations given in equation \eqref{gi4dso3} are gauge-invariant. As expected, the linearized Einstein equation \eqref{eq2.5}  constrains only these gauge invariant fluctuations. Anything else is a gauge degree of freedom. We next cast the metric in the form given in equation \eqref{hgi4} in terms of the gauge-invariant fluctuations. This is further substituted in the gravitational energy-momentum tensor \eqref{getc}. The radiated power (radiated energy per unit time) through a region $M$ whose boundary is asymptotically  far away from sources can be computed 
using:
\ba
P=-\frac{dE}{dt}=\int_{\partial M}  d^{d-2}x \sqrt{-\bar g} n_i k_\mu {\cal T}^{\mu i},\label{radpow10}
\ea
where $k_\mu$ is a time-like background Killing vector and $n^i$ is a unit vector transverse to the boundary $\partial M$. 
Furthermore, assuming that the sources are generating gravitational waves and that the period of the gravitational waves is $T$,  we defined the averaged radiated power through the boundary $\partial M$:
\ba
\langle P\rangle = \frac 1T \int_{0}^T dt\int_{\partial M} d^{d-2}x \sqrt{-\bar g} n_i k_\mu {\cal T}^{\mu i}.\label{radpow11}
\ea
We verified that the radiated power reduces to the known expression in equation \eqref{eq:power-4d}.  In Appendix \ref{only tensor for non-compact geometry} we also verified that had we performed a different SVT decomposition,  with respect to the  $SO(1,2)$ isometry subgroup, we would arrive at the same expression for the power radiated away by a binary source as with the previous $SO(3)$ SVT decomposition. In section \ref{sec:5d_flat}, we further checked that the same procedure of retaining only the gauge-invariant linearized fluctutations of the 5d metric, with a flat background $R^{3,1}\times S^1$ and a compact fifth dimension, yields the expected results. We decomposed the metric with respect to the isometry group $SO(1,3)$, and retained only the gauge-invaraint fluctuations as in equation \eqref{hgi5}.  We simplified the formula for the power  radiated by a binary located at the same position along the fifth dimension (the binary sources are on the same brane) and further evaluated the expression in \eqref{5dradpow}. In Appendix \ref{allmodescompact}, we reproduced once more known results.  

After having tested our proposal for the gravitational wave energy momentum tensor and our ability to perform the calculation of the radiated power starting from the gauge-invariant fluctuations in a variety of set-ups, we turned our attention to the Randall-Sundrum large extra-dimension scenario in section \ref{sec:RS}. We performed the SVT decomposition of the linearized metric fluctuations with respect to the $SO(1,3)$ background isometry group and we wrote the equations which constrain their dynamics, by analyzing the linearized Einstein equations. Using these gauge-invariant fluctuations we arrived at the expression for the gravitational wave energy-momentum tensor in equations (4.53)-(4.57).
In section \ref{sec:GW}, after reviewing the construction of various Green's functions in flat and AdS backgrounds,   we derived the profile of the gravitational  waves,  which propagate in the Randall-Sundrum geometry,  when sourced by a localized periodic signal. 
From the form of the Green's function \eqref{sphrs} we inferred that the gauge-invariant fluctuations are spherical waves. This allowed us to evaluate the gravitational wave energy momentum tensor expressions $\mathcal T_{0i}$ used in the computation of the radiated power and arrive  at the simplified expressions \eqref{caltrs}.
 A complete evaluation of the radiated power \eqref{powerrs} is forthcoming.

\acknowledgements{The work of D.V. and Y.D. was supported in part by the U.S. Department of Energy under Grant No. DE-SC0007984. K.Y. acknowledges support
from NSF Grant PHY-1806776, PHY-2207349, a Sloan
Foundation Research Fellowship and the Owens Family
Foundation. The authors are grateful to Shammi Tahura for her contribution to Appendix A and many fruitful discussions while this work was in the early stages.

\appendix
\section{The second order expansion of the Ricci tensor}\label{appendixa}
In this appendix we perform explicitly the second order expansion of the Einstein equations,  given a metric 
 $g_{\mu\nu}$ which differs from a background metric $\bar g_{\mu\nu}$ by a small fluctuation
\be
g_{\mu\nu}=\bar g_{\mu\nu}+h_{\mu\nu}.
\ee
Expanding order-by-order in the fluctuation $h_{\mu\nu}$,  the inverse metric and Christoffel symbols are
\ba
&&g^{\mu\nu}=\bar g^{\mu\nu}-h^{\mu\nu}+[h^2]^{\mu\nu}-[h^3]^{\mu\nu}+\dots\,,\nn\\
&&\Gamma_{\mu\nu}^\rho=\bar \Gamma_{\mu\nu}^\rho + \delta^{(1)} \Gamma_{\mu\nu}^\rho+ \delta^{(2)} \Gamma_{\mu\nu}^\rho+\dots\,,\nn\\
&& \delta^{(1)} \Gamma_{\mu\nu}^\rho=\tfrac 12 \bar g^{\rho\sigma}(-h_{\mu\nu;\sigma}+h_{\sigma\nu;\mu}
+h_{\mu\sigma;\nu})\,,\nn\\
&& \delta^{(2)} \Gamma_{\mu\nu}^\rho=-\tfrac 12 h^{\rho\sigma}(-h_{\mu\nu;\sigma}+h_{\sigma\nu;\mu}
+h_{\mu\sigma;\nu})\,,
\ea
where $[h^2]^{\mu\nu}= h^{\mu\rho}h^{\nu\sigma}\bar g_{\rho\sigma}$ etc,  and all the indices are raised and lowered with the background metric.
Consequently,  the Ricci tensor and Ricci scalar, expanded to second order in the fluctuation $h_{\mu\nu}$ are 
\ba
&&R_{\mu\nu}=\bar R_{\mu\nu}+\delta^{(1)} R_{\mu\nu}+\delta^{(2)} R_{\mu\nu}+\dots\,,\nn\\
&&\delta^{(1)} R_{\mu\nu}=\bar\nabla_\rho \delta^{(1)}\Gamma_{\mu\nu}^\rho-\bar\nabla_\mu
\delta^{(1)}\Gamma^\rho_{\rho\nu}\nn\\
&&\qquad\qquad\!\!\!=\tfrac 12(-{h_{\mu\nu;\rho}}^{;\rho}-h_{;\mu;\nu}+{h_{\rho\nu;\mu}}^{;\rho}+{h_{\mu\rho;\nu}}^{;\rho})\,,\nn\\
&&\delta^{(2)} R_{\mu\nu}=\bar\nabla_\rho \delta^{(2)}\Gamma_{\mu\nu}^\rho-\bar\nabla_\mu
\delta^{(2)}\Gamma^\rho_{\rho\nu}+\delta^{(1)}\Gamma^{\rho}_{\mu\nu}\delta^{(1)}\Gamma^{\sigma}_{\sigma\rho}
-\delta^{(1)}\Gamma^{\rho}_{\mu\sigma}\delta^{(1)}\Gamma^{\sigma}_{\rho\nu}\nn\\
&&\qquad\qquad=\tfrac 14 \bar\nabla_\mu\bar\nabla_\nu(h^{\rho\sigma}h_{\rho\sigma})
-\tfrac 14 {h^{\rho\sigma}}_{;\mu} h_{\rho\sigma;\nu}-
\tfrac12 h^{\rho\sigma}(h_{\sigma\nu;\mu;\rho}+
h_{\mu\sigma;\nu;\rho}-h_{\mu\nu;\sigma\rho})\nn\\
&&\qquad\qquad-\tfrac 12 {h^{\rho\sigma}}_{;\rho}(h_{\sigma\nu;\mu}+h_{\mu\sigma;\nu}-h_{\mu\nu;\sigma})
-\tfrac 12({h_{\mu\sigma}}^{;\rho}{h_{\rho\nu}}^{;\sigma}-{h_{\mu\sigma}}^{;\rho} {h^\sigma}_{\nu;\rho})\nn\\
&&\qquad\qquad+\tfrac 14 h^{;\sigma}(h_{\sigma\nu;\mu}+h_{\mu\sigma;\nu}-h_{\mu\nu;\sigma})\,,\nn\\
&&R=g^{\mu\nu}R_{\mu\nu}=\bar R+\delta^{(1)} R+\delta^{(2)} R+\dots\,,\nn\\
&&\delta^{(1)}R=-h^{\mu\nu}\bar R_{\mu\nu}-{h_{;\rho}}^{;\rho}+{h_{\rho\sigma;}}^{;\sigma;\rho}\,,\nn\\
&&\delta^{(2)} R=[h^2]^{\mu\nu}\bar R_{\mu\nu}-h^{\mu\nu}\delta^{(1)}R_{\mu\nu}+\bar g^{\mu\nu}\delta^{(2)}R_{\mu\nu}\,,\label{big}
\ea
where  $h=h_{\mu\nu}\bar g^{\mu\nu}$,  and the background covariant derivatives are denoted either through $\bar\nabla_\mu$ or are implied by an index  preceded by a semicolon ($;\mu$).


\section{The contribution of the scalar and vector modes to the radiated power in 4d flat space,  with an $SO(3)$ SVT decomposition}\label{appendix: 4dSVanalysis}
\noindent First, consider the scalar mode $\Psi$ which satisfies the equations
\begin{eqnarray}
	G_{00} & = & - 2 \delta^{i j} \partial_i \partial_j \Psi =-2\nabla^2 \Psi= T_{00}\,,\label{G00}\\ 
	G^{(L)}_0 & = & - 2 \partial_0 \Psi =T^{(L)}_0\,.\label{GL}
\end{eqnarray}
Due to the compact support of the source, equation \eqref{G00} tells us that far away from the source, $\Psi$ is at most of order $\mathcal O(1/r)$ with its phase depending on time $t$ instead of retarded time $(t-r)$. Thus $\partial_i \Psi$ will be at most of order $\mathcal O(1/r^2)$, which means any term in $\mathcal{T}_{0i}$ that contains $\partial_i \Psi$ will not contribute to the calculation of radiated power.

Let us now move on to equation \eqref{GL}. $T_0^{(L)}$ can be solved as
\ba
T_0^{(L)} = -\frac{\partial^i}{\nabla^2}T_{0i}\,.
\ea
Since the source has compact support, at large distance $r$, we can see that $T_0^{(L)}$ will be at most of $\mathcal O(1/r^2)$, which means any term in $\mathcal{T}_{0i}$ that contains $\partial_0 \Psi$ will not contribute to the calculation of radiated power.

Next, consider the scalar mode $\Phi$ which relates to $\Psi$ by the equation
\begin{eqnarray}
	G^{(L L)} & = & \Phi - \Psi = T^{(L L)}\,.\label{GLL}
\end{eqnarray}
$T^{(LL)}$ can be solved as
\ba
T^{(LL)} = \frac{3}{2}\frac{\partial^i \partial^j}{(\nabla^2)^2}T_{ij}-\frac{1}{2\nabla^2}\delta^{ij}T_{ij}\,,
\ea
which means when far away from the source, similar to $\Psi$, $T^{(LL)}$ will also be at most of order $\mathcal O(1/r)$ with its phase depending on time $t$ instead of retarded time $(t-r)$. Thus any term in $\mathcal{T}_{0i}$ that contains $\partial_i \Phi$ will not contribute to the calculation of radiated power.

The analysis for the vector mode ${\bf S}_i$ is the same as that for $\Psi$, which leads to the conclusion that any term that contains $\partial_j {\bf S}_i$ or $\partial_0 {\bf S}_i$ will not contribute to the calculation of radiated power.

In addition, since we need to take a time averaging over the period of the gravitational wave to compute the radiated power, any term in $\mathcal{T}_{0i}$ that is a total derivative of time will not give any contribution. 

Therefore none of the terms in \eqref{T0i4d} that involve any of the scalar modes, $\Phi$ and $\Psi$, or the vector mode, ${\bf S}_i$ will contribute to the calculation of radiated power.

\section{The non-locality of the SVT decomposition}
\label{app:non-local}

Whenever we perform a decomposition of SVT type, the resulting metric components are extracted with projectors that are non-local in position space. The same operators when applied to a delta-function localized energy-momentum source will also result in a non-local set of SVT $T_{\mu\nu}$ components. To gain a better understanding of the action of the non-local projectors and the consequences of a non-local energy-momentum source we consider a flat background,  a static source and perform an $SO(1,3)$ SVT decomposition both for the metric and the Einstein equations. Let us begin with the metric fluctuations: 
\be
h_{\mu\nu}=\eta_{\mu\nu} \Psi +\partial_\mu h_\nu^{(LT)} + \partial_\nu h_\mu ^{(LT)} + \partial_\mu\partial_\nu h^{(LL)}+ h_{\mu\nu}^{(TT)},
\label{04svt}\ee
where $h_\mu^{(LT)} $ is a transverse vector and $h_{\mu\nu}^{(TT)}$ is a transverse traceless tensor
\be
\partial_\mu h^{(LT)\,\mu} = 0,\qquad \eta_{\mu\nu} h_{\mu\nu}^{(TT)} =0, \qquad \partial_\mu h^{(TT)\, \mu\nu}=0.
\ee
The indices are raised with the background (Minkowski) metric.

Given $h_{\mu\nu}$ we can solve for each of $\Psi, h^{(LL)}, h_\mu^{(LT)} $ and $h^{(TT)}_{\mu\nu}$:
\be
\Psi=-\frac{\partial_\mu\partial_\nu}{d-1}\frac{1}{\Box} h^{\mu\nu} + \frac{h}{d-1}, 
\ee
where $h=h_{\mu\nu} \eta^{\mu\nu}$,
\be
h^{(LL)}=\frac{d}{d-1}\frac{\partial_\mu\partial_\nu}{\Box^2} h^{\mu\nu} - \frac {1}{(d-1)}\frac 1\Box h,
\ee
\be
h^{(LT)\,\nu}=\frac{1}{\Box}\bigg(\partial_\mu h^{\mu\nu} - \partial^\nu \frac{\partial^\rho\partial^\sigma}{\Box} h_{\rho \sigma}\bigg),
\ee
and where $d=4$ here.  Lastly, $h_{\mu\nu}^{(TT)}$ is obtained substituting the previous expressions into \eqref{04svt}.
While a bit more cumbersome than the usual $SO(3)$ SVT decomposition, this $SO(1,3)$ decomposition arises naturally in the context of a 5d spacetime that has 4d Lorentz symmetry.  

Consider next a static source and solve for the linearized fluctuation in the usual fashion (define the trace-reversed metric fluctuations,  impose the Lorentz gauge,  and solve the resulting decoupled equations):
\be
T_{00}=M \delta^3(\vec r),  \qquad h_{\mu\nu} dx^\mu dx^\nu=\frac {2 MG}{r} dt^2 + \delta_{ij} \frac{2 MG}{r}dx^i dx^j.
\ee
Then we can extract the SVT metric components by applying the non-local projectors as above:
\ba
&&\Psi=\frac{2 MG}{3r}, \qquad h^{(LL)}=-\frac{MG}{3} r, \qquad h^{(LT)}_\mu=0,  \nn\\
&&h_{00}^{(TT)}=\frac{8MG}{3r}, \qquad h_{0i}^{(TT)}=0, \qquad
h_{ij}^{(TT)}=\frac{2 MG}{3r}(\delta_{ij}+n_i n_j),\label{linstat}
\ea
where $ n^i=\frac{x^i}r $.
The matter energy-momentum tensor is decomposed in a similar fashion:
\be
T_{\mu\nu}=\eta_{\mu\nu}T^{(\Psi)}+\partial_\mu T_{\nu}^{(LT)}+\partial_\nu T_{\mu}^{(LT)}+\partial_{\mu}\partial_{\nu} T^{(LL)} + T_{\mu\nu}^{(TT)},
\ee
with
\ba
&&T^{(\Psi)}=-\frac{MG}3\delta^3(\vec r), \qquad T^{(LL)}=\frac{MG}{12 \pi r}, \qquad T_\mu^{(LT)}=0, \nn\\
&&T_{00}^{(TT)}=\frac{2 MG}{3} \delta^3(\vec r), \qquad T_{ij}^{(TT)}=\frac{2MG}9\delta^3(\vec r)\delta_{ij}
-\frac{MG}{12\pi r^3}(\delta_{ij}-3 n_i n_j).\label{tstat}
\ea

As a check, we verify that the SVT fluctuations obey decoupled linearized Einstein equations\footnote{Use that $\Box \frac 1r = -4\pi \delta^3(\vec r), \,\Box \frac{n_i n_j}{r}= -\frac{4\pi}{3}\delta_{ij}\delta^3(\vec r) -\frac{1}{r^3}(6 n_i n_j-2\delta_{ij}).$} 
\be
\Box \Psi= -8 \pi G T^{(\psi)},  \qquad \Box h_{\mu\nu}^{(TT)}=-16\pi G T_{\mu\nu}^{(TT)}\label{svteq}.
\ee
There is one more linearized Einstein equation that involves $\Psi$:
\ba
\partial_\mu\partial_\nu \Psi=-8\pi G \partial_\mu\partial_\nu T^{(LL)}.
\ea
However this equation is satisfied due the transversality of $T_{\mu\nu}$ which implies that $\Box T^{(LL)}+T^{(\Psi)}=0$ and 
$T_\nu^{(LT)}=0$.

Notice also that the Einstein equations constrain only the gauge-independent fluctuations, $\Psi$ and $h_{\mu\nu}^{(TT)}$.  The other two fluctuations $h_{\mu}^{(LT)}$ and $h^{(LL)}$ are pure gauge.
Given the expressions for $\Psi$  and $h_{\mu\nu}^{(TT)}$ in \eqref{linstat} and the SVT energy-momentum tensor components in \eqref{tstat},  we can proceed to verify that the equations \eqref{svteq} are satisfied.

We can now take stock of what we have learned.  While source terms for the equations \eqref{svteq} obeyed by the decoupled SVT fluctuations are non-local,  the only consequence of this non-locality is that the linearized SVT metric fluctuations, which still fall off as $1/r$,  acquire a dependence on $n_i n_j$.



\section{Zero mode sector}\label{ZM}
\subsection{Maxwell field}
The SVT decomposition requires a slight modification in the case of null eigenvectors of the d'Alembertian. 
As a warm-up we consider first the Maxwell field  in a $d$-dimensional flat space.  The scalar-vector (SV) decomposition with respect to  the $SO(1,d-1)$ Lorentz group
\be
A_\mu=\partial_\mu A^{(L)} + A^{(T)}_\mu, \qquad \eta^{\mu\nu}\partial_\mu A_\nu^{(T)}=0,
\ee
or in terms of Fourier modes, 
\be
A_\mu = i k_\mu A^{(L)} +\sum_{p=1}^{d-1} \epsilon_\mu^{(p)} A^{p},\qquad \epsilon^{(p)} \cdot k=\eta^{\mu\nu} \epsilon_\mu^{(p)} k_\nu=0,  \qquad \epsilon^{(p)}\cdot\epsilon^{(q)}=\delta^{pq},
\ee
maps the $d$ components of the vector field into a longitudinal vector $\partial_\mu A^{(L)}$ and a $d-1$ component transverse vector $A^{(T)}_\mu$.  The latter components are gauge-invariant.  However,  this breaks down when $A^{(L)}$ is a null eigenvector of the d'Alembertian, i.e. $\Box  A^{( L)}=0$,   since in this case $\partial_\mu A^{( L)}$ is transverse (or,  in Fourier space,  $k_\mu$ is null).  In this case, we proceed with
\ba
&&\!\!\!\!\!\!\!\!A_\mu=i k_\mu a+i \tilde k_\mu \tilde a + \sum_{p=1}^{d-2} \epsilon_\mu^{(p)} A^p, \nonumber\\
&&\!\!\!\!\!\!\!\!k_\mu=(k_0,\vec k), \qquad \tilde k_\mu=(k_0,-\vec k), \qquad \!\!\!\!\!k\cdot k=\tilde k\cdot\tilde k=  \epsilon^{(p)} \cdot k =\epsilon^{(p)} \cdot \tilde k=0, \qquad \epsilon^{(p)}\cdot\epsilon^{(q)}=\delta^{pq}.
\ea
The gauge-invariant components are $\tilde a $ and $A^{ p}$.  Furthermore, Maxwell's equations set $\tilde a=0$.
For an on-shell gauge field, we can write then
\be
A_\mu=\partial_\mu A^{(L)} + A^{(T)}_\mu, 
\ee
where the gauge-invariant components are transverse $A^{(T)}_\mu=(0,\vec A^{(T)}),  \;\vec\nabla\cdot\vec  A^{(T)}=0$.

Next, let us consider a Maxwell field in $d+1$ dimensions and perform an SV decomposition with respect to the $SO(1,d-1)$ Lorentz group 
\ba
A_M&=&(A_\mu, A_{d+1})\\
A_\mu&=&\partial_\mu A^{(L)} + A^{(T)}_\mu, \qquad \eta^{\mu\nu}\partial_\mu A_\nu^{(T)}=0,
\label{eq:Maxwell_decomp}
\ea
where $M$ is a $d+1$-index and $\mu=0,1,2,\dots, d-1$.
Such an expansion would be appropriate if we are working with one compact dimension, $x^{d+1}$. 
In terms of Fourier modes $\exp(i k_\mu x^\mu)$ (scalar eigenfunctions of the $d$-dimensional d'Alembertian) we can write
\ba
A_M =\left(i k_\mu A^{( L)} +\sum_{p=1}^{d-1} \epsilon_\mu^{(p)} A^{p}\,, A_{d+1} \right), \qquad \epsilon^{(p)} \cdot k\equiv \eta^{\mu\nu} k_\mu\epsilon_\nu^{(p)}=0.
\ea 
This assumes that $k\cdot k=\eta^{\mu\nu} k_\mu k_\nu\neq 0$. 
The gauge-invariant components are $A_{d+1} -\partial_{d+1} A^{(L)}$ and $A^{(T)}$.  Furthermore Maxwell's equations set the scalar gauge-invariant combination $A_{d+1} -\partial_{d+1} A^{( L)}$ to zero and require that $(-k\cdot k +\partial_{d+1}^2)A^{p}=0$.

If the Fourier momenta are null ($k\cdot k= 0$), then we proceed as we did earlier,  with
\ba
A_\mu =i  k_\mu  a+ i\tilde k_\mu \tilde a + \sum_{p=1}^{d-2} \epsilon_\mu^{(p)} A^{p}, 
\ea
with the polarization vectors $\epsilon_\mu ^{(p)}$ transverse to both null vectors $k_\mu$ and $\tilde k_\mu$. Maxwell's equations set $\tilde a=0$ and require that the scalar gauge-invariant combinations $A_{d+1}-\partial_{d+1} a$ and $A^{p}$ be independent of $x^{d+1}$.  
To conclude,  for an on-shell Maxwell field we can write 
\ba
A_\mu=\partial_\mu A^{( L)}+ A_\mu^{(T)},
\ea
where $A_\mu^{(T)}$ satisfies $\partial^\mu A_\mu^{(T)}=0$ if the Fourier momentum is not null, or $A_\mu^{(T)}=(0,\vec A^{(T)})$ with $\vec\nabla  \cdot \vec A^{(T)}=0$ if the Fourier momentum is null.  The additional physical degree of freedom is the scalar  $A_{d+1}-\partial_{d+1} a$.  

\subsection{GR}

For concreteness, we begin by considering 4d GR in a flat background and perform an SVT decomposition with respect to the Lorentz isometry group $SO(1,3)$.  We decompose the metric fluctuations in terms of eigenvectors of the 4d d'Alembertian and focus on the zero eigenvalues sector (e.g. the scalar eigenvectors satisfy $\Box_{4d} e^{ik\cdot x}=0$ etc.)
After Fourier-transforming, the metric fluctuations are decomposed as
\ba
h_{\mu\nu} = 2 \psi \eta_{\mu\nu} - k_\mu k_\nu E-(k_\mu\tilde k_\nu+k_\nu\tilde k_\mu) \tilde E -\tilde k_\mu\tilde k_\nu\tilde{\tilde E}+ i (k_\mu F_\mu+k_\nu F_\mu)+ i (\tilde k_\mu \tilde F_\nu+\tilde k_\nu \tilde F_\mu)+ f_{\mu\nu}\,,\nonumber\\
\ea
where $ \mu,\nu=0,1,2,3, $ the momenta $k_\mu$ and $\tilde k_\mu$ are null: $k\cdot k=\tilde k\cdot\tilde k=0$ and where $F_\mu=\sum_{p=1,2} \epsilon_\mu^{(p)} F^{p}, \tilde F_\mu=\sum_{p=1,2}\epsilon_\mu^{(p)} \tilde F^{p}, 
f_{\mu\nu}=\sum_{p,q=1,2}\epsilon_\mu^{(p)}\epsilon_\nu^{(q)} f^{pq}, $ $\sum_{p=1,2} f^{pp}=0, $ and $\epsilon_\mu^{(p)} k^\mu=\epsilon_\mu^{(p)} \tilde k^\mu=0, \epsilon^{(p)}\cdot \epsilon^{(q)}=\delta^{pq}$.
The gauge-invariant fluctuations are $\psi, f_{\mu\nu}, \tilde F_\mu, \tilde{\tilde E}$. The rest are gauge-dependent: $\delta F_\mu=\xi^\perp_\mu, \delta E=2\xi, \delta \tilde E=\tilde \xi$,  where we decomposed the gauge parameter in a similar way: $\xi_\mu= \xi_\mu^{(T)}+i k_\mu \xi+i\tilde k_\mu\tilde \xi$, with $\xi_\mu^{(T)}=\sum_{p=1,2} \epsilon_\mu^{(p)} \xi^p$. 
The equations of motion set $\tilde{\tilde E}=0, \tilde F_\mu=0, \psi=0.$ The two degrees of freedom of the on-shell graviton are contained in the transverse  and traceless tensor $f^{\mu\nu}$. \footnote{ Since the vectors  $\epsilon^{(p)}_\mu$ are transverse to both $k_\mu$ and $\tilde k_\mu$, this means that $\epsilon^{(p)}_\mu=(0,\vec \epsilon^{(p)})$.  So the non-zero components of the tensor $f_{\mu\nu}$  are purely spatial,  and as a result,   $f_{\mu\nu}$ is transverse with respect to the 3d gradient $\vec\nabla$.} Of course, there are no solutions to the equations of motion for non-null momenta.

We consider next a 4d {flat} background with one compact dimension and we perform the SVT decomposition with respect to the $SO(1,2)$  Lorentz isometry group. This is the same scenario we will discuss further in Appendix \ref{so12}.  Here we focus only on the zero-mode sector of  the 3d d'Alembertian. After Fourier-transforming and restricting to null 3d momenta ($k\cdot  k\equiv \eta^{\mu\nu}k_\mu k_\nu=0$) we proceed with
\ba
&&h_{\mu\nu} = 2 \psi \eta_{\mu\nu} - k_\mu k_\nu E-(k_\mu\tilde k_\nu+k_\nu\tilde k_\mu) \tilde E -\tilde k_\mu\tilde k_\nu\tilde{\tilde E}+ i (k_\mu F_\mu+k_\nu F_\mu)+ i (\tilde k_\mu \tilde F_\nu+\tilde k_\nu \tilde F_\mu)
\nonumber\\
&&h_{3\mu}=i k_\mu B+i \tilde k_\mu \tilde B+S_\mu, \qquad
h_{33}=2\phi,
\ea
where $\mu,\nu=0,1,2$, $S_\mu=\epsilon_\mu S, \;F_\mu=\epsilon_\mu F, \;\tilde F_\mu=\epsilon_\mu \tilde F,  \; \;\epsilon_\mu\cdot k=\epsilon_\mu\cdot \tilde k=0$, and we recall that $\tilde k\cdot\tilde k=0$.  Note that there is no transverse traceless tensor contribution $f_{\mu\nu}$ since in 3d there is only one vector $\epsilon_\mu$, perpendicular to both $k_\mu$ and $\tilde k_\mu$.
The six gauge-invariant combinations are: $\psi, \Phi=\phi-\partial_3(B-\partial_3 E/2), {\bf S}_\mu=S_\mu-\partial_3 F_\mu, \tilde F_\mu, \tilde{\tilde E},{ \bf \tilde B}=\tilde B-\partial_3\tilde E$.  The vacuum linearized equations of motion impose the following conditions $\tilde{\tilde E}=0, \;\tilde F=0, \;\partial_3^2 \psi=0, \;\partial_3{\bf \tilde B}=0, \;4\partial_3\psi +(k\cdot\tilde k){\bf\tilde B}=0, \;\Phi=0, \;\partial_3{\bf S}_\nu=0.$
The two degrees of freedom of the 4d graviton are contained in the scalar $\psi$ and the transverse gauge-invariant vector ${\bf S}_\mu$, which are both null eigenvectors of the 3d d'Alembertian and $x^3$-independent.

Similarly, for the case of a $d+1$ flat background with one compact dimension and $d>3$ we make the decomposition
\ba
&&h_{\mu\nu} = 2 \psi \eta_{\mu\nu} - k_\mu k_\nu E-(k_\mu\tilde k_\nu+k_\nu\tilde k_\mu) \tilde E -\tilde k_\mu\tilde k_\nu\tilde{\tilde E}+ i (k_\mu F_\mu+k_\nu F_\mu)+ i (\tilde k_\mu \tilde F_\nu+\tilde k_\nu \tilde F_\mu)+ f_{\mu\nu}
\nonumber\\
&&h_{d+1 \mu}=i k_\mu B+i \tilde k_\mu \tilde B+S_\mu, \qquad
h_{d+1\,d+1}=2\phi,
\ea
where $\mu,\nu=0,1,2,\dots d-1$, $S_\mu=\sum_{p=1}^{d-2} \epsilon_\mu^{(p)} S^p, \;F_\mu=\sum_{p=1}^{d-2} \epsilon_\mu^{(p)} F^p, \;\tilde F_\mu=\sum_{p=1}^{d-2} \epsilon_\mu^{(p)} \tilde F^p,  \; \;\epsilon^{(p)}\cdot k=\epsilon^{(p)}\cdot \tilde k=0,\epsilon^{(p)}\cdot \epsilon^{(q)}=\delta^{pq}$,  $f_{\mu\nu}=\sum_{p,q=1}^{d-2}\epsilon_\mu^{(p)}\epsilon_\nu^{(q)} f^{pq}, $ $\sum_{p=1}^{d-2} f^{pp}=0, $ and  $k\cdot k=\tilde k\cdot\tilde k=0$.  

 The  gauge-invariant fluctuations are $\psi,\,\tilde {\tilde E}, \,\tilde F_\mu, \,\Phi=\phi-\partial_{d+1} ( B-\partial_{d+1}E/2), {\bf\tilde B}=\tilde B-\partial_{d+1} \tilde E, \,{\bf S_\mu}=S_\mu-\partial_{d+1} F_\mu$ and $f_{\mu\nu}$. The vacuum linearized equations of motion set $\tilde{\tilde E}=0, $ $\Phi=-(d-2)\psi, \tilde F=0, \partial_{d+1}^2\psi=0, \;\partial_{d+1} {\bf\tilde B}=0, \;2(d-1) \partial_{d+1}\psi+ (k\cdot\tilde k) {\bf\tilde B}=0,\; \partial_{d+1} {\bf S_\mu}=0,  \partial_{d+1} f_{\mu\nu}=0$.  The $(d-1)(d-2)/2-1$ degrees of freedom of the $d+1$ dimensional graviton are parametrized by the transverse (in a $(d-1)$-spatial sense since $e_\mu^{(p)}=(0,\vec e_{\mu}^{(p)})$ ) and traceless tensor $f_{\mu\nu}$,  the transverse vector ${\bf S}_\mu$ and the scalar $\psi$, all of which are independent of the compact coordinate $x^{d+1}$.  This is what we expect to see when performing a Kaluza-Klein reduction of $d+1$ gravity, in the massless sector. 

Thus,  in the zero-mode sector,  for the on-shell linearized fluctuation, we can reach a gauge where all the gauge-dependent terms are zero and write
\ba
&&h_{\mu\nu} |_{g.i.}= 2 \psi \eta_{\mu\nu} + f_{\mu\nu}
\nonumber\\
&&h_{d+1 \mu}|_{g.i.} ={\bf S}_\mu, \qquad
h_{d+1\,d+1}|_{g.i.}=2\Phi.
\ea
which is of the form used in  \eqref{so12hgf}.
The one caveat is that the transverse tensors $f_{\mu\nu}$ are transverse in a $d-1$ sense for the zero modes,  while for the massive modes ($k_{d+1}\neq 0$ which implies $k\cdot k\neq0$)  the tensor fluctuations are transverse in a $d$-sense (and the scalar and vector fluctuations are zero).

\section{The $SO(1,2)$ SVT decomposition of the  metric fluctuations about a 4d flat background  } \label{so12}

In this appendix we want to use the familiarity of 4d gravity to study a less standard way to decompose the metric fluctuations, namely using the $SO(1,2)$ background isometry rather than the rotational isometry $SO(3)$.    As we will see,  unlike the $SO(3)$ case analyzed in Section IV A,   the tensor modes are not the only dynamical ones,  and both scalar and vector modes contribute together with the tensor modes  to the gravitational energy-momentum tensor.  

We begin by writing the metric perturbation as
\ba
h_{\mu\nu } = \left(\begin{array}{cc}
     2 \psi \eta_{\bar\mu \bar\nu} + \partial_{\bar \mu} \partial_{\bar \nu} E + \partial_{\bar\mu}
     F_{\bar\nu} + \partial_{\bar\nu} F_{\bar\mu} + f_{\bar\mu \bar \nu} & \partial_{\bar\nu} B +
     S_{\bar\nu}\\
     \partial_{\bar\mu} B + S_{\bar\mu} & 2 \phi
   \end{array}\right),
   \label{so12h}\,
\ea
where we denote the 4d indices by $\mu, \nu = 0, 1, 2, 3$,  while $\bar\mu, \bar \nu = 0, 1, 2$. The gauge
invariant pieces are $\Phi,\Psi, {\bf S}_{\bar\mu}$ and $f_{\bar\mu\bar\nu}$, where
\begin{eqnarray}
  \Phi & = & \phi - \partial_3 (B - \tfrac 12\partial_3 E)\nn\\
  \Psi & = & \psi\nn\\
  {\bf S}_{\bar\mu} & = & S_{\bar\mu} - \partial_3 F_{\bar\mu}\,.\label{so12gi}
\end{eqnarray}
As in \cite{Abramo:1997hu} we can restrict to the gauge-invariant fluctuations by going to the gauge:
\ba
h_{\mu\nu }|_{g.i.} = \left(\begin{array}{cc}
     2 \Psi \eta_{\bar\mu\bar \nu} + f_{\bar\mu\bar \nu} & 
    {\bf  S}_{\bar\nu}\\
      {\bf S}_{\bar\mu} & 2 \Phi
   \end{array}\right) \label{so12hgf}.
\ea
We apply the same SVT $SO(1,2)$  decomposition to the Einstein equations $G_{\mu\nu} = T_{\mu\nu }$:
\begin{eqnarray}
 G_{\mu\nu} & = & \left(\begin{array}{cc}
  2 G^{(Y)} \eta_{\bar\mu \bar\nu} + \partial_{\bar\mu} \partial_{\bar\nu} G^{(\tmop{LL})} +
    \partial_{\bar\mu} G^{(L)}_{\bar\nu} + \partial_{\bar\nu} G^{(L)}_{\bar\mu} + G^{(TT)}_{\bar\mu
   \bar \nu} & \;\;\;\partial_{\bar\nu} G^{(L)}_3 + G^{(T)}_{\bar\nu 3}\\
    \partial_{\bar\mu} G^{(L)}_3 + G^{(T)}_{\bar\mu 3} & \;\;\;2 G_{33}\,,
  \end{array}\right)\\
  T_{\mu\nu} & = & \left(\begin{array}{cc}
    2 T^{(Y)} \eta_{\bar\mu \bar\nu} + \partial_{\bar\mu} \partial_{\bar\nu} T^{(\tmop{LL})} +
    \partial_{\bar\mu} T^{(L T)}_{\bar\nu} + \partial_{\bar\nu} T^{(L T)}_{\bar\mu} +
    T^{(TT)}_{\bar\mu \bar\nu} &\;\;\; \partial_{\bar\nu} T^{(L)}_3 + T^{(T)}_{\bar\nu 3}\\
    \partial_{\bar\mu} T^{(L)}_3 + T^{(T)}_{\bar\mu 3} & \;\;\;2 T_{33}
  \end{array}\right).
\end{eqnarray}
The linearized equations of motion for the scalar fluctuations $\Phi$ and $\Psi$ come from the components $  \delta^{(1)}G_{33}$,
$ \delta^{(1)}G^{(L)}_3$, $ \delta^{(1)}G^{(Y)}$and $ \delta^{(1)}G^{(\tmop{LL})}$:
\begin{eqnarray}
2   \delta^{(1)}G_{33} & = & 2 \eta^{\bar\alpha \bar\beta} \partial_{\bar\alpha} \partial_{\bar\beta} \Psi =
2 T_{33}\nn\\
  \delta^{(1)} G^{(L)}_3 & = & - 2 \partial_3 \Psi = T^{(L)}_3\nn\\
2   \delta^{(1)}G^{(Y)} & = & 2 \partial_3^2 \Psi + \eta^{\bar\alpha \bar\beta} \partial_{\bar\alpha}
  \partial_{\bar\beta} \Phi + \eta^{\bar\alpha \bar\beta} \partial_{\bar\alpha}
  \partial_{\bar\beta} \Psi = 2 T^{(Y)}\nn\\
 \delta^{(1)}  G^{(L L)} & = & - \Phi - \Psi =T^{(\tmop{LL})}\,.
\end{eqnarray}
The equations of motion for the transverse vector ${\bf S}_{\bar\mu}$ come from the components $ \delta^{(1)}G^{(T)}_{\bar\mu 3}$
and $ \delta^{(1)}G^{(L T)}_{\bar\mu}$:
\begin{eqnarray}
 \delta^{(1)}  G^{(T)}_{\bar\mu 3} & = & - \frac{1}{2} \eta^{\bar\alpha\bar \beta} \partial_{\bar\alpha}
  \partial_{\beta} {\bf S}_{\bar\mu} = T^{(T)}_{\bar\mu 3}\nn\\
 \delta^{(1)}  G_{\bar\mu}^{(L T)} & = & \frac{1}{2} \partial_3 {\bf S}_{\bar\mu} = T^{(L
  T)}_{\bar\mu}\,.
\end{eqnarray}
Lastly,  the equation of motion  for the transverse traceless tensor ${f}_{\bar\mu \bar\nu}$ comes from $ \delta^{(1)}G^{(TT)}_{\bar\mu \bar\nu}$,
\ba
 \delta^{(1)}G^{(TT)}_{\bar\mu \bar\nu} = - \frac{1}{2} \partial_3^2 {f}_{\bar\mu \bar\nu} -
   \frac{1}{2} \eta^{\bar\alpha\bar \beta} \partial_{\bar\alpha} \partial_{\bar\beta}
   {f}_{\bar\mu\bar \nu} =  T^{(TT)}_{\bar\mu\bar \nu} \,. 
\ea
Next, we count the degrees of freedom by considering the vacuum equations of motion:
\begin{eqnarray}
 && \eta^{\bar\alpha \bar\beta} \partial_{\bar\alpha} \partial_{\bar\beta} \Psi  =  0\,, 
 \qquad \partial_3 \Psi  =  0\,,\qquad
\qquad  \Phi  =  - \Psi\nn\\
  &&\eta^{\bar\alpha \bar\beta} \partial_{\bar\alpha} \partial_{\bar\beta} {\bf S}_{\bar\mu} = 
  0\,, \qquad
  \partial_3 {\bf S}_{\bar\mu} =  0\,, \nn\\
&&  \partial_3^2 {f}_{\bar\mu\bar \nu} + \eta^{\bar\alpha \bar\beta} \partial_{\bar\alpha}
  \partial_{\bar\beta} {f}_{\bar\mu\bar \nu}  =  0\,.
\end{eqnarray}
Due to the constraints $\partial_3 \Psi = 0$ and $\partial_3 {\bf S}_{\bar\mu} =
0$, it is natural to separately consider the case $p_3 = 0$ and the case $p_3
\neq 0$.
When $p_3 = 0$: ${\Psi}$ describes a 3d massless scalar, which has one degree of freedom;
  ${\bf S}_{\bar\mu}$ describes a 3d massless vector, which has one degree
  of freedom;
 ${f}_{\bar\mu \bar\nu}$ describes a 3d massless graviton, which has zero
  degrees of freedom.
When $p_3 \neq 0$: 
${\Psi}$ has no solution,  hence its degree of freedom is 0;
  ${\bf S}_{\bar\mu}$ has no solution and its degree of freedom is 0;
  ${f}_{\bar\mu \bar\nu}$ describes a 3d massive graviton, which has two
  degrees of freedom.  This again adds up to the correct number of degrees of freedom of the 4d graviton.  However, our analysis was not rigorous. If $p_3=0$ then the equations of motion require that the 3-momentum $p_\mu$ is null ($\eta^{\bar\mu\bar\nu} p_{\bar\mu} p_{\bar\nu}=0$) which means that the SVT decomposition starting point \eqref{so12h} is invalid. Nonetheless, as we showed in Appendix \ref{ZM}, the conclusion reached here stands: the two degrees of freedom of the 4d graviton are the massless scalar $\Psi$ and the massless vector ${\bf S}_{\bar\mu}$.

The point to be made is that unlike
the $SO(3)$ SVT decomposition, where only the $SO(3)$-tensor fluctuation is dynamical,  when performing an $SO(1,2)$ decomposition all types of fluctuations (scalar, vector and tensor)  are
dynamical. Thus, the energy-momentum tensor can receive contributions from all
three types of fluctuations.  If the $x^3$ dimension were compact, the Fourier spectrum along $x^3$ is discrete and indeed all three types of fluctuations do contribute, with the scalar and vector modes part of the massless sector of a Kaluza-Klein reduction. If on the other hand the $x^3$ dimension is non-compact,  the Fourier spectrum along $x^3$ is continuous and only the tensor modes contribute to the radiated power as we will see in Appendix \ref{only tensor for non-compact geometry}.

\section{The $SO(1,2)$ SVT modes sourced by a binary in 4d flat space and the luminosity of the gravitational waves }\label{only tensor for non-compact geometry}
In this appendix we derive concrete expressions for the $SO(1,2)$ SVT modes in 4d flat space,  asymptotically far away from sources,  and verify that we correctly reproduce known results  for the radiated power using the results from
Section \ref{sec:GW_EM_tensor}.

We begin with a known form of the 4d metric perturbation far away from a binary source (see for example eqn. (5.24) in \cite{Du:2020rlx}).  Keeping only 
the time-dependent parts, $h_{\mu\nu}$, which are the relevant pieces for the computation of the radiated power, we write
\begin{equation}\label{generalForm}
    h_{\mu \nu} =\tilde h_{\mu\nu}-\tfrac 12 \eta_{\mu\nu}\eta^{\alpha\beta}\tilde h_{\alpha\beta}= \left(\begin{array}{cccc}
    \frac{1}{2}  \widetilde{ h}_{00} & - n_1  \tilde{h}_{11} - n_2 
    \tilde{h}_{12} & - n_1 \tilde{h}_{12} - n_2  \tilde{h}_{22} & 0\\
    - n_1  \tilde{h}_{11} - n_2  \tilde{h}_{12} & \tilde{h}_{11} + \frac{1}{2}
    \tilde{h}_{00} & \tilde{h}_{12} & 0\\
    - n_1 \tilde{h}_{12} - n_2  \tilde{h}_{22} & \tilde{h}_{12} &
    \tilde{h}_{22} + \frac{1}{2} \tilde{h}_{00} & 0\\
    0 & 0 & 0 & \frac{1}{2}  \tilde{h}_{00}
  \end{array}\right)\,,
\end{equation}
where $\tilde h_{\mu\nu}$ is the so-called trace-reversed metric fluctuation,  and $n_i=x^i/r$ are the Cartesian components of a radial pointing, unit vector $\vec n=\vec r/r$.  In writing 
\eqref{generalForm} we have used that for the binary solution $\tilde h_{3\mu}=0$ and $\tilde h_{11}+\tilde h_{22}=0$.  Note that the $x^3$ direction is perpendicular to the plane of the binary. 
This metric perturbation satisfies the harmonic gauge $\eta^{\mu\nu} \partial_\mu \tilde h_{\nu\rho}=0$,  which implies that $\tilde{h}_{00}$ can be written as
\begin{eqnarray}
  && \tilde{h}_{00}  =  (n_1)^2  \tilde{h}_{11} + 2 n_1 n_2  \tilde{h}_{12} +
  (n_2)^2  \tilde{h}_{22}\,.
\end{eqnarray}
Each of the components $\tilde{h}_{00}$, $\tilde{h}_{11}$, $\tilde{h}_{12}$, $\tilde{h}_{22}$ are in the form of spherical waves.  For example,  $\tilde h_{12}=\mu r_{12}^2\Omega^2\sin(2\Omega(t-r))/(2\pi r)$, where $\mu$ is the reduced binary mass, $r_{12}$ is the separation distance between the binary components, $\Omega $ is the angular frequency of the binary,  and we set $8\pi G=1$.

Next  let us consider the $\tmop{SO} (1, 2)$ decomposition of the perturbation \eqref{so12h} and the gauge-invariant fluctuations given in \eqref{so12gi}. We recall that $\bar\mu,\bar\nu=0,1,2$.  Later
we will also use indices $\bar{i}, \bar{j} = 1, 2$.
From  \eqref{generalForm}, we can see that $
  B  = 0$ and 
  $S_{\bar{\mu}} =  0.$
In terms of the SVT decomposed fields, the harmonic gauge condition $\partial_\mu\tilde h^{\mu\bar\nu}=\partial_{\bar\mu} \tilde h^{\bar\mu\bar\nu}=\partial_{\bar\mu} (h^{\bar\mu\bar\nu}+(1/2) \eta^{\bar\mu\bar\nu} \tilde h)=0$ becomes
\ba
\partial_{\bar{\nu}} (\Box_{\text{3d}} E +2\Psi -2 \phi) +\Box_{\text{3d}} F_{\bar{\nu}}\label{oone}
   = 0, \ea\,
where $\Box_{\text{3d}} = - (\partial_0)^2 + (\partial_1)^2 + (\partial_2)^2$.  We also used that 
\ba
2\phi=\tfrac 12 \tilde h_{00}, \qquad \tilde h=\tilde h_{\mu\nu}\eta^{\mu\nu}=-\tilde h_{00}\,.
\ea
By tracing $h_{\bar\mu\bar \nu}$ with $\eta^{\bar\mu\bar\nu}$ we learn
\ba
\Box_{\text{3d}}E+6\Psi=2\phi\label{otwo}\,.
\ea
From \eqref{oone} and \eqref{otwo} we infer that $\Box_{\text{3d}}F_{\bar \nu}=4\partial_{\bar \nu}\Psi$, which given the transversality of $F_{\bar \nu}$ implies $\Box_{\text{3d}}\Psi=0$. However, given the spherical symmetry of the problem we are led to conclude $\Psi=0$, and $F_{\bar\nu}=0$.

Thus, the $\tmop{SO} (1, 2)$ decomposition becomes
\begin{eqnarray}
  h_{\mu \nu} & = & \left(\begin{array}{cc}
     \partial_{\bar{\mu}} \partial_{\bar{\nu}} E + f_{\bar{\mu} \bar{\nu}} &
    0\\
    0 & 2 \phi
  \end{array}\right)\,.
\end{eqnarray}
Next we solve for $E$ and $f_{\bar{\mu} \bar{\nu}}$ by
further assuming $E$ is in the form of a spherical wave.  We can verify our
assumption later by checking the consistency of our solution. This method
works when the solution is supposed to be unique.  With this assumption for $E$ and keeping everything to
leading order in $1/r$, we have
\begin{eqnarray}
  \Box_{\text{3d}} E \simeq (n_3)^2 \omega^2 E
   \simeq  \frac{1}{2}  \tilde{h}_{00}\,.
\end{eqnarray}
Thus, to leading oder in $1/r$
\begin{eqnarray}
  E & = & \frac{1}{2 (n_3)^2 \omega^2}  \tilde{h}_{00}\,,\nn\\
  f_{\bar{\mu} \bar{\nu}} & = & h_{\bar{\mu} \bar{\nu}} - 
  \partial_{\bar{\mu}} \partial_{\bar{\nu}} E
   =  h_{\bar{\mu} \bar{\nu}} + \frac{n_{\bar{\mu}} n_{\bar{\nu}}}{2
  (n_3)^2}  \tilde{h}_{00}\,,
\end{eqnarray}
where we have defined 
\ba
n_{\bar{\mu}} = (- 1, n_1, n_2)\,.
\ea
Since $(n_1)^2+(n_2)^2+(n_3)^2=1$ we have
\ba
\eta^{\bar\mu\bar\nu} n_{\bar\mu} n_{\bar\nu}=-(n_3)^2\,.
\ea
We are now checking that the gauge-invariant tensor fluctuation $f_{\bar{\mu} \bar{\nu}}$ is indeed transverse and traceless.
For tracelessness, we have
\begin{eqnarray}
  \eta^{\bar{\mu} \bar{\nu}} f_{\bar{\mu} \bar{\nu}} & = & \eta^{\bar{\mu}
  \bar{\nu}} h_{\bar{\mu} \bar{\nu}} + \frac{\eta^{\bar{\mu} \bar{\nu}}
  n_{\bar{\mu}} n_{\bar{\nu}}}{2 (n_3)^2}  \tilde{h}_{00}\nn\\
  & = & \frac{1}{2}  \tilde{h}_{00} - \frac{1}{2}  \tilde{h}_{00}
   =  0\,.
\end{eqnarray}
For transversality, 
\begin{eqnarray}
  \partial^{\bar{\mu}} f_{\bar{\mu} \bar{\nu}} & = & \partial^{\bar{\mu}}
  h_{\bar{\mu} \bar{\nu}} + \frac{n_{\bar{\mu}} n_{\bar{\nu}}}{2 (n_3)^2}
  \partial^{\bar{\mu}} \tilde{h}_{00}\nn\\
  & = & - n^{\bar{\mu}}  \dot{h}_{\bar{\mu} \bar{\nu}} - \frac{n_{\bar{\mu}}
  n_{\bar{\nu}}}{2 (n_3)^2} n^{\bar{\mu}}  \dot{\tilde{h}}_{00}\nn\\
  & = & - \dot{h}_{0 \bar{\nu}} - n^{\bar{i}}  \dot{h}_{\bar{i} \bar{\nu}} +
  \frac{1}{2} n_{\bar{\nu}}  \dot{\tilde{h}}_{00}\label{transvfbm}
\end{eqnarray}
must vanish.
Substituting  $\bar{\nu} = 0$ in \eqref{transvfbm} yields
  \begin{eqnarray}
    \partial^{\bar{\mu}} f_{\bar{\mu} 0} & = & - \dot{h}_{00} - n^{\bar{i}} 
    \dot{h}_{\bar{i} 0} + \frac{1}{2} n_0  \dot{\tilde{h}}_{00\nn}\\
    & = & - \frac{1}{2}  \dot{\tilde{h}}_{00} - n^{\bar{i}} 
    \dot{\tilde{h}}_{0 \bar{i}} - \frac{1}{2}  \dot{\tilde{h}}_{00}
     =  0\,,
  \end{eqnarray}
where in the last step we used the transversality of  $\tilde h_{\mu\nu}$ and the spherical wave nature of the fluctuations.  Similarly, if $\bar{\nu} = \bar{j}$ we find
  \begin{eqnarray}
    \partial^{\bar{\mu}} f_{\bar{\mu} \bar{j}} & = & - \dot{h}_{0 \bar{j}} -
    n^{\bar{i}}  \dot{h}_{\bar{i} \bar{j}} + \frac{1}{2} n_{\bar{j}} 
    \dot{\tilde{h}}_{00}\nn\\
    & = & - \dot{\tilde{h}}_{0 \bar{j}} - n^{\bar{i}}  \left(
    \dot{\tilde{h}}_{\bar{i} \bar{j}} + \frac{1}{2} \delta_{\bar{i} \bar{j}} 
    \dot{\tilde{h}}_{00} \right) + \frac{1}{2} n_{\bar{j}} 
    \dot{\tilde{h}}_{00}
    =  0\,.
  \end{eqnarray}
This concludes the check on our asymptotic solution for $f_{\bar\mu\bar\nu}$. 

We  compute next  the
gauge-invariant scalar $\Phi$. We find that it vanishes
\begin{eqnarray}
  \Phi =  \phi - \partial_3 (B - \tfrac 12\partial_3 E)
   =  \phi +\tfrac 12 (\partial_3)^2 E
   =  0\,.
\end{eqnarray}
We see that the relation which the two gauge-invariant scalars obey in vacuum 
$\Phi = - \Psi$ is satisfied by our asymptotic solution.
The vanishing of scalar and vector modes may not be a coincidence but could be a
general feature of uncompactified flat spacetime. Since in the analysis performed in this appendix the
extra dimension ($x^3$) is non-compact,  the spectrum of Fourier modes of
the gauge-invariant fluctuations is continuous. Because the scalar and vector modes
consist of only zero modes of the 3d d'Alembertian,  they vanish in such cases.

At last,  we can compute the radiated power, substituting the gauge-invariant fluctuations in the formula for the gravitational energy-momentum tensor.  We need ${\cal T}_{0\bar i}$ and  ${\cal T}_{03}$ which can be easily extracted from \eqref{2nd} substituting
\ba
h_{\mu\nu}\bigg|_{g.i.}=\left(\begin{array}{cc}f_{\bar\mu\bar\nu}&0\\0&0\end{array}\right).
\ea
To compute the radiated power we start by integrating over a 3d sphere as in \eqref{radpow} and use the same simplifications of turning spatial-derivatives into time-derivatives (to leading order in $1/r$) when acting on spherical waves at spatial infinity.  Explicitly,  we write:
\begin{eqnarray}
  \langle P \rangle
& = & \frac{1}{T} \int_0^T \tmop{dt} \int d \Omega_2 R_{\infty}^2
  n^{{i}} \mathcal{T}_{0 {i}} |_{r = R_{\infty}}\nn\\
  & = & \frac{1}{T} \int_0^T \tmop{dt} \int d \Omega_2 R_{\infty}^2 
  \frac{1}{4} \dot{f}_{\bar{\mu} \bar{\nu}}  \dot{f}^{\bar{\mu}
  \bar{\nu}}\nn\\
  & = & \frac{1}{4 T} \int_0^T \tmop{dt} \int d \Omega_2 R_{\infty}^2
  \left( \dot{h}_{\bar{\mu} \bar{\nu}} + \frac{n_{\bar{\mu}} n_{\bar{\nu}}}{2
  (n_3)^2}  \dot{\tilde{h}}_{00} \right)^2\nn\\
  & = & \frac{1}{4 T} \int_0^T \tmop{dt} \int d \Omega_2 R_{\infty}^2
  \left( \dot{\tilde{h}}_{\bar{\mu} \bar{\nu}} + \frac{1}{2} \eta_{\bar{\mu}
  \bar{\nu}}  \dot{\tilde{h}}_{00} + \frac{n_{\bar{\mu}} n_{\bar{\nu}}}{2
  (n_3)^2}  \dot{\tilde{h}}_{00} \right)^2\nn\\
  & = & \frac{1}{4 T} \int_0^T \tmop{dt} \int d \Omega_2
  R_{\infty}^2
   \left( \dot{\tilde{h}}_{\bar{\mu} \bar{\nu}} 
  \dot{\tilde{h}}^{\bar{\mu} \bar{\nu}} + \frac{3}{4}  \dot{\tilde{h}}_{00} 
  \dot{\tilde{h}}_{00} + \frac{1}{4}  \dot{\tilde{h}}_{00}
  \dot{\tilde{h}}_{00} + \eta^{\bar{\mu} \bar{\nu}} 
  \dot{\tilde{h}}_{\bar{\mu} \bar{\nu}}  \dot{\tilde{h}}_{00} +
  \frac{n^{\bar{\mu}} n^{\bar{\nu}}}{(n_3)^2}  \dot{\tilde{h}}_{\bar{\mu}
  \bar{\nu}}  \dot{\tilde{h}}_{00} - \frac{1}{2}  \dot{\tilde{h}}_{00} 
  \dot{\tilde{h}}_{00} \right)\nn\\
  & = & \frac{1}{4 T} \int_0^T \tmop{dt} \int d \Omega_2 R_{\infty}^2 
  \left( \frac{1}{2}  \dot{\tilde{h}}_{00}  \dot{\tilde{h}}_{00} - 2
  \dot{\tilde{h}}_{0 \bar{i}}  \dot{\tilde{h}}_{0 \bar{i}} +
  \dot{\tilde{h}}_{\bar{i} \bar{j}}  \dot{\tilde{h}}_{\bar{i} \bar{j}}
  \right)\nn\\
  & = & \frac{R_{\infty}^2}{20 G} \langle \dot{\tilde{h}}_{\bar{i} \bar{j}} 
  \dot{\tilde{h}}_{\bar{i} \bar{j}} \rangle= \frac{32}{5}G \mu^2 \Omega^6 (r_{12})^4\,,
\end{eqnarray}
where we used the transversality of $\tilde h_{\mu\nu}$ and that the fluctuations are spherical waves to set $n^{\bar \mu} \dot{\tilde h}_{\bar \mu\bar \nu}=0$. In the last step we reintroduced the dependence on Newton's constant (recall that we have been working with $8\pi G=1$). We have thus recovered a well known 4d result (see for example footnote 10, with $D=4$, leading further to eqn. (6.11) in \cite{Du:2020rlx}).

\section{The $SO(1,3)$ SVT modes sourced by a binary in 5d flat space with compact $x^5$ and the luminosity of the gravitational waves}
\label{allmodescompact}

In this appendix we aim to recover the power radiated away by gravitational waves in a 5d flat space, with one compact dimension $x^5\sim x^5+l$ \cite{Du:2020rlx}. As discussed in \cite{Du:2020rlx},  for a small extra dimension,  the contribution from the 5d graviton modes with $p_5\neq0$ can be safely ignored far away from the sources, since it is exponentially suppressed.  
This fact can be easily understood from a 4d perspective where these modes appear massive, with the 4d mass proportional to the $p_5$ momentum. Thus, the radiated power receives its dominant contribution from 5d graviton modes with $p_5=0$.  This translates in 5d fluctuations which are independent of $x^5$.  Far away from a binary source, the metric fluctuations are given in the equation (5.24) in \cite{Du:2020rlx}, and the radiated power, which was computed using Isaacson's averaging scheme, is given in the equation (6.8) and (6.10) in \cite{Du:2020rlx}.

  Given the symmetry of the problem, we proceed with performing an $SO(1,3)$ SVT decomposition of the metric fluctuations. 
\begin{eqnarray}
  h_{\mu \nu} & = & 2 \psi \eta_{\mu \nu} + \partial_{\mu} \partial_{\nu} E +
  \partial_{\mu} F_{\nu} + \partial_{\nu} F_{\mu} + f_{\mu \nu}\,,\nn\\
h_{\mu 5}&=&\partial_\mu B+S_\mu\,,\nn\\
h_{55}&=&2\phi\,.\label{5dsvt}
\end{eqnarray}
The analog of the 4d trace-reversed fluctuations in 5d is $\tilde h_{MN}$ where
\begin{eqnarray}
  \tilde{h}_{M N} & = & h_{M N} - \frac{1}{2} \eta_{M N} h\,, \qquad h=h_{MN} \eta^{MN}\,,\\
  h_{M N} & = & \tilde{h}_{M N} - \frac{1}{3} \eta_{M N}\tilde{h}\,\qquad \tilde h=\tilde h_{MN} \eta^{MN} \,.
\end{eqnarray}
The 5d solution in   \cite{Du:2020rlx} is of the form
\begin{eqnarray}
  \tilde{h}_{M N} & \sim & \left(\begin{array}{cc}
    \tilde{h}_{\mu \nu} & 0\\
    0 & 0
  \end{array}\right) \sim
 \frac{3}{4} \text{} \left(\begin{array}{cc}
    \tilde{h}^{(4 d)}_{\mu \nu} & 0\\
    0 & 0
  \end{array}\right)\,,
\end{eqnarray}
where we used the squiggle line to  indicate that we only take into account components that
are explicitly time dependent and drop the static (Coulombic) metric fluctuation,  just as we did in the previous Appendix  \ref{only tensor for non-compact geometry}. The time-independent terms are irrelevant in the computation of the radiated power.

Assuming no $x^5$-coordinate dependence, the harmonic gauge condition
\begin{eqnarray}
  \partial^M \tilde{h}_{M N} & = & 0
\end{eqnarray}
reduces to
\begin{eqnarray}
  \partial^{\mu} h_{\mu \nu} & = & \frac{1}{2} \partial_{\nu} h\,.
\end{eqnarray}

After comparing \eqref{5dsvt}  with the solution in   \cite{Du:2020rlx} we identify 
\ba
\Phi=\phi=\tfrac 12 h_{55}=-\tfrac 16\tilde h\,.
\ea
The harmonic gauge condition becomes
\begin{eqnarray}
  \partial_{\nu} \left( \frac{1}{2} \Box_{\text{4d}} E - \frac{1}{2} h_{55} - 2 \psi
  \right) +\Box_{\text{4d}} F_{\nu} & = & 0\,.\label{harm5d}
\end{eqnarray}

Next,  while remaining in the harmonic gauge,  we use the residual gauge freedom to set $\Box_{\text{4d}} E$ to be 0.  We use the gauge
parameter
\ba
  \xi_M & = & (\xi_0, 0, 0, 0, 0)\,.
\ea
Since the harmonic gauge constrains $\xi^M$ by $\Box_{\text{4d}}\xi^M=0$,  then far away from the sources we take
with $\xi_0$  to be of the form of a (4d) spherical wave.
After performing  this gauge transformation, the new metric is $h_{MN}^{\tmop{new}}$:
\begin{eqnarray}
  h^{(\tmop{new})}_{M N} & = & h_{M N} + \partial_M \xi_N + \partial_N \xi_M\,.
\end{eqnarray}
We consider the trace of the 4d part of the metric perturbation
\begin{eqnarray}
  \eta^{\mu \nu} h^{(\tmop{new})}_{\mu \nu} & = & \eta^{\mu \nu} h_{\mu \nu} -
  2 \partial_0 \xi_0
  =  8 \psi^{(\tmop{new})} +\Box_{\text{4d}} E^{(\tmop{new})}\label{xi0eqn}
\end{eqnarray}
and we require that $\Box E^{(\tmop{new})} = 0$.  We recall that since $\psi$  is gauge-independent then $\psi^{(\tmop{new})}=\psi=\Psi$.   We use the vacuum equations for the scalar fluctuations to relate $\Psi$ and $\Phi$: $\Psi=-(1/2) \Phi$ and solve for the gauge-parameter from \eqref{xi0eqn}
\begin{eqnarray}
  2 \partial_0 \xi_0 & = & \eta^{\mu \nu} h_{\mu \nu} + 2 h_{55}
   =  \eta^{\mu \nu} \tilde{h}_{\mu \nu} + 2 \tilde{h}_{55} - 2 \tilde{h}=-\tilde h\,.
\end{eqnarray}

We proceed to compute the gauge-transformed metric perturbation:
\begin{eqnarray}
  h^{(\tmop{new})}_{00} & = & h_{00} + 2 \partial_0 \xi_0
   =   \tilde{h}_{00} - \frac{2}{3}  \tilde{h}\,,\nn\\
  h_{0 i}^{(\tmop{new})} & = & h_{0 i} + \partial_i \xi_0
   =  \tilde{h}_{0 i} + \frac{n_i}{2} \tilde{h}\,,\nn\\
  h_{i j}^{(\tmop{new})} & = & h_{i j}
   =  \tilde{h}_{i j} - \frac{1}{3} \delta_{i j}  \tilde{h}\,,\nn\\
  h^{(\tmop{new})}_{5 \mu} & = & 0\,,\nn\\
  h^{(\tmop{new})}_{55} & = & h_{55}
   =  - \frac{1}{3} \tilde{h}\,.
\end{eqnarray}
Now, we have the gauge invariant pieces
\begin{eqnarray}
  \Phi & = & - \frac{1}{6} \tilde{h}=\frac 16 \tilde h_{00}\,,\\
  \Psi & = & \frac{1}{12}  \tilde{h}=-\frac 1{12} \tilde h_{00}\,,\\
  {\bf S}_{\mu} & = & 0\,,\\
  f_{\mu \nu} & = & h_{\mu \nu}^{(\tmop{new})} - \frac{1}{6}  \tilde{h}
  \eta_{\mu \nu}\,.\label{gaugeinv5dcomp}
\end{eqnarray}
We can explictly check that $f_{\mu\nu}$ is indeed transverse and traceless\footnote{Since we are looking at zero modes, i.e.  gravitational waves with a null 4-momentum, the results of Appendix \ref{ZM} apply.  The tensor mode $f_{\mu\nu}$ given in \eqref{gaugeinv5dcomp} contains a 4d transverse traceless gauge-dependent term $\partial_\mu F_\nu+\partial_\nu F_\mu$, with $F_\mu$ such that $\Box_{\text{4d}} F_\mu=0$ which follows from \eqref{harm5d}.  To obtain the gauge-invariant tensor, which is transverse and traceless in a 3d sense as explained in Appendix \ref{ZM}, we ought to remove this term. Nonetheless in computing the radiated power such terms are harmless and yield no contribution, so we are free to leave them packaged in $f_{\mu\nu}$ given in \eqref{gaugeinv5dcomp}.}.
Finally we can compute the power radiated by gravitational waves far  away, at a distance $R_\infty$,  from a binary source using \eqref{5dradpow}:
\begin{eqnarray}
  \langle P \rangle & = & \frac{l R_\infty^2}{T}\int_0^T   dt
\int {d\Omega_2} \left( 6 \dot{\Psi}  \dot{\Psi}
  + \frac{1}{4} \dot{f}_{\mu \nu}  \dot{f}^{\mu \nu} \right)\\
  & = &  \frac{l R_\infty^2}{4T}\int_0^T   dt
\int {d\Omega_2} \bigg[\left( \frac{4\cdot 6}{144} \,
  \dot{\tilde{h}}_{00}  \dot{\tilde{h}}_{00} \right)
 +\left( \dot{\tilde{h}}_{00} -
  \frac{1}{2}  \dot{\tilde{h}} \right) \left( \dot{\tilde{h}}_{00} -
  \frac{1}{2}  \dot{\tilde{h}} \right)\nn\\
  &  & - 2 \left(
  \dot{\tilde{h}}_{0 i} + \frac{n_i}{2}  \dot{\tilde{h}} \right) \left(
  \dot{\tilde{h}}_{0 i} + \frac{n_i}{2}  \dot{\tilde{h}} \right) 
+ \left( \dot{\tilde{h}}_{i j} -
  \frac{1}{2} \delta_{i j}  \dot{\tilde{h}} \right) \left( \dot{\tilde{h}}_{i
  j} - \frac{1}{2} \delta_{i j}  \dot{\tilde{h}} \right)\bigg]\nn\\
  & = & \frac{l R_\infty^2}{4T}\int_0^T   dt
\int {d\Omega_2}\bigg[
+ \left( \frac{1}{6} +
  \frac{9}{4} \right)  \dot{\tilde{h}}_{00}  \dot{\tilde{h}}_{00} \nn\\
  &  & + \left( - 2 \dot{\tilde{h}}_{0
  i} \dot{\tilde{h}}_{0 i} - 2 \dot{\tilde{h}}_{00}  \dot{\tilde{h}}_{00} -
  \frac{1}{2}  \dot{\tilde{h}}_{00}  \dot{\tilde{h}}_{00} \right)
 + \left( \dot{\tilde{h}}_{i j}
  \dot{\tilde{h}}_{i j} + \frac{3}{4}  \dot{\tilde{h}}_{00} 
  \dot{\tilde{h}}_{00} \right)\bigg]\nn\\
  & = &  \frac{l R_\infty^2}{4T}\int_0^T   dt
\int {d\Omega_2}   \left( \frac{2}{3} 
  \dot{\tilde{h}}_{00}  \dot{\tilde{h}}_{00} - 2 \dot{\tilde{h}}_{0 i} 
  \dot{\tilde{h}}_{0 i} + \dot{\tilde{h}}_{i j}  \dot{\tilde{h}}_{i j}
  \right)\nn\\
  & = &  \frac{19 \,l \,R_\infty^2}{360 \,G_{\text{5d}}} \langle \dot{\tilde{h}}_{i j} 
  \dot{\tilde{h}}_{i j} \rangle\,,
\end{eqnarray}
where we recall that $l$ is the length of the compact dimension and where in the last step we have restored the dependence on the 5d gravitational constant $G_\text{5d}$ and used the transversality of the trace-reversed metric $\tilde h_{\mu\nu}$.  After accounting for an overall negative sign which we introduced in our earlier definition of the radiated power \eqref{radpow}, we have thus
recovered the previous 5d result given in the equation (6.8) in \cite{Du:2020rlx}.  We also notice  that the contributions from scalar 
and tensor fluctuation match the corresponding parts in Einstein-Maxwell-dilaton
theory respectively \cite{Julie:2017rpw}.

\end{document}